\documentclass{article}

\usepackage[preprint]{neurips_2025}

\usepackage[utf8]{inputenc} 
\usepackage[T1]{fontenc}    
\usepackage{hyperref}       
\usepackage{url}            
\usepackage{booktabs}       
\usepackage{amsfonts}       
\usepackage{nicefrac}       
\usepackage{microtype}      
\usepackage{xcolor}         
\usepackage{multirow}
\usepackage{graphicx}
\usepackage{todonotes}
\usepackage{pifont}
\usepackage{tcolorbox}
\usepackage[normalem]{ulem}
\usepackage{tabularx}  
\usepackage{placeins}

\usepackage{enumitem}

\title{Linguistic properties and model scale in brain encoding: from small to compressed language models}

\author{}

\author{
  Subba Reddy Oota$^{1}$, 
  Vijay Rowtula$^2$, 
  Satya Sai Srinath Namburi$^3$, 
  Khushbu Pahwa$^{4}$, \\
  \textbf{Anant Khandelwal}$^{5}$,
  \textbf{Manish Gupta}$^6$, 
  \textbf{Tanmoy Chakraborty}$^7$, 
  \textbf{Bapi S. Raju}$^{2}$ \\
  $^1$TU Berlin, Germany, $^2$IIIT-Hyderabad, India, $^3$GE HealthCare, $^4$AWS AI Labs, Amazon \\ 
  $^5$Microsoft Research, Bangalore, India, $^6$Microsoft, Hyderabad, India, $^7$IIT Delhi, India\\
  \texttt{\small subba.reddy.oota@tu-berlin.de, vijay.r@research.iiit.ac.in, namburisrinath@gmail.com} \\
  \texttt{\small \{anantk, gmanish\}@microsoft.com, tanchak@iitd.ac.in, raju.bapi@iiit.ac.in}
}

\begin{document}

\maketitle

\begin{abstract}
Recent work has shown that scaling large language models (LLMs) improves their alignment with human brain activity, yet it remains unclear what drives these gains and which representational properties are responsible. Although larger models often yield better task performance and brain alignment, they are  increasingly difficult to analyze mechanistically.
This raises a fundamental question: \emph{what is the minimal model capacity required to capture brain-relevant representations?}
To address this question, we systematically investigate how constraining model scale and numerical precision affects brain alignment.
We compare full-precision LLMs, small language models (SLMs), and compressed variants (quantized and pruned) by predicting fMRI responses during naturalistic language comprehension.
Across model families up to 14B parameters, we find that 3B SLMs achieve brain predictivity indistinguishable from larger LLMs, whereas 1B models degrade substantially, particularly in semantic language regions.
Brain alignment is remarkably robust to compression: most quantization and pruning methods preserve neural predictivity, with GPTQ as a consistent exception.
Linguistic probing reveals a dissociation between task performance and brain predictivity: compression degrades discourse, syntax, and morphology, yet brain predictivity remains largely unchanged. 
Overall, brain alignment saturates at modest model scales and is resilient to compression, challenging common assumptions about neural scaling and motivating compact models for brain-aligned language modeling.
\end{abstract}

\section{Introduction}
\label{sec:introduction}
Transformer-based language models (e.g., GPT*, BERT), although trained only on text, predict human brain activity to a remarkable degree, capturing neural responses during natural language comprehension across diverse cortical regions~\citep{toneva2019interpreting,schrimpf2021neural,goldstein2022shared,oota2022neural,lamarre2022attention,caucheteux2020language,antonello2021low,tuckute2022many,oota2024speech}. 
Recent work has further shown that this alignment improves as models scale, suggesting neural scaling laws analogous to those observed in language modeling~\citep{kaplan2020scaling,hoffmann2022training,li2024datacomp,matsuyama2023applicability,antonello2024scaling,alkhamissi2025language}.
However, these studies have not examined why scaling improves alignment or which representational properties are responsible.
Moreover, scaling comes at a steep computational cost~\citep{faizllmcarbon,diaz2024scaling,villalobos2024position}, making mechanistic analysis and controlled comparisons increasingly difficult.
This raises a fundamental question: \emph{if the human language system is compact and efficient, what model capacity is actually necessary to predict its neural responses?}

\begin{figure*}[t]
    \centering
    \includegraphics[width=\linewidth]{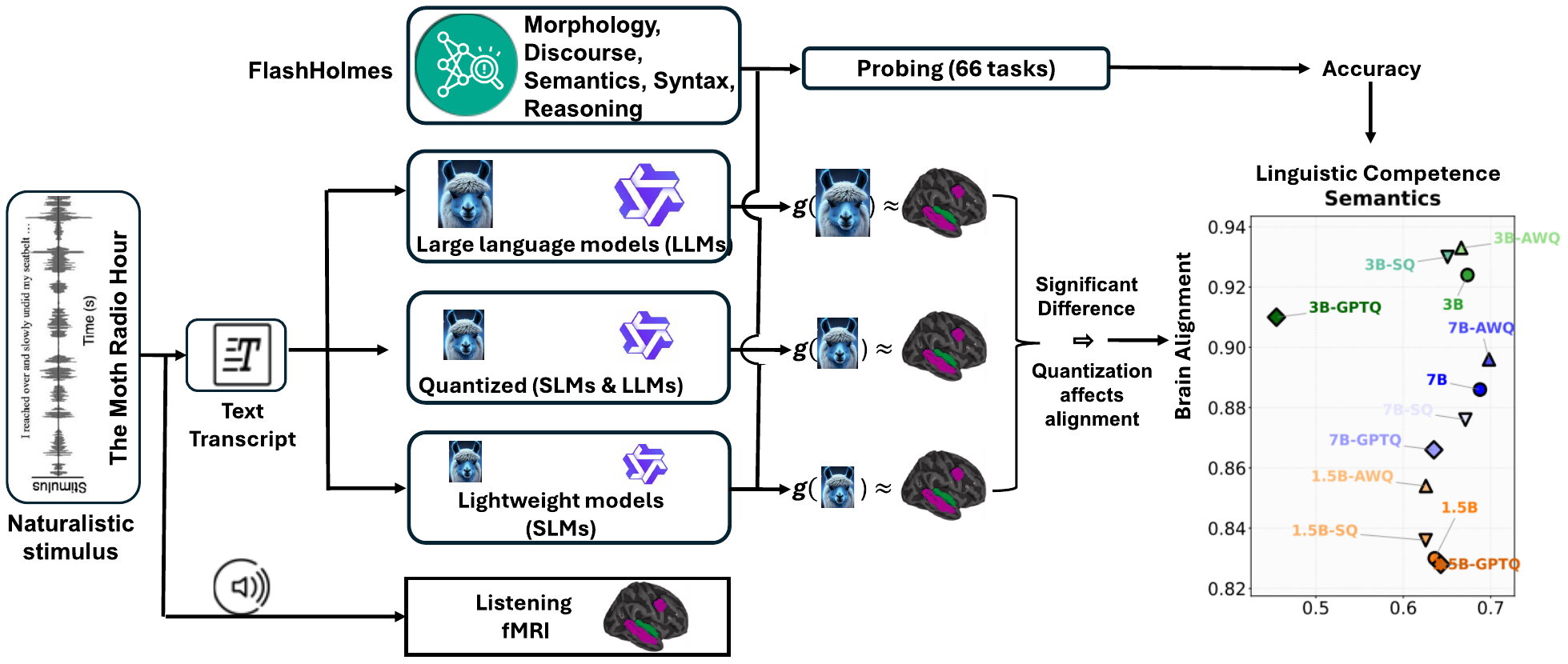}
    \caption{\textbf{Does linguistic competence drive brain–model alignment?} Participants listened to naturalistic English narratives while fMRI responses were recorded. Language models such as large, small, and their compressed variants, processed the same transcripts, and their internal representations were mapped to brain activity to quantify alignment. In parallel, the models were evaluated on the FlashHolmes benchmark to measure linguistic competence across morphology, syntax, semantics, discourse, and reasoning. By jointly comparing brain alignment and task performance across model scale and compression, we test whether reductions in linguistic competence induced by compression systematically degrade brain predictivity.}
    \label{fig:workflow}
\end{figure*}

Within NeuroAI, a growing body of work examines how language is processed in the brain and how these processes compare to representations learned by language models.
Although larger models often improve task performance and brain alignment, they pose increasing challenges for mechanistic analysis. 
We therefore use model size and numerical precision as controlled constraints to probe how representational changes impact brain alignment. 
In the AI community, two widely adopted strategies provide such controls: (i) compressing pretrained LLMs via pruning, distillation, or quantization, and (ii) training small language models (SLMs) that achieve competitive performance with far fewer parameters~\citep{touvron2023LLaMA,yang2024qwen2,guo2025deepseek}.

Despite these advances, studies on brain encoding have overwhelmingly focused on large, full-precision models~\citep{antonello2024scaling,alkhamissi2025language}, while compression is typically evaluated using engineering benchmarks rather than neuroscientific criteria. 
As a result, it remains unclear whether small or compressed models preserve the brain-relevant representational geometry needed for accurate brain predictivity, making it uncertain whether they can be used as controlled probes of brain–model correspondence.
Moreover, SLMs and compression methods may selectively impair linguistic competencies (e.g., discourse or syntax) in ways that are not reflected in aggregate task accuracy, raising the possibility of dissociation between linguistic competence and brain alignment. Hence, we ask the following research questions (RQs):

\noindent (1) What is the minimal model capacity required to achieve brain alignment comparable to larger LLMs, in both brain encoding and decoding settings?\\    
\noindent (2) How do compression methods (quantization and pruning) affect brain alignment for both SLMs and LLMs?\\
\noindent (3) Which linguistic properties are preserved or degraded across small, large, and compressed models, and do these changes correlate with brain alignment?

To address these questions, we systematically investigate how model scale and compression jointly shape brain alignment and linguistic competence. 
Using fMRI recordings collected while participants listened to naturalistic stories from the Moth Radio Hour dataset~\citep{deniz2019representation}, we evaluate language models from three model families (LLaMA-3.2~\citep{touvron2023LLaMA}, Qwen2.5~\citep{yang2024qwen2}, and DeepSeek-R1~\citep{guo2025deepseek}) spanning \(\approx\)1B–14B parameters.
We quantify brain alignment in both (i) \textbf{encoding} (predicting voxel-wise fMRI responses from model representations~\citep{oota2025deep}) and (ii) \textbf{decoding} (reconstructing linguistic representations and downstream text from fMRI~\citep{oota2025deep}), thus showing that 3B models are sufficient not only for voxel-wise prediction, but also for reconstructing semantically coherent text from fMRI.
We compare full-precision models against compressed variants, including multiple post-training quantization schemes (AWQ, GPTQ, SmoothQuant) and unstructured pruning. 
To assess linguistic competence independently of brain predictivity, we benchmark all models on FlashHolmes~\citep{waldis2024holmes}, a streamlined suite probing morphology, syntax, semantics, discourse, and reasoning across 66 linguistic phenomena.

This unified framework enables us to test whether compact and compressed models match large LLMs in brain alignment, identify which linguistic properties are preserved or disrupted, and assess whether these changes impact neural predictivity. 
We evaluate models along two complementary dimensions: voxel-wise fMRI encoding performance and targeted linguistic probing.


Our findings lead to three overarching insights.

    (1) Brain alignment saturates early: across model families, 3B SLMs match 7B–14B LLMs in neural predictivity across the whole brain and all major language-relevant regions, including semantic and integrative cortices, with consistent effects across subjects, whereas 1B models degrade substantially, especially in semantic regions. 
    These results suggest that $\sim$3B parameters are sufficient to reach the saturation regime of brain alignment, while $\sim$1B remains below the required capacity for robust brain alignment. 
    In decoding, 3B SLMs enable stable brain-to-language reconstruction with semantic fidelity comparable to larger models, whereas 1B models show marked degradation.
    
    (2) Brain alignment is remarkably robust to compression: most quantization and pruning methods preserve brain predictivity, except GPTQ, which consistently reduces it and produces widespread voxel-level degradation, especially in semantic brain regions (e.g., angular gyrus). 
    Pruning remains stable up to moderate sparsity levels (10-25\%), beyond which brain alignment degrades sharply, particularly for smaller models.
    
    (3) Linguistic probing reveals a nuanced dissociation: compression disproportionately degrades discourse, syntax, and morphology-related competencies, yet these impairments do not consistently translate into reduced brain alignment, highlighting a divergence between task-level performance and brain-relevant representations.

Together, our results refine previous claims about neural scaling by demonstrating an early saturation of brain alignment at modest model scales and a striking robustness to compression. 
These findings position compact and efficiently compressed language models not only as engineering compromises but as principled and cognitively grounded alternatives for brain-aligned language modeling.

A detailed discussion of related work on brain–language model alignment, neural scaling, and model compression is provided in Appendix~\ref{app:relatedwork}.

\section{Methodology}
\vspace{-0.1cm}

\subsection{Naturalistic Brain Imaging Dataset}
\label{sec:braindataset}
\vspace{-0.1cm}

We use a publicly available fMRI dataset~\citep{deniz2019representation} collected while nine participants listened to narrative stories from the Moth Radio Hour. 
The dataset comprises 3,737 training and 291 testing samples  (TRs: Repetition Time). These were specifically selected for their ability to elicit unique auditory and high-level linguistic responses in the brain. 
Following~\citet{deniz2019representation}, we examine this dataset using the Glasser Atlas multi-modal parcellation of the cerebral cortex, targeting 180 ROIs per hemisphere~\citep{glasser2016multi}.

This includes one early sensory processing region (early auditory) and eight language-relevant regions spanning semantic, syntactic, and discourse-level processing, including the angular gyrus (AG), lateral temporal cortex (ATL and PTL), inferior frontal gyrus (IFG and IFGOrb), middle frontal gyrus (MFG), posterior cingulate cortex (PCC) and dorsomedial prefrontal cortex (dmPFC), based on Fedorenko's language parcels~\citep{milton2021parcellation,desai2022proper}. 
These regions are central to the semantic and integrative effects examined in our analysis.
More details about dataset and ROI functionality are in Appendix~\ref{naturalistic_dataset} and Table~\ref{rois_description}. 

\noindent\textbf{Estimating cross-subject prediction accuracy.}
To account for intrinsic noise in biological measurements, we adapt the method proposed by ~\citet{schrimpf2021neural,oota2024speech,alkhamissi2025language} to estimate the ceiling value for a model's performance for the Subset-Moth-Radio-Hour fMRI dataset. 
Note that the estimated cross-subject prediction accuracy is based on the assumption of a perfect model, which might differ from real-world scenarios, yet offers valuable insights into model's performance.
We present the average cross-subject prediction accuracy across voxels for the \emph{listening fMRI} dataset in Appendix~\ref{cross_subject_prediction}.

\setlength{\tabcolsep}{2pt}
\begin{table}[t]
\caption{Pretrained Transformer-based SLMs and LLMs, and their post-training compressed variants. SQ=SmoothQuant. Config shows \#layers and \#parameters. Other columns show sizes in GBs.} 
\label{neural_models} 
\scriptsize
\centering
\begin{tabular}{p{1cm}|ccccc|ccccc} 
\hline
\multirow{2}{0.8cm}{\textbf{Model Family}} & \multicolumn{5}{c|}{\textbf{SLMs}} & \multicolumn{5}{c}{\textbf{LLMs}} \\
\cline{2-11}
& \textbf{Config} & \textbf{Orig} & \textbf{AWQ} & \textbf{GPTQ}& \textbf{SQ} & \textbf{Config} & \textbf{Orig} & \textbf{AWQ} & \textbf{GPTQ}& \textbf{SQ} \\
\hline
\multirow{2}{0.8cm}{LLaMA\newline3.2} & 16L, 1B&2.47 & 1.56& 1.02&  2.02 & \multirow{2}{*}{28L, 8B} &\multirow{2}{*}{16.1} & \multirow{2}{*}{5.73} & \multirow{2}{*}{5.70}& \multirow{2}{*}{9.08} \\
&28L, 3B&6.4 &3.04& 2.26&  4.40&  &&&& \\
\hline
\multirow{2}{*}{Qwen 2.5} & 28L, 1.5B & 3.1 & 1.61& 1.15&  2.25  & \multirow{2}{*}{28L, 7B} & \multirow{2}{*}{15.1} & \multirow{2}{*}{5.57} & \multirow{2}{*}{5.58}& \multirow{2}{*}{8.67}\\
& 36L, 3B &6.1 & 2.69& 2.10&  4.02&  &&&&\\
\hline
\multirow{2}{*}{DeepSeek} & 28L, 1.5B & 3.55 & 1.62& 1.61&  2.25  & \multirow{2}{*}{28L, 7B} & \multirow{2}{*}{15.23} & \multirow{2}{*}{5.57} & \multirow{2}{*}{5.58}& \multirow{2}{*}{8.71}\\
& 28L, 3B &6.43 & 2.69& 2.37&  4.02&  &&&&\\
\hline
\end{tabular} 
\end{table}

\subsection{SLMs and their larger counterparts}
\label{slms_llms_models}


To investigate whether small and compressed language models align with human language processing in the brain, we consider multiple modern model families spanning scales from 1B to 14B parameters which are publicly available on Huggingface~\citep{wolf2020transformers}, including LLaMA-3.2~\citep{touvron2023LLaMA}, Qwen-2.5~\citep{yang2024qwen2} and DeepSeek-R1~\cite{guo2025deepseek}. 
We evaluate both SLMs; 1B–3B and their larger counterparts (7B–14B), along with post-training compressed variants.
We report the model parameters and layer details in Table~\ref{neural_models}.
All models are base (non–instruction-tuned) checkpoints to avoid task-specific fine-tuning effects. 
For each model, we extract representations from all transformer layers and select the single best-performing layer per model, ensuring comparability across architectures with different depths.

\noindent\textbf{Extracting text representations.}
For text transcripts from the Moth Radio Hour dataset, we follow previous work in extracting hidden-state representations from each layer of the language models for a fixed-length input~\citep{toneva2019interpreting,aw2022training,oota2022joint,oota2024speech}. 
To obtain the stimulus features from these pretrained models, we constrain the tokenizer to use a maximum context of 20 words. 
Given the constrained context length, each word is successively input to the network with at most $C$ (=20) previous words. 
For instance, given a story of $M$ words and considering the context length of 20, while the third word's vector is computed by presenting (w$_1$, w$_2$, w$_3$) as input to the network, the last word’s vector w$_M$ is computed by presenting the network with (w$_{M-20}$, $\dots$, w$_M$). 
The pretrained Transformer model outputs token representations at different layers. 
We use the \#words $\times d$ hidden-state representations from each layer, where $d$ is the model-specific hidden dimension, to obtain word-level representations from each pretrained Transformer language model.

The preprocessing and HRF delays are detailed in Appendix~\ref{naturalistic_dataset}.
Following prior work, we evaluate representations from all layers and report results for the best-performing layer per model, ensuring fair comparison across architectures with differing depths and hidden dimensions.


\subsection{Post-training Compression: Quantization and Pruning}
\label{quantization_methods}
Quantization can dramatically reduce memory usage and accelerate inference by mapping model weights and activations to lower-precision formats such as INT8, INT4, or FP8 arithmetic.  
In this work, our goal is not only to assess efficiency, but also to examine how compression, including quantization and unstructured pruning, impacts linguistic competence and brain alignment.
To achieve this, 
we perform three widely-used quantization techniques: (1) Activation-aware Weight Quantization (AWQ)~\citep{lin2023awq}, which adjusts weight scales using activation statistics to enable highly accurate 4‑bit or 8-bit compression with minimal quality loss; (2) GPTQ~\citep{frantar-gptq}: applies the post‑training, gradient‑guided weight quantization method that delivers near-lossless INT8/INT4 speed‑ups; and (3) SmoothQuant~\citep{liu2024kivi,xiao2023smoothquant}, which jointly quantizes weights and activations by equalizing variance to reduce memory usage and latency while preserving accuracy. 
We provide more details, including model parameters after quantization in Table~\ref{neural_models}. 

\noindent\textbf{Unstructured pruning.}
To study the effect of parameter removal on brain alignment, we apply post-training unstructured magnitude-based pruning. 
Specifically, we prune individual weights with the smallest absolute values (L1 norm) across all linear layers, following standard practice in language model compression. 
We evaluate multiple sparsity levels (10\%, 25\%, and 50\%), without retraining or fine-tuning after pruning, to isolate the effect of parameter removal on representational geometry.



\subsection{Flash-Holmes Benchmark}
\label{holmes_benchmark}
FlashHolmes is a streamlined version of the Holmes benchmark~\citep{waldis2024holmes}, designed to efficiently evaluate the linguistic competence of language models. 
The FlashHolmes benchmark covers nearly 200 probing datasets that span 66 linguistic tasks. 
The linguistic tasks are grouped into five major categories: (1) Morphology (19 tasks, e.g., subject–verb agreement and irregular word forms)~\citep{warstadt2020blimp,huebner2021babyberta}, (2) Syntax (75 tasks, e.g., constituent labeling and filler-gap dependencies)~\citep{conneau2018you,warstadt2020blimp}, (3) Semantics (67 tasks, e.g., semantic role labeling and natural language inference)~\citep{wang2018glue}, (4) Discourse (28 tasks, e.g., coreference resolution and discourse relation prediction)~\citep{webber2019penn}, and (5) Reasoning (19 tasks, e.g., paraphrasticity with negation and antonyms)~\citep{vahtola2022not}. 
Overall, these linguistic tasks allow FlashHolmes to probe a wide spectrum of linguistic phenomena, making it a suitable tool for evaluating both SLMs and LLMs and their quantized variants. 
\section{Experimental Setup}
\label{sec:modelArch}

\noindent\textbf{Models and compression.}
We evaluate three modern Transformer-based model families: Qwen2.5, LLaMA-3.2, and DeepSeek-R1, spanning scales from 1B to 14B parameters. 
For each family, we include small language models (SLMs; 1B–3B) and larger counterparts (7B–14B), all using pretrained base checkpoints.
In addition to full-precision models, we evaluate post-training compressed variants using quantization and pruning. 
All compression methods, sparsity levels, and implementation details are described in Section 3.3. 
Unless otherwise stated, compressed models are evaluated without retraining using the same encoding and probing pipelines as dense models.

\noindent\textbf{Voxel-wise encoding model.} 
To perform voxel-wise encoding, we train an fMRI encoding model using bootstrap ridge regression~\citep{tikhonov1977solutions} to predict the fMRI recording associated with each voxel as a function of the stimulus representations obtained from the language models. 
Before the bootstrap ridge regression, we first z-score each feature channel separately for training and testing. 
This is done to match the features to the fMRI responses, which were also z-scored for training and testing.
Formally, at the time step (t), we encode the stimuli as $X_{t}\in \mathbb{R}^{N \times D}$ and brain region voxels $Y_{t}\in \mathbb{R}^{N \times V}$, where $N$ is the number of training examples, $D$ denotes the dimension of the concatenation of delayed 4 TRs, and $V$ denotes the number of voxels.
To find the optimal regularization parameter for each feature space, we use a range of regularization parameters that is explored using cross-validation. 
The main goal of each fMRI encoding model is to predict brain responses associated with each brain voxel given a stimulus. 
Following prior work, we train encoding models using representations from all layers and report results for the best-performing layer per model, ensuring fair comparison across architectures with different depths.
The detailed hyperparameter settings and statistical significance tests are provided in Appendix~\ref{app:hyperparameters_details} and~\ref{app:statistical_significance}.

\noindent\textbf{Brain decoding.} In addition to encoding, we perform brain decoding experiments that reconstruct linguistic representations and text from fMRI using the same model representations and alignment framework, as discussed in Appendix~\ref{app:decoding_reconstruction}.

\noindent\textbf{Normalized alignment.}
The final measure of a model's performance is obtained by calculating Pearson's correlation between the model's predictions and brain recordings. 
This correlation is then divided by the estimated cross-subject prediction accuracy and averaged across voxels, resulting in a standardized measure of performance referred to as normalized alignment.
For normalized alignment, we restrict analyses to voxels with cross-subject prediction accuracy $\ge$ 0.05, ensuring that comparisons focus on reliably stimulus-driven responses.

\begin{figure*}[t]
    \centering
    \includegraphics[width=\linewidth]{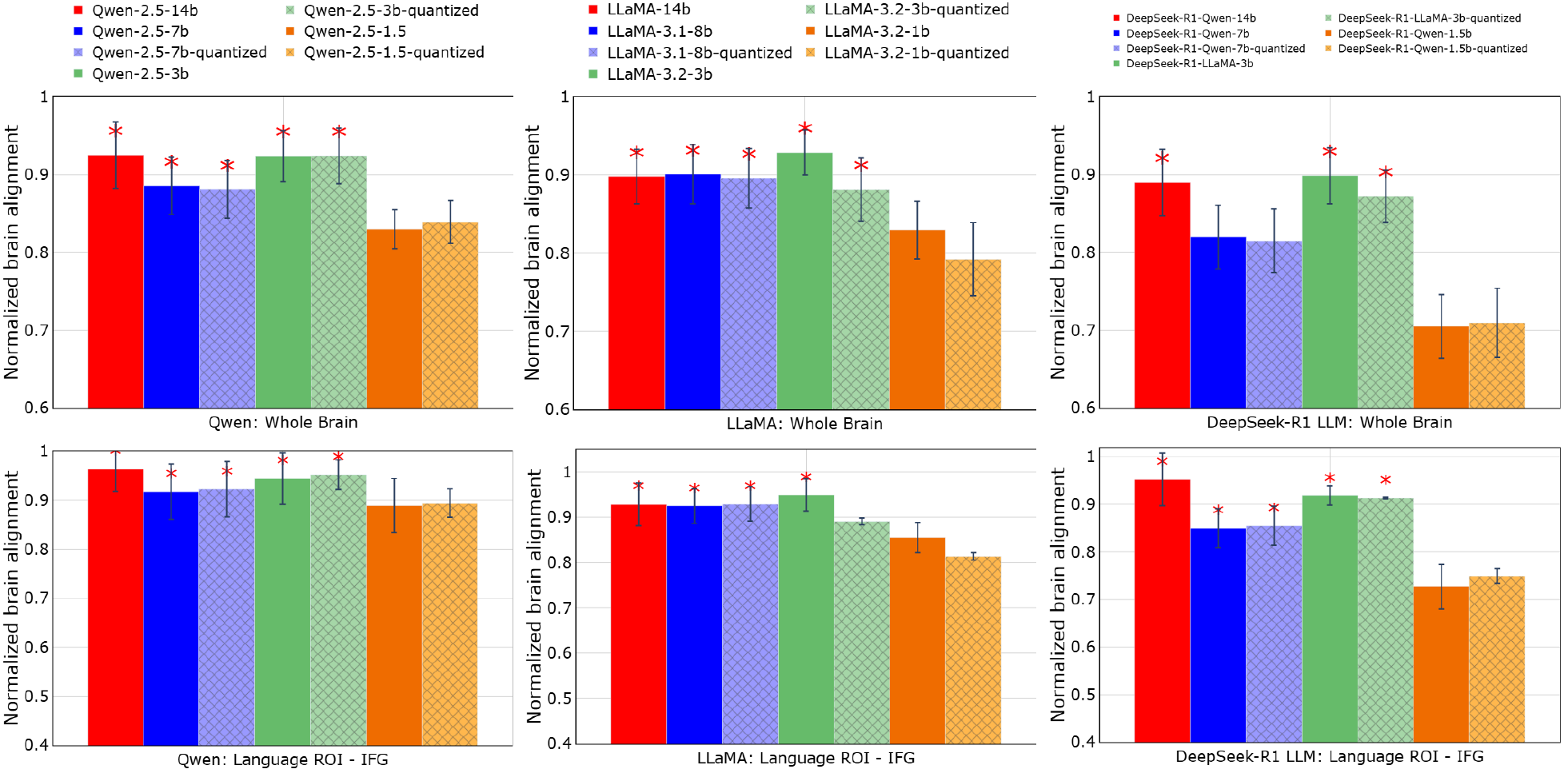}
    \caption{Qwen2.5, LLaMA, and DeepSeek-R1: Normalized brain alignment was computed by averaging across participants, layers, and voxels. \textcolor{red}{Red: }14b, \textcolor{blue}{Blue: }7b, \textcolor{green}{Green: } 3b, \textcolor{orange}{Orange:} 1.5b, Solid: full-precision SLMs/LLMs, Patterned: quantized models. \textcolor{red}{*} at a particular bar indicates that the model's prediction performance is significantly better than 1b/1.5b SLMs. The top row shows whole-brain normalized alignment, while the bottom row focuses on a language-selective ROI (IFG).}
    \label{fig:vem_wholebrain_language}
\end{figure*}

\noindent\textbf{Encoding and probing of Flash-Holmes benchmark.}
To evaluate the linguistic competence of language models, we use the FlashHolmes benchmark~\citep{waldis2024holmes}, as detailed in Section~\ref{holmes_benchmark}. 
For each task in FlashHolmes, we apply classifier-based probing to the internal representations of SLMs, LLMs, and quantized models. 
Our hypothesis is that SLMs or quantized models may show reduced linguistic competence on specific tasks, even if overall brain alignment remains intact. 
In this case, FlashHolmes can reveal which linguistic properties (e.g., syntax or discourse) are disproportionately affected by compression. 
Thus, FlashHolmes plays a complementary but essential role: it allows us to determine whether reductions in linguistic competence under compression correspond to reductions in brain alignment, or whether these two measures dissociate.
To perform probing, following~\citet{waldis2024holmes}, we train a simple linear classifier on the representations obtained from SLMs, LLMs, and quantized models. 
Model performance is then aggregated across tasks within each linguistic category, producing scores that reflect the accessibility of different types of linguistic information in the representations.

Overall, by treating compression as a controlled intervention on internal representations, probing analyses help isolate which linguistic properties are preserved or disrupted without directly manipulating model parameters.

\section{Results}

\subsection*{\textbf{[RQ1]:} Effects on Brain Encoding Performance: 3B SLMs match similar brain alignment as 7B–14B models across model families}

To examine whether SLMs achieve brain encoding performance comparable to larger models, we compare the voxelwise encoding performance of three families of language models (Qwen-2.5, LLaMA-3.2, and DeepSeek-R1), at different scales. For both SLMs and LLMs, we also apply post-training quantization and measure normalized brain predictivity at both the whole-brain level and within language-specific regions. Fig.~\ref{fig:vem_wholebrain_language} shows average normalized brain predictivity across participants and layers. 


\noindent\textbf{Whole-brain analysis.}
Across whole brain (Fig.~\ref{fig:vem_wholebrain_language}, top row), 3B SLMs in all three families (Qwen, LLaMA, DeepSeek) achieve normalized brain alignment comparable to their 7B–14B counterparts, indicating saturation of encoding performance beyond the 3B scale. 
Focusing on Qwen2.5 (Table~\ref{qwen2.5_qunatitative_analysis}), subject-wise paired \(t\)-tests (\(n=9\)) show no reliable difference between 3B and 14B (\(\Delta=0.000\), \(t(8)=-0.03\), \(p=1.0\)), while both 3B and 14B significantly outperform 1.5B (3B vs.\ 1.5B: \(\Delta=0.07\), \(t(8)=4.89\), \(p=0.004\); 14B vs.\ 1.5B: \(\Delta=0.07\), \(t(8)=3.16\), \(p=0.025\)). 
We also observe a small but significant advantage of 3B and 14B over 7B in best-layer alignment (\(\Delta\approx0.04\), \(p\approx0.02\)--\(0.04\)). Overall, these results support our main claim in this regime: scaling beyond \(\sim\)3B yields at most modest gains in brain alignment, whereas \(\sim\)1--1.5B models are reliably worse. 
We observe the same qualitative pattern for LLaMA-3.2 and DeepSeek-R1 (Tables~\ref{llama_quantitative_analysis} and~\ref{deepseek_quantitative_analysis} in Appendix~\ref{app:quantitative_analysis}).

Fig.~\ref{fig:vem_wholebrain_language} aggregates post-training quantization results by averaging across AWQ, GPTQ, and SmoothQuant, and reports whole-brain alignment averaged across the three model families. Overall, quantization largely preserves whole-brain alignment for both LLMs and 3B SLMs, whereas the smallest 1B--1.5B models show a significant drop in alignment across compression settings (\(p<0.01\)). 

\begin{table}[t]
\centering
\scriptsize
\caption{Pairwise differences in Qwen2.5 best-layer encoding scores across 9 subjects. 
For each subject and model, we take the maximum encoding score across evaluated layers 
and then compute paired $t$-tests between models. $\Delta$ is the mean difference A--B over subjects.}
\label{qwen2.5_qunatitative_analysis}
\resizebox{0.9\columnwidth}{!}{\begin{tabular}{|l|c|c|c| l|}
\hline
Comparison (A--B) & $\Delta$ & $t(8)$ & $p$ (two-sided, approx.) & Interpretation \\
\hline
3B -- 14B   & 0.000 & -0.00 & 1.00   & 3B $\approx$ 14B (no difference) \\
3B -- 7B    & 0.028 &  2.43 & 0.06   & 3B $>$ 7B (small effect, trend) \\
3B -- 1.5B  & 0.073 &  4.89 & 0.004  & 3B $>$ 1.5B (clear, significant) \\
14B -- 7B   & 0.028 &  2.02 & 0.10   & 14B $>$ 7B (small effect, n.s.) \\
14B -- 1.5B & 0.073 &  3.16 & 0.025  & 14B $>$ 1.5B (clear, significant) \\
7B -- 1.5B  & 0.045 &  2.67 & 0.045  & 7B $>$ 1.5B (moderate, significant) \\
\hline
\end{tabular}}
\end{table}

\noindent\textbf{Language-ROI analysis.}
Within language-selective regions, 3B SLMs and larger LLMs achieve high alignment that is largely preserved under quantization (Fig.~\ref{fig:vem_wholebrain_language}, bottom). In contrast, 1B--1.5B models show pronounced drops, indicating that sub-3B capacity is insufficient even within core language circuitry. Scale sensitivity is strongest in integrative/semantic regions including angular gyrus, PCC, and dmPFC, and weaker in ATL, PTL, and MFG, where alignment varies comparatively little with model size. Additional ROI results are shown in Figs.~\ref{fig:qwen_merged_quantized}, ~\ref{fig:llama_merged_quantized}, and~\ref{fig:deepseek_merged_quantized}.



\begin{table}[t]
\centering
\caption{Overall performance metrics for brain-to-text decoding.}
\scriptsize
\label{brain_decoding_metrics}
\resizebox{0.65\columnwidth}{!}{\begin{tabular}{|l|c|c|c|c|c|}
\hline
Model           & BLEU-1  & WER    & METEOR & BERT-F1 & Samples \\
\hline
LLaMA-3-8B           & 0.0699  & 5.7839 & 0.0550 & 0.8108  & 784   \\
LLaMA-3.2-3B    & 0.1198  & 4.2237 & 0.1101 & 0.8252  & 784    \\
LLaMA-3.2-1B    & 0.1105  & 4.4869 & 0.0990 & 0.8237  & 784     \\
\hline
\end{tabular}}
\end{table}

\noindent\textbf{Decoding performance: 3B SLMs enable stable brain-to-language reconstruction }

We perform end-to-end text stimulus reconstruction from fMRI brain activity. We follow the BrainLLM methodology inspired from~\citet{ye2025generative}, where we use the same Subset-Moth-Radio-Hour dataset (11 stories) with the same train/test split, where ten stories are used for training and one held-out story is used for generation. Concretely, we train a brain-to-text decoder and report standard text-generation metrics-BLEU-1, WER, METEOR, and BERT-F1-for three models: LLaMA-3-8B, LLaMA-3.2-3B, and LLaMA-3.2-1B (Table~\ref{brain_decoding_metrics}). Across reconstructed segments per model on test dataset, LLaMA-3.2-3B achieves the best performance on all four metrics (BLEU-1 = 0.120, WER = 4.22, METEOR = 0.110, BERT-F1 = 0.825), slightly outperforming LLaMA-3.2-8B and clearly improving over the LLaMA-3.2-1B baseline (BLEU-1 = 0.070, METEOR = 0.055, BERT-F1 = 0.811). These BERT-F1 scores in the 0.81–0.83 range indicate that the decoded text reliably preserves the semantic content of the original stimulus, while BLEU-1 in the 0.07–0.12 range is in line with prior work where exact word-level recovery from fMRI is known to be challenging. We also include qualitative examples comparing ground-truth text and decoded outputs (see Appendix~\ref{app:decoding_reconstruction} Table.~\ref{brain_decoding_examples}). These examples illustrate that the decoder often recovers the overall meaning, emotional tone, and discourse context, even when individual words differ-e.g., reconstructions that correctly express embarrassment, uncertainty, or interactions with children, despite not matching every token verbatim.

\noindent\textbf{Encoding performance on the Reading fMRI dataset.} 
To test generalization across paradigms, we additionally perform voxel-wise encoding on the Subset-Moth-Radio-Hour \emph{Reading} fMRI dataset~\citep{deniz2019representation} (same nine subjects, different task i.e. Reading). Using Qwen2.5 models (1.5B/3B/7B/14B), Fig.~\ref{fig:reading_fmri} shows that 3B SLMs achieve alignment comparable to 7B and 14B, while 1.5B exhibits a clear drop. Overall, the 3B saturation pattern generalizes to the reading setting.

\begin{figure}[t]
    \centering
    \includegraphics[width=0.7\linewidth]
      {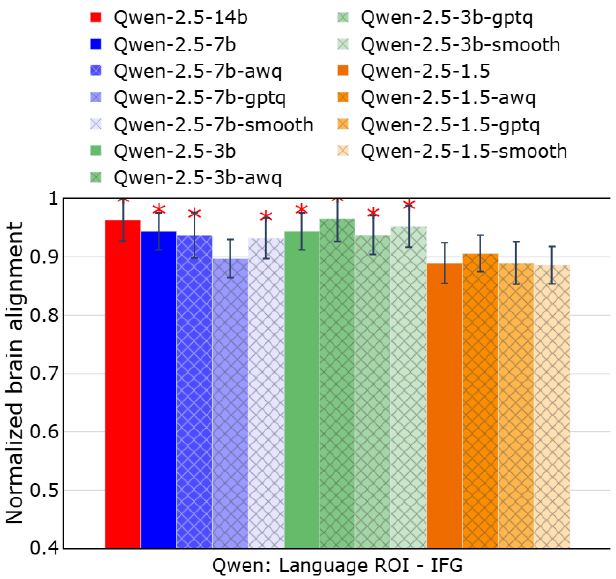}
    \caption{
    Normalized brain alignment averaged across participants and voxels, using the best-performing layer for Qwen2.5 model. \textcolor{red}{Red:} 14b, \textcolor{blue}{Blue: }7b/8b, \textcolor{green}{Green: } 3b, \textcolor{orange}{Orange:} 1.5b, Solid: full-precision SLMs/LLMs, Patterned: quantized models. \textcolor{red}{*} at a particular bar indicates that the model's prediction performance is significantly better than 1b/1.5b SLMs. 
    Plots for other model familes and regions are in Figs.~\ref{fig:qwen_vem_language} and~\ref{fig:llama_vem_language} in Appendix~\ref{app:alignmentLangROIs}.}
    \label{fig:vem_qwen_language_ifg}
\end{figure}

\begin{figure*}[t] 
\centering
\includegraphics[width=\linewidth]{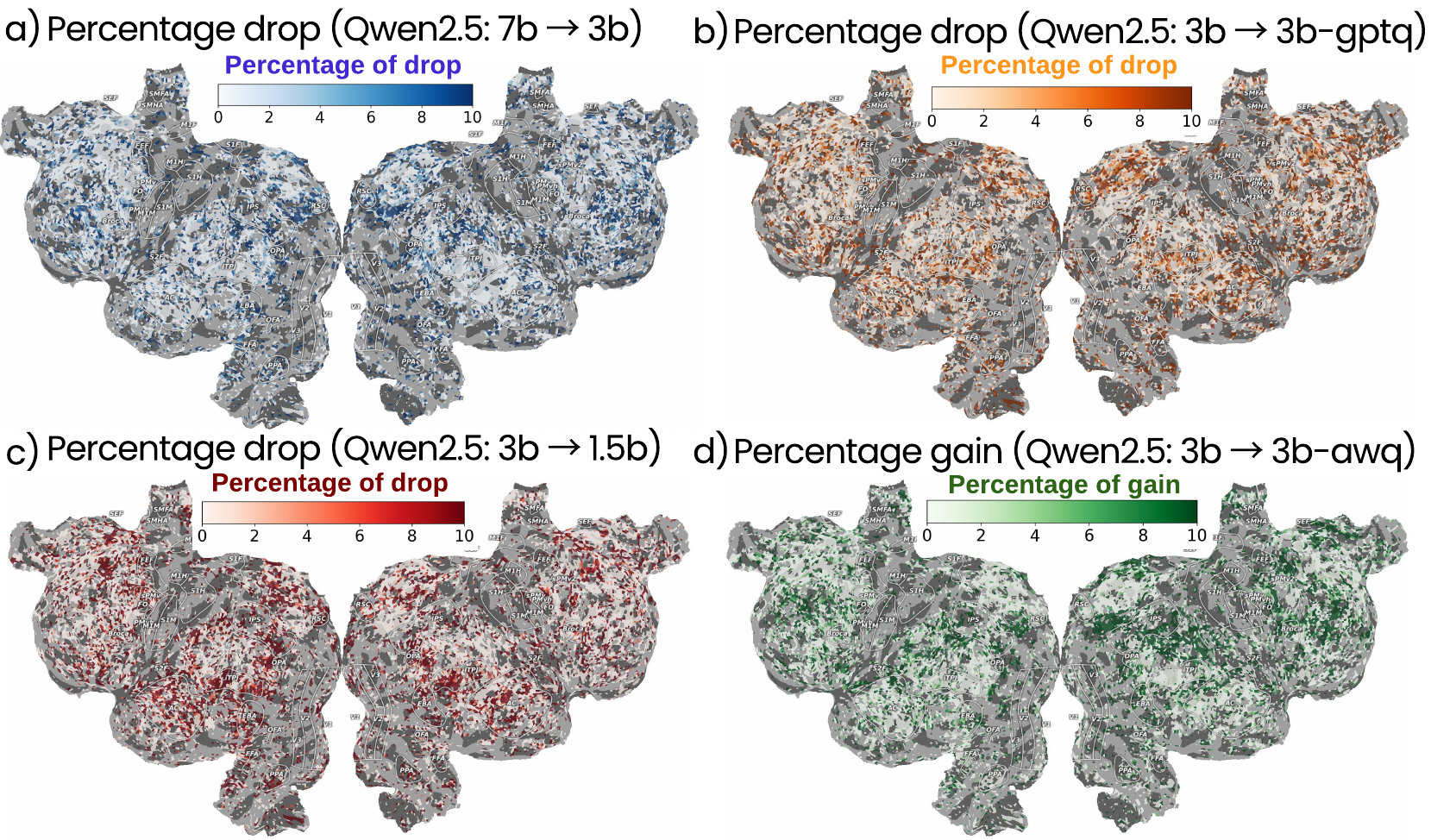}
\caption{Qwen2.5: Percentage change in brain alignment across model scales and quantization methods, shown on the flattened cortical surface of a representative subject (subject-5). Blue, orange, and red voxels indicate regions of information loss ((a) LLMs $\rightarrow$ 3B SLMs, (b) 3B SLMs $\rightarrow$ 3B SLMs GPTQ, (c) 3B SLMs $\rightarrow$ 1.5B SLMs, respectively), (d) while green voxels highlight regions of improvement for 3B SLMs AWQ over 3B SLMs. White voxels denote regions with no change. Results for other participants for Qwen2.5 and LLaMA models are in Appendix~\ref{app:flatMaps}.}
\label{fig:qwen_flatmap_subject05}
\end{figure*}

\begin{table}[t]
\centering
\scriptsize
\caption{Pairwise comparisons of brain-alignment differences across quantization methods for Qwen2.5 model. The Table reports mean differences ($\Delta$), $t$-statistics, and two-sided significance tests for 7B (left), 3B (right), and 1.5B (bottom).}
\label{statistical_results_qwen2.5_main}
\centering
\resizebox{0.5\columnwidth}{!}{\begin{tabular}{|l|c|c|c|}
\hline
Comparison (A–B) & $\Delta$ & $t(8)$ & Sig. \\
\hline
Qwen2.5-7B–AWQ         & -0.020 & -6.10 & $p<0.001$ \\
Qwen2.5-7B–GPTQ        &  0.020 &  6.20 & $p<0.001$ \\
Qwen2.5-7B–SmoothQuant & -0.005 & -3.50 & $p<0.016$ \\
AWQ–GPTQ         &  0.040 &  7.10 & $p<0.001$ \\
AWQ–SmoothQuant  &  0.015 &  4.20 & $p<0.008$ \\
GPTQ–SmoothQuant & -0.025 & -4.90 & $p<0.004$ \\
\hline
\end{tabular}}
\end{table}

\subsection*{\textbf{[RQ2]:} Most quantization and pruning methods preserve brain alignment, except GPTQ,}

\noindent\textbf{Effect of quantization on brain encoding.}
Fig.~\ref{fig:vem_qwen_language_ifg} shows Qwen2.5 encoding in IFG across sizes and quantization methods. At 3B SLMs, post-training quantization (AWQ, GPTQ, SmoothQuant) largely preserves alignment with only marginal changes from FP16, whereas 1B--1.5B models remain under-aligned, reinforcing that their limited capacity, rather than representational redundancy, is the primary bottleneck.
We quantify these effects with subject-wise best-layer scores and paired \(t\)-tests across methods (Table~\ref{statistical_results_qwen2.5_main}). For 7B, AWQ and SmoothQuant significantly outperform FP16 and GPTQ, and GPTQ is significantly worse than FP16. For 3B, quantized variants do not differ significantly from FP16, but AWQ/SmoothQuant significantly outperform GPTQ, indicating that well-designed quantization preserves alignment while GPTQ induces modest degradation. For 1.5B, AWQ improves over FP16, whereas GPTQ and SmoothQuant do not differ reliably from FP16 (and differences among quantized variants are not significant after correction). We observe the similar quantization effects for 
LLaMA-3.2 and DeepSeek-R1 (Tables~\ref{statistical_results_qwen2.5} 
in Appendix~\ref{statistical_variability}).

\noindent\textbf{Qualitative voxel-wise changes.}
Fig.~\ref{fig:qwen_flatmap_subject05} visualizes voxel-wise percentage changes in brain alignment across scale and quantization. (i) Scaling down from LLMs to 3B SLMs (Fig.~\ref{fig:qwen_flatmap_subject05} (a)): reductions in brain alignment are negligible in the bilateral temporal lobe and remain under 5\% in the parietal cortex and IFGorb. Large cortical regions remain white, indicating that 3B SLMs preserve brain-relevant representations comparable to LLMs. The blue-marked voxels are sparse and localized, suggesting only limited information loss. (ii) 3B SLMs to 3B SLMs GPTQ (Fig.~\ref{fig:qwen_flatmap_subject05} (b)): applying GPTQ leads to widespread orange-marked voxels, reflecting consistent losses across distributed cortical regions. While some areas remain preserved (white voxels), the extent of information loss is greater than that observed with downscaling alone, confirming that GPTQ disproportionately disrupts brain-relevant alignment. (iii) 3B SLMs to 1.5B SLMs (Fig.~\ref{fig:qwen_flatmap_subject05} (c)): relative to the 3B baseline, we observe extensive red-marked voxels, especially in temporal and language-related regions, indicating larger drops in alignment. This demonstrates the limits of scaling, as ultra-small models fail to capture brain-relevant representations. (iv) 3B SLMs $\rightarrow$ 3B SLMs AWQ (Fig.~\ref{fig:qwen_flatmap_subject05} (d)): AWQ produces localized green-marked voxels, indicating regions of improved alignment relative to the uncompressed 3B baseline, particularly in the IFG and AG. Most of the cortex remains unchanged (white), suggesting that AWQ maintains representational fidelity while offering modest regional gains. Results for other participants for Qwen2.5 and LLaMA models are in Appendix~\ref{app:flatMaps}.

\begin{table}[t]
\centering
\scriptsize
\caption{Comparison of quantization and pruning for Qwen2.5-3B.}
\label{pruning_effect_qwen2.5_3b}
\resizebox{0.9\columnwidth}{!}{\begin{tabular}{|l| l| c| c|}
\hline
Model variant              & Method        & Sparsity & Normalized Brain Alignment \\
\hline
Qwen-2.5-3B                & FP16 (baseline) & 0\%   & 0.924 $\pm$0.033 \\
Qwen-2.5-3B-AWQ            & Quantization (AWQ)        & 0\%   & 0.933 $\pm$ 0.035 \\
Qwen-2.5-3B-GPTQ           & Quantization (GPTQ)       & 0\%   & 0.910 $\pm$ 0.037 \\
Qwen-2.5-3B-Smooth         & Quantization (SmoothQuant)& 0\%   & 0.930 $\pm$ 0.035 \\
Qwen-2.5-3B-0.1            & Pruning                    & 10\%  & 0.910 $\pm$ 0.032 \\
Qwen-2.5-3B-0.25           & Pruning                    & 25\%  & 0.908 $\pm$ 0.033 \\
Qwen-2.5-3B-0.5            & Pruning                    & 50\%  & 0.907 $\pm$ 0.043 \\
\hline
\end{tabular}}
\end{table}

\begin{figure*}[t]
    \centering
    \includegraphics[width=\textwidth]{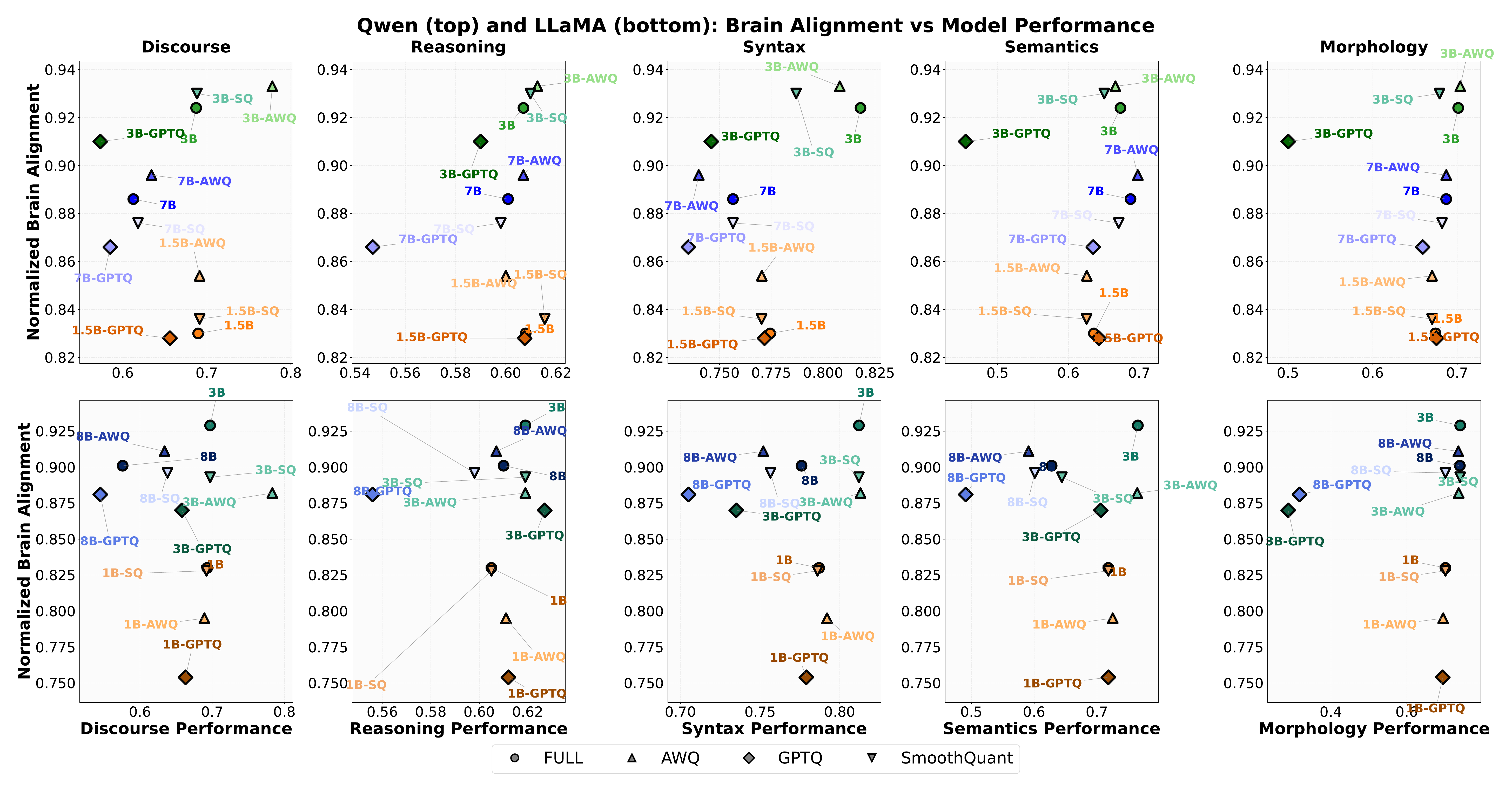}
    \caption{Tradeoff between normalized brain alignment and linguistic competence performance on FlashHolmes Tasks (Qwen and LLaMA Model Families). \textcolor{blue}{Blue: }7b/8b, \textcolor{green}{Green: } 3b, \textcolor{orange}{Orange:} 1.5b.
    }
    \label{fig:brain_alignment_flashholmes}
\end{figure*}

Taken together, these results support Hypothesis 1: 3B SLMs achieve brain alignment comparable to larger LLMs and remain robust under AWQ and SmoothQuant quantization. By contrast, 1B–1.5B models consistently underperform, and GPTQ disproportionately disrupts brain-relevant alignment. Importantly, this pattern holds across both the Qwen and LLaMA model families.

\noindent\textbf{Pruning preserves alignment up to moderate sparsity (10--25\%), but degrades at high sparsity for 1.5B.}
We evaluate unstructured magnitude pruning of linear layers at sparsity levels 0.10, 0.25, and 0.50 and report normalized brain alignment (Table~\ref{pruning_effect_qwen2.5_3b}). For Qwen2.5-3B, alignment remains in a narrow range under pruning up to 50\% (0.907--0.910; $\pm$  0.032--0.043), comparable to FP16 and post-training quantization (AWQ/SmoothQuant: 0.930--0.933 vs.\ FP16: 0.924; GPTQ: 0.910). In contrast, for Qwen2.5-1.5B (see Table~\ref{pruning_effect_qwen2.5_1.5b} in Appendix~\ref{app:unstructured_pruning}), pruning at 10--25\% largely preserves alignment, whereas 50\% pruning yields a marked drop. Overall, these results complement our quantization findings: moderate pruning and quantization can preserve brain alignment surprisingly well, but aggressive sparsification can harm alignment, especially in smaller models, and may introduce trade-offs with linguistic competence.

\subsection*{\textbf{[RQ3]:} Impact of linguistic competence in language models and brains}
\vspace{-0.2cm}
While previous analyses show that 3B SLMs match LLMs in brain alignment, alignment drops for \(\sim\)1B models and under GPTQ. To further investigate the linguistic competence of SLMs, LLMs, and their quantized counterparts, and to examine whether linguistic competence influences brain alignment, we benchmark these models on FlashHolmes and analyze brain alignment trends across linguistic tasks.

Figs.~\ref{fig:brain_alignment_flashholmes} illustrate how scaling and quantization affect the relationship between linguistic competence and brain alignment across SLMs, LLMs, and their quantized variants. We make the following observations: For LLMs and 3B SLMs, brain alignment is largely preserved under AWQ and SmoothQuant, while GPTQ consistently degrades discourse/reasoning/morphology and also reduces brain alignment, suggesting disproportionately degrades higher-order linguistic skills. In contrast, 3B SLMs generally maintain (and sometimes improve) FlashHolmes performance while keeping alignment comparable to FP16, indicating that well-designed quantization can preserve both linguistic competence and brain-relevant representations. Finally, 1B models (and their quantized variants) retain task performance across linguistic categories but show markedly lower brain alignment than 3B, revealing a clear dissociation between task accuracy and neural alignment.

Overall, our analysis reveals a dissociation between linguistic competence and brain alignment across SLMs, LLMs, and their quantized variants. AWQ and SmoothQuant largely preserve both neural predictivity and probing performance, whereas GPTQ tends to reduce both—particularly higher-order competencies such as discourse, reasoning, and syntax/morphology. In contrast, \(\sim\)1B SLMs illustrate the limits of scale: they can maintain task performance while failing to capture brain-relevant representational structure. Tables~\ref{tab:qwen_flashholmes_quant_7b} and~\ref{tab:LLaMA_flashholmes_quant_8b} in Appendix~\ref{app:linguisticCompetencellama} report detailed FlashHolmes results across five competence categories for SLMs/LLMs and their quantized variants.



\vspace{-0.1cm}
\section{Discussion and Conclusion}

We comprehensively evaluate large and small language models, along with compressed variants, for fMRI-based brain encoding during naturalistic language comprehension. Across model families and scales, brain alignment saturates at modest sizes: 3B SLMs match 7B--14B LLMs, whereas 1B--1.5B models consistently underperform. Notably, this saturation largely persists under post-training compression. 

Our compression analyses show that brain alignment is largely robust to post-training efficiency methods. AWQ and SmoothQuant preserve near-baseline alignment, whereas GPTQ produces consistent losses, especially in semantic and discourse-related regions. This underscores that compression methods are not interchangeable: they perturb brain-relevant representations in qualitatively different ways.
Finally, combining voxel-wise encoding with large-scale linguistic probing reveals a dissociation between linguistic competence and brain alignment. Compression can degrade specific linguistic skills without impacting brain alignment, while ultra-small models may retain task performance yet fail to capture brain-relevant representations.

\section{Limitations}
\label{app:limitations}

While prior work has evaluated brain encoding with LLMs up to 72B parameters~\citep{antonello2024scaling}, our analysis of efficiency-oriented regimes extends up to 14B parameters and includes additional architectures such as DeepSeek. These results confirm that the observed performance plateau at the 3B scale generalizes across model families; however, evaluating extremely large models (e.g., 70B+) under post-training compression remains computationally challenging and is an important direction for future work.
Second, although our primary focus is on quantization, we validate our conclusions using unstructured pruning, showing that moderate sparsity preserves brain alignment comparable to quantization. A broader comparison with other compression strategies, such as structured pruning or knowledge distillation, would further clarify how different efficiency interventions affect neural representations.
Third, our experiments focus primarily on fMRI data collected during naturalistic listening, supplemented by validation on a naturalistic reading dataset. While this demonstrates robustness across tasks, future work incorporating higher temporal resolution modalities such as MEG or ECoG could better capture the dynamics of language processing.
Finally, while our main analyses emphasize brain encoding, we include complementary decoding experiments that assess stimulus reconstruction from brain activity. These results suggest that encoding alignment does not always guarantee decoding fidelity, motivating future work that more directly links neural alignment to downstream brain–computer interface performance.

\section*{Impact Statement}

This work advances brain–language model alignment research by identifying the model scale and compression regimes sufficient for capturing brain-relevant language representations during naturalistic comprehension. We show that brain alignment saturates at modest model sizes and remains robust to most post-training quantization methods and moderate levels of pruning, challenging the assumption that increasingly large, full-precision models are necessary for modeling neural language processing.

Scientifically, these findings motivate the use of smaller and more interpretable models as principled tools for studying brain–language computations, and position compression as a controlled intervention for probing which linguistic properties are essential for neural alignment. Practically, our results support compact and efficiently compressed language models as accessible foundations for fMRI-based brain encoding and decoding analyses, potentially lowering computational barriers and improving reproducibility in NeuroAI research.

This work does not involve the collection of new neural data and relies exclusively on the publicly available Subset-Moth-Radio-Hour dataset (\url{https://gin.g-node.org/denizenslab/narratives_reading_listening_fmri}). We do not foresee direct harmful applications arising from this work.

\bibliography{iclr2026_conference}
\bibliographystyle{icml2026}

\newpage
\appendix
\noindent{\Large\textbf{Overview of Appendix Sections}}
\begin{itemize}
\item Appendix~\ref{app:relatedwork}: Related Work
\item Appendix~\ref{naturalistic_dataset}: Naturalistic Listening fMRI Dataset
\item Appendix~\ref{cross_subject_prediction}: Cross-subject Prediction Accuracy
\item Appendix~\ref{app:hyperparameters_details}: Hyperparameter Details 
\item Appendix~\ref{app:statistical_significance} Statistical Significance
\item Appendix~\ref{app:alignmentLangROIs}: Normalized Brain Alignment for Language ROIs
\item Appendix~\ref{app:flatMaps}: Contrast of Estimated Model Prediction Accuracy for Various Subjects Across the Two Model Families
\item Appendix~\ref{app:linguisticCompetencellama}: Impact of Linguistic Competence for Qwen2.5 and LLaMA-3.2 Models
\item Appendix~\ref{app:quantitative_analysis}: Quantitative Analysis across Model Families
\item Appendix~\ref{statistical_variability}: Statistical Validation of Quantization Effects
\item Appendix~\ref{app:roi_analysis_layer_subject_variablity}: ROI-Specific Analysis, Best Layer Selection and Subject Variability.
\item Appendix~\ref{app:reading_fmri}: Encoding Performance on Naturalistic Reading fMRI Dataset
\item Appendix~\ref{app:decoding_reconstruction}: Decoder gap: Brain Decoding (Stimulus Reconstruction)
\item Appendix~\ref{app:unstructured_pruning}: Effect of Pruning
\item Appendix~\ref{app:model_size_brain_alignment}: Model Size vs. Brain Alignment
\end{itemize}

\section{Related work}
\label{app:relatedwork}

\noindent\textbf{Small and compressed language models.}
Scaling laws characterize how task performance improves with model size, data, and compute~\citep{kaplan2020scaling,hoffmann2022training,diaz2024scaling}. In parallel, recent work has produced strong small language models (SLMs) and efficient model variants (e.g., LLaMA and Qwen families, Gemma, DeepSeek) that achieve competitive NLP performance at modest parameter counts~\citep{touvron2023LLaMA,yang2024qwen2,team2024gemma,guo2025deepseek}. For NeuroAI, these models offer a complementary opportunity: constraining capacity provides a controlled way to probe which representational properties are sufficient for brain alignment. 

A related line of work studies compression, including pruning, distillation, and post-training quantization~\citep{gupta2022compression}. Post-training quantization methods such as AWQ, GPTQ, and SmoothQuant can substantially reduce memory and compute while often preserving benchmark accuracy~\citep{lin2023awq,frantar-gptq,xiao2023smoothquant,namburi2023cost,kuzmin2023pruning}. However, compression is typically evaluated with engineering metrics, and its consequences for brain alignment and neural predictivity remain underexplored. Our work fills this gap by systematically testing model scale and compression (quantization and pruning) as controlled perturbations and evaluating their effects on neural predictivity and linguistic competence.

\noindent\textbf{Scaling laws for brain encoding.}
A growing body of work has demonstrated that representations from large language models can predict brain activity evoked by text and speech with high fidelity~\citep{antonello2024scaling,matsuyama2023applicability,alkhamissi2025language}. 
In a seminal study, \citet{antonello2024scaling} compared small and large models (from OPT-125M to OPT-175B and LLaMA-66B), showing that larger models yield substantial gains in fMRI encoding performance. 
Subsequent work by \citet{alkhamissi2025language} identified language-selective networks within LLMs that mirror functional specialization in the human brain.
Together, these studies establish that scale matters for brain encoding under full-precision settings.
Our work departs from this line of research by showing that brain alignment does not grow monotonically with scale: instead, it saturates at relatively modest model sizes and remains robust under substantial compression. 
By extending scaling analysis to efficiency-oriented models, including compressed LLMs and modern SLMs across multiple families, we reveal qualitative differences in how scale and compression affect brain-relevant representations.


\noindent\textbf{Linguistic competence and neural alignment.}
Prior work has also examined which linguistic properties of language models are predictive of brain activity~\citep{oota2022joint,oota2024speech}. 
Some studies adopt direct interventions on model representations, such as residualization or feature ablation, to estimate the causal contribution of specific linguistic features to neural alignment~\citep{toneva2022same,oota2022joint,oota2024speech}. 
Other studies follow an indirect approach, first measuring a model’s brain predictivity and then relating it to the performance on linguistic tasks~\citep{schrimpf2021neural,goldstein2022shared}. 
Our work is complementary to both paradigms. Rather than manipulating representations directly, we use compression as a natural intervention on representational geometry and examine its effects on both brain alignment and linguistic competence across a wide spectrum of tasks. 
This approach allows us to uncover the dissociations between task-level linguistic performance and neural predictivity, revealing which linguistic competencies are critical for brain alignment and which can degrade without disrupting it.

\section{Naturalistic Listening fMRI Dataset}
\label{naturalistic_dataset}
We use the publicly available naturalistic story listening fMRI dataset provided by~\citep{deniz2019representation}. The dataset consists of 11 stories, 9 participants, and all participants listened to all the stories. The speech stimuli consisted of 10- to 15 min stories taken from The Moth Radio Hour and used previously~\citep{huth2016natural}. The 10 selected stories cover a wide range of topics and are highly engaging. The model validation dataset consisted of one 10 min story. All stimuli were played at 44.1 kHz using the pygame library in Python. The audio of each story was down-sampled to 11.5 kHz and the Penn Phonetics Lab Forced Aligner was used to automatically align the audio to the transcript. Finally the aligned transcripts were converted into separate word and phoneme representations using Praat's TextGrid object. The word representation of each story is a list of pairs (W, t), where W is a word and t is the onset time in seconds.

The total number of words in each story as follows:
Story1: 2174;
Story2: 1469;
Story3: 1964;
Story4: 1893;
Story5: 2209;
Story6: 2786;
Story7: 3218;
Story8: 2675;
Story9: 1868;
Story10: 1641;
Story11: 1839 (test dataset)

To align the stimulus presentation rate with the slower fMRI data acquisition rate (TR = 2.0045 sec), where multiple words correspond to a single TR, we downsample the stimulus features to match fMRI recording times using a 3-lobed Lanczos filter~\citep{duchon1979lanczos}, thus creating chunk-embeddings for each TR. To account for the slowness of the hemodynamic response (HRF), we model HRF using a finite response filter (FIR) per voxel and for each subject separately with 4 temporal delays corresponding to 8 secs.

\section{Cross-subject prediction accuracy}
\label{cross_subject_prediction}

\begin{figure*}[!ht]
    \centering
    \includegraphics[width=0.49\linewidth]{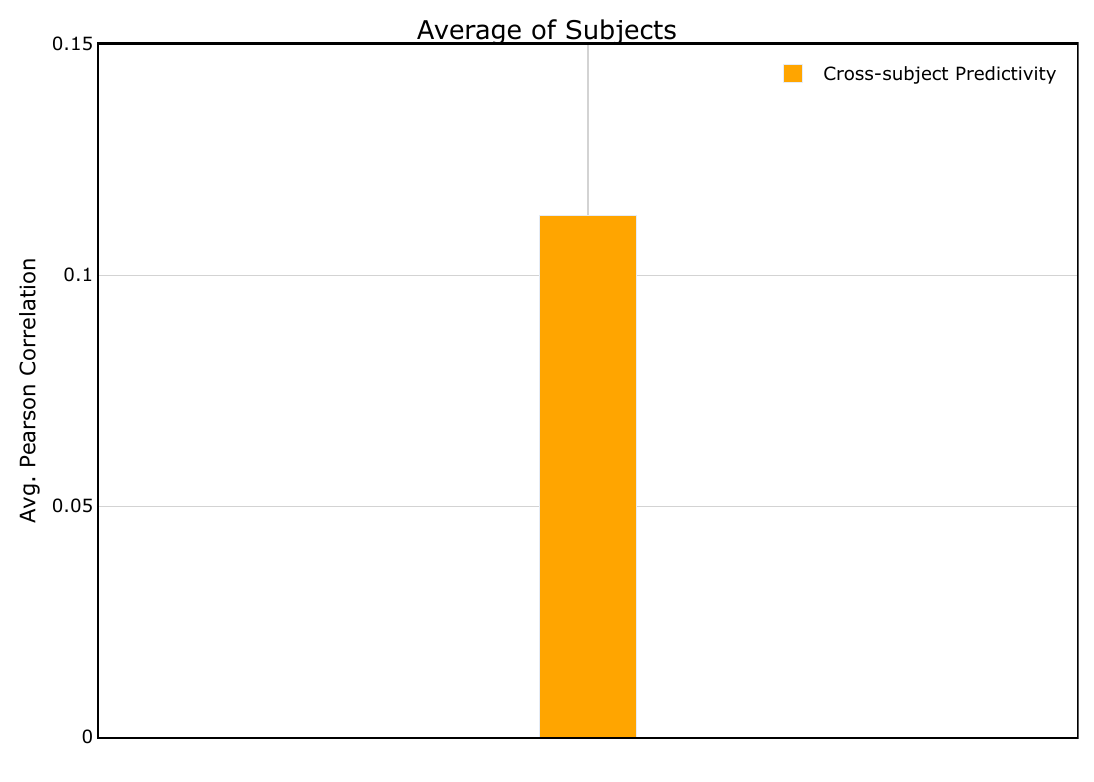}
    \includegraphics[width=0.49\linewidth]{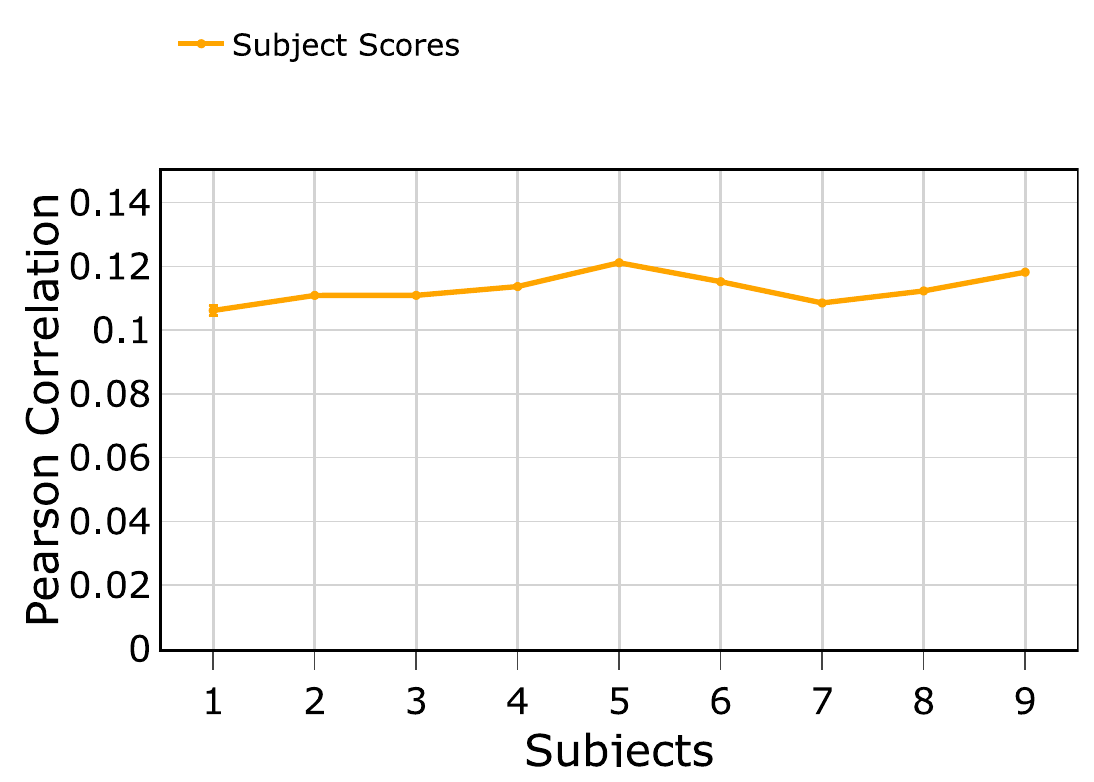}
    \caption{The estimated cross-subject prediction accuracy was computed across all participants for the Subset-Moth-Radio-Hour naturalistic listening fMRI dataset. The average cross-subject prediction accuracy is shown across predicted voxels where each voxel ceiling value is $\ge$ 0.05.}
    \label{fig:cross_subject_predictivity}
\end{figure*}

We present the average estimated cross-subject prediction accuracy across voxels for the \emph{naturalistic listening fMRI} dataset in Fig.~\ref{fig:cross_subject_predictivity}.
We observe that the average estimated cross-subject prediction accuracy across voxels for the listening dataset is higher across subjects. 

\begin{figure*}[!ht] 
\centering
\begin{minipage}{\textwidth}
\centering
    \includegraphics[width=0.81\linewidth]{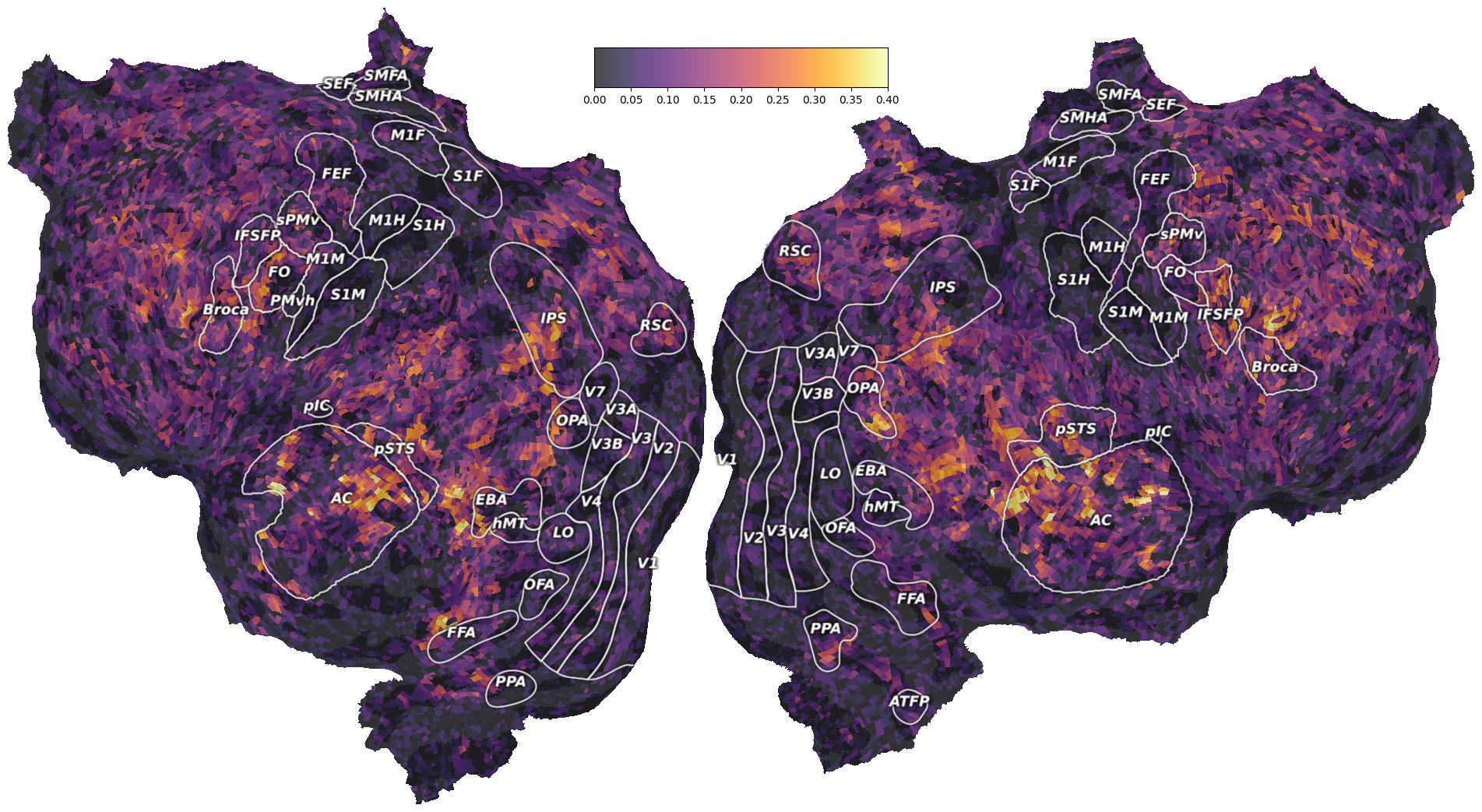}
    \\(a) Subject-01 \\
\end{minipage}
\begin{minipage}{\textwidth}
\centering
    \includegraphics[width=0.81\linewidth]{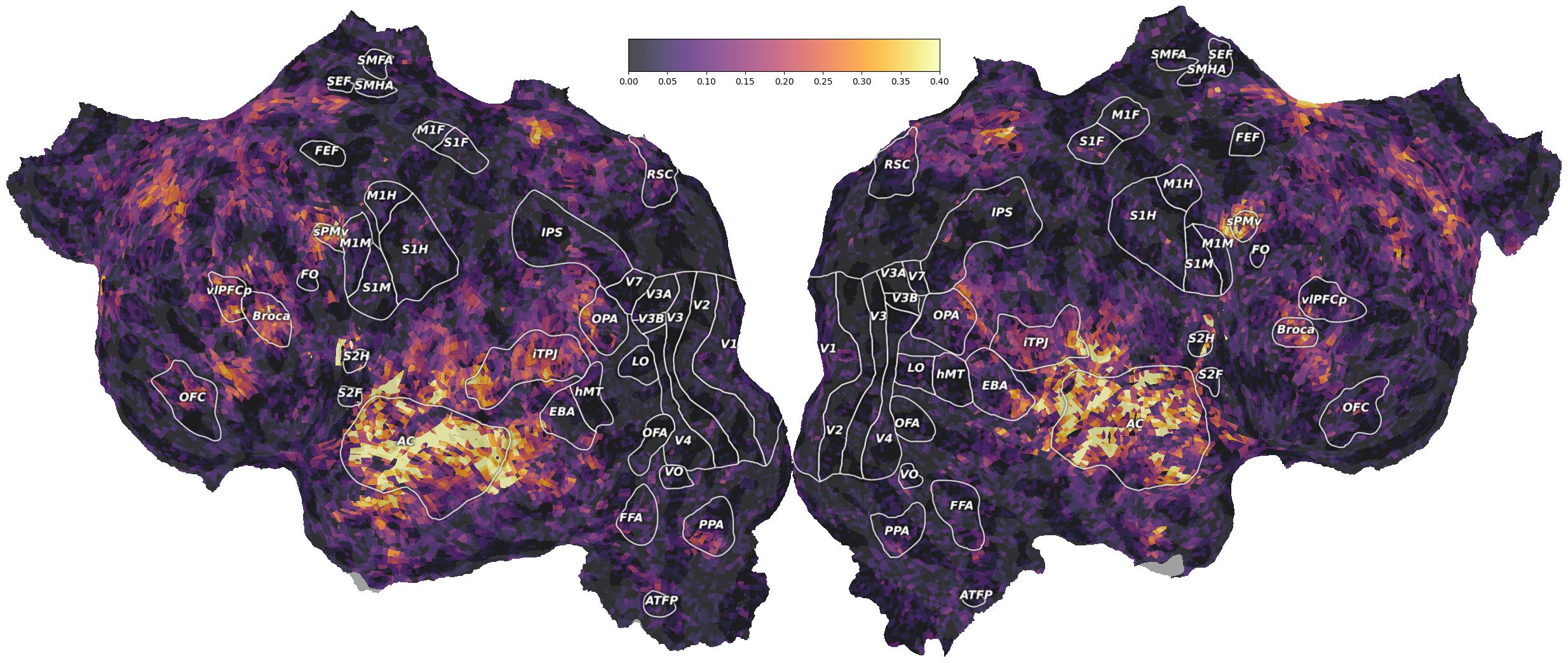}
    \\(a) Subject-02 \\
\end{minipage}
\begin{minipage}{\textwidth}
\centering
    \includegraphics[width=0.81\linewidth]{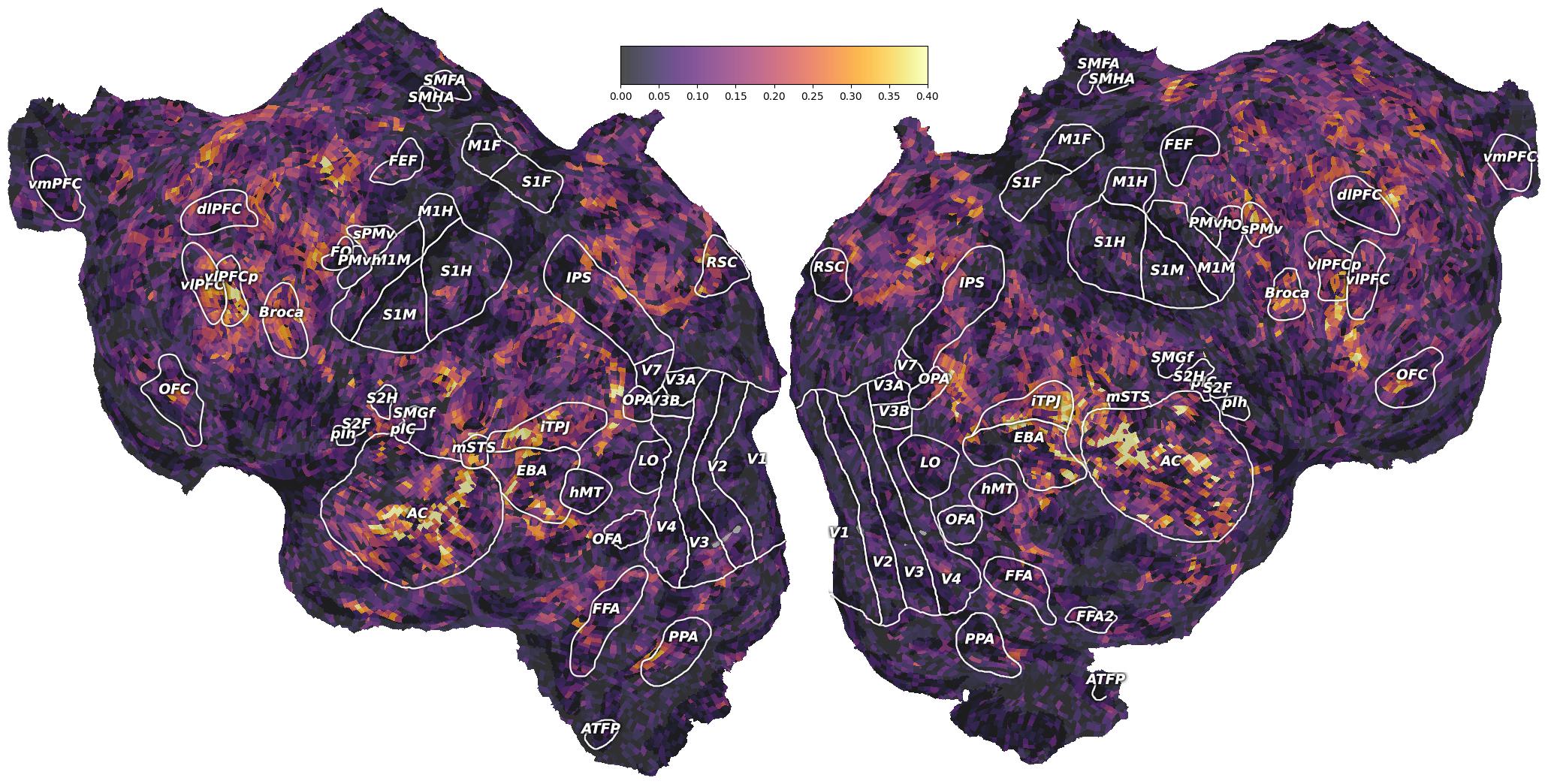}
    \\(a) Subject-03 \\
\end{minipage}
\caption{Contrast of estimated cross-subject prediction accuracy for the participants for the listening condition. The color bar denotes Pearson Correlation.}
\label{fig:noise_ceiling_subject01}
\end{figure*}

\begin{figure*}[!ht] 
\centering
\begin{minipage}{\textwidth}
\centering
    \includegraphics[width=0.81\linewidth]{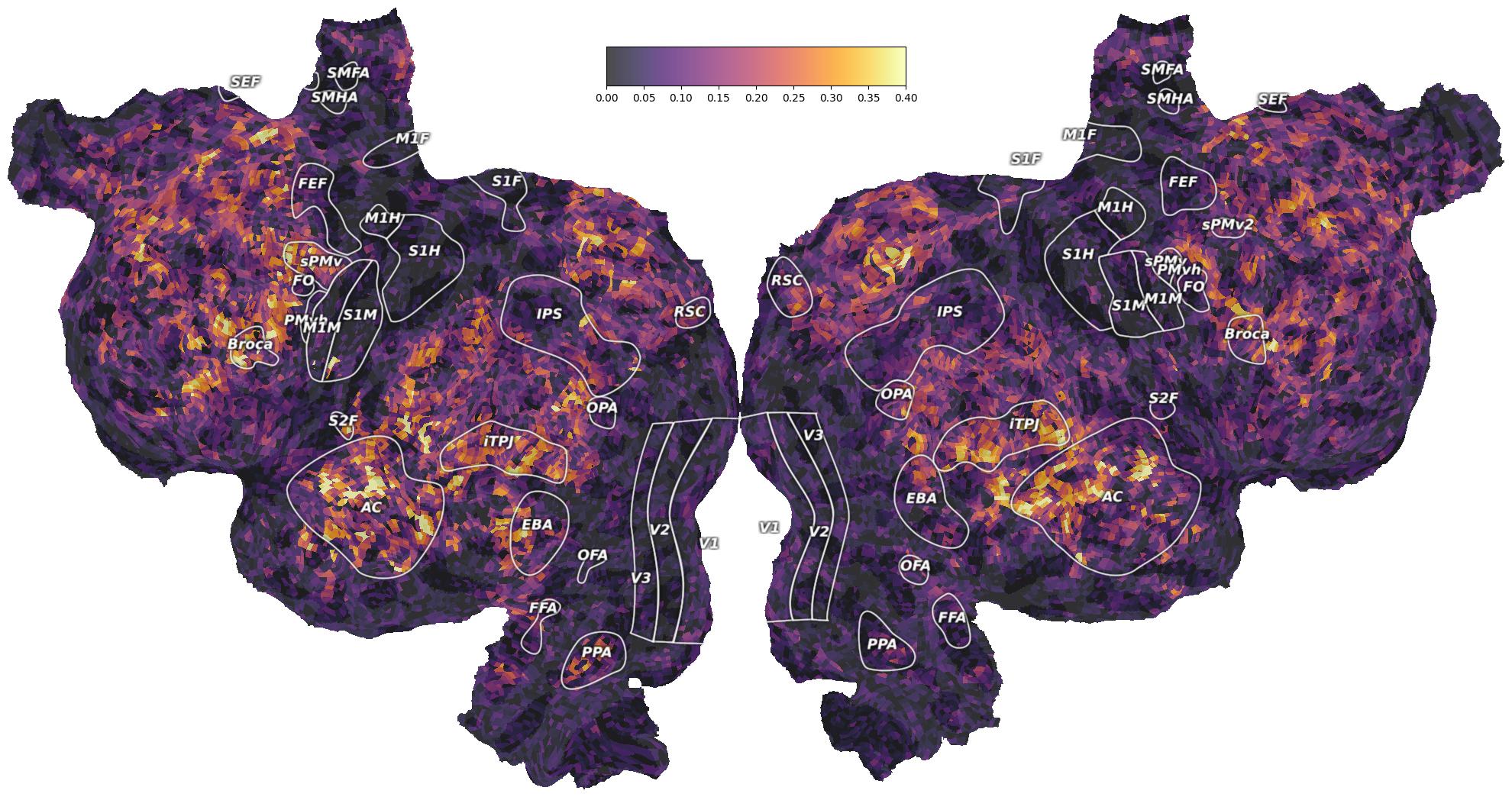}
    \\(a) Subject-05 \\
\end{minipage}
\begin{minipage}{\textwidth}
\centering
    \includegraphics[width=0.81\linewidth]{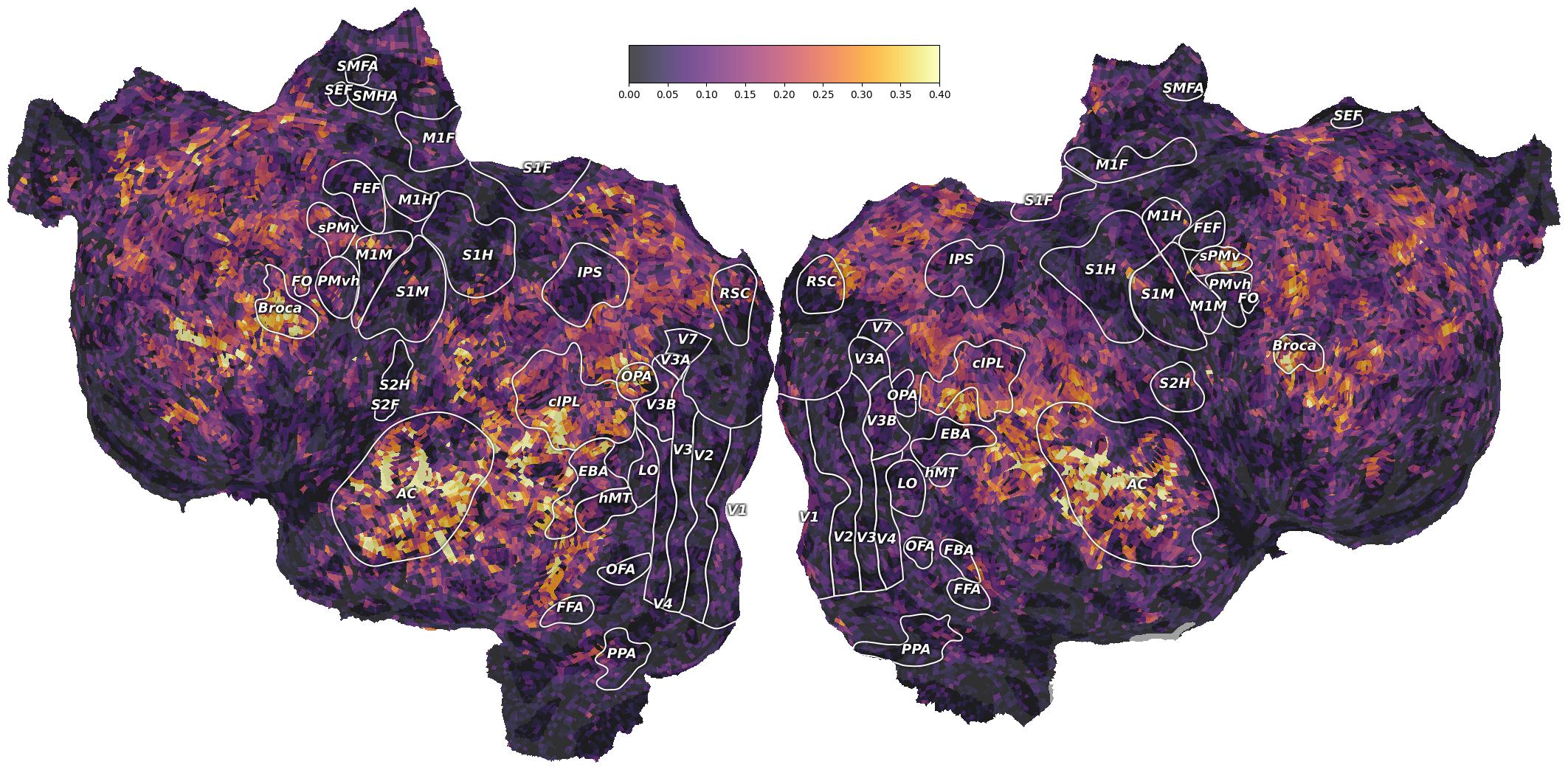}
    \\(a) Subject-07 \\
\end{minipage}
\begin{minipage}{\textwidth}
\centering
    \includegraphics[width=0.81\linewidth]{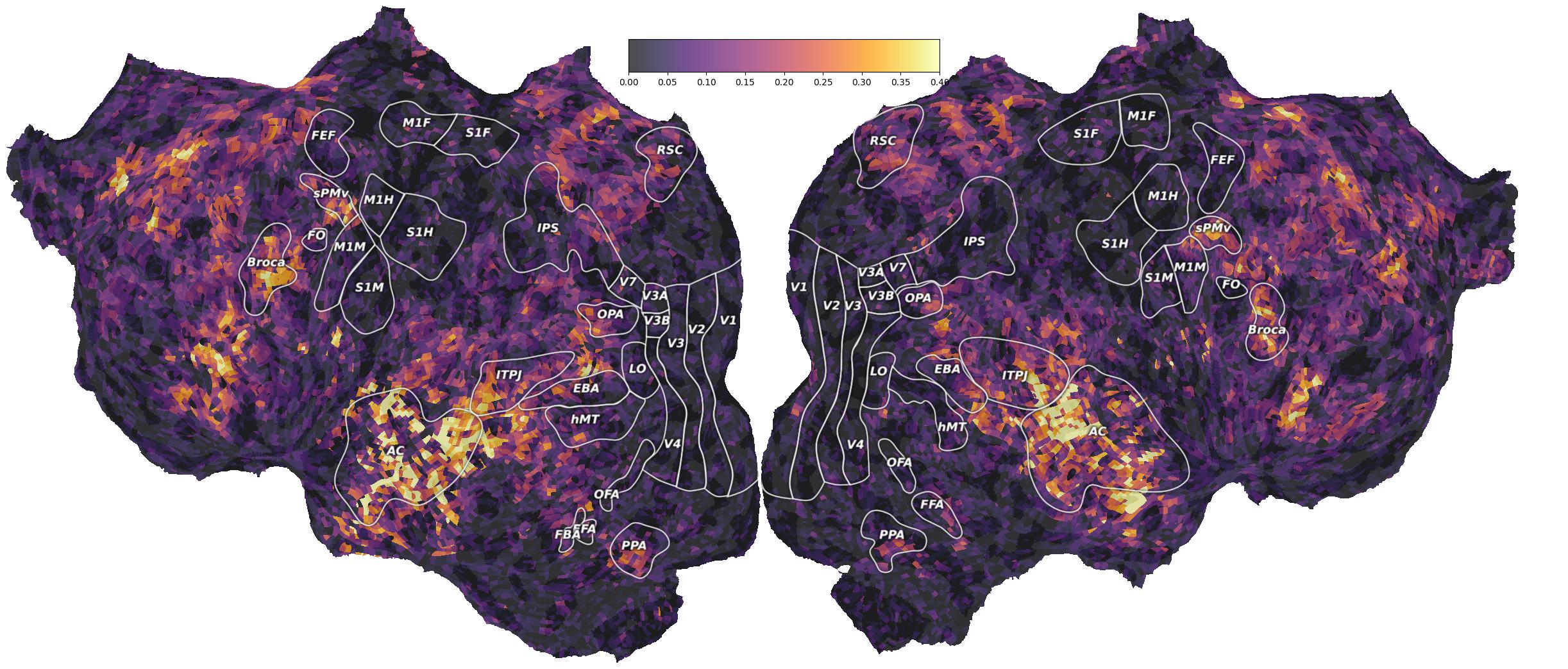}
    \\(a) Subject-08 \\
\end{minipage}
\caption{Contrast of estimated cross-subject prediction accuracy for the participants for the listening condition. The color bar denotes Pearson Correlation.}
\label{fig:noise_ceiling_subject05}
\end{figure*}



\setlength{\tabcolsep}{3pt}
\begin{table*}[!h]
\caption{Detailed functional description of various brain regions.}
\centering
\begin{tabular}{p{1in}p{4in}}
\toprule
Early auditory & The early auditory region is the earliest cortical region for speech processing. This region is specialized for processing elementary speech sounds, as well as other temporally complex acoustical signals, such as music.\\
\midrule
Late Language & Late language regions contribute to various linguistic processes. Regions 44 and 45 (Broca's region) are vital for speech production and grammar comprehension~\citep{friederici2011brain}. The IFJ, PG, and TPOJ clusters are involved in semantic processing, syntactic interpretation, and discourse comprehension~\citep{deniz2019representation,toneva2022combining}. STGa and STS play roles in phonological processing and auditory-linguistic integration~\citep{vaidya2022self,millet2022toward,gong2023phonemic}. TA2 is implicated in auditory processing, especially in the context of language. \\
\bottomrule
\end{tabular}
\label{rois_description}
\end{table*}


\section{Hyperparameter Details}
\label{app:hyperparameters_details}

\noindent\textbf{Implementation details for reproducibility.}
All experiments were conducted on a machine with 2 NVIDIA A100 GPUs with 40GB GPU RAM. We used bootstrap ridge-regression with the following parameters: MSE loss function, and L2-decay ($\lambda$) varied from  10$^{1}$ to 10$^{3}$; best $\lambda$ was chosen by tuning on validation data that comprised a randomly chosen 10\% subset from train set used only for hyper-parameter tuning.

\section{Statistical significance}
\label{app:statistical_significance}
To determine if normalized predictivity scores are significantly higher than chance, we use block permutation tests. We employ the standard implementation of a block permutation test for fMRI data, which is to split the fMRI data into blocks of 10 contiguous TRs and permute the order of these blocks, while maintaining the original order of the TRs within each block. 
By permuting predictions 5000 times, we create an empirical distribution for chance performance, from which we estimate the p-value of the actual performance.
To estimate the statistical significance of performance differences, such as between the model's predictions and chance or quantized model predictions and chance, we utilized the Wilcoxon signed-rank test, applying it to the mean normalized predictivity for the participants.
In all cases, we denote significant differences with an asterisk \textcolor{red}{*}, indicating cases where p$\leq 0.05$.

\section{Normalized brain alignment for language ROIs}
\label{app:alignmentLangROIs}

\begin{figure*}[!ht]
    \centering
    \includegraphics[width=\linewidth]{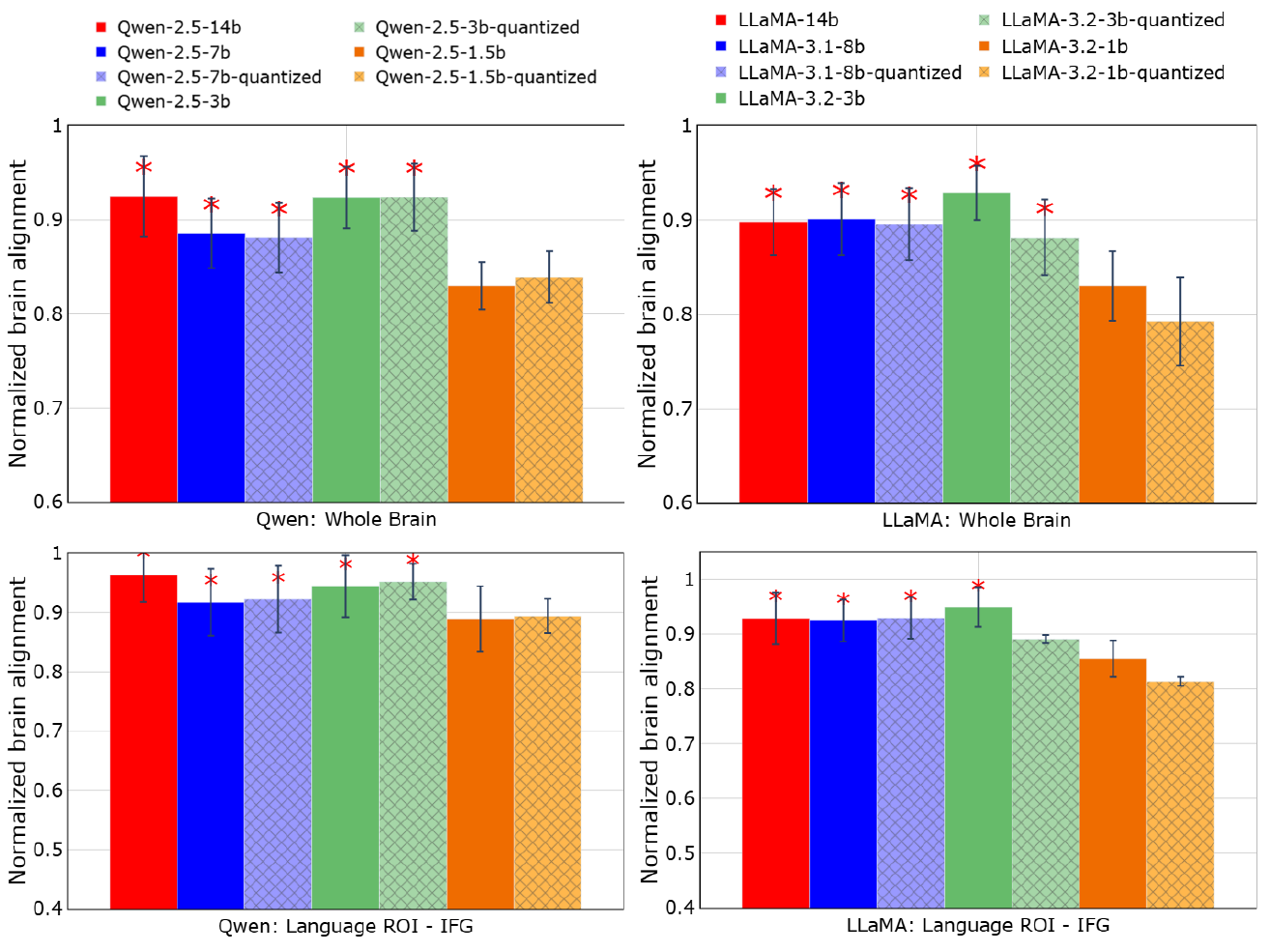}
    \caption{Qwen2.5 and LLaMA: Normalized brain alignment was computed by averaging across participants, layers, and voxels. \textcolor{red}{Red: }14b, \textcolor{blue}{Blue: }7b, \textcolor{green}{Green: } 3b, \textcolor{orange}{Orange:} 1.5b, Solid: full-precision SLMs/LLMs, Patterned: quantized models. \textcolor{red}{*} at a particular bar indicates that the model's prediction performance is significantly better than 1b/1.5b SLMs. The top row shows whole-brain normalized alignment, while the bottom row focuses on a language-selective ROI (IFG).}
    \label{fig:qwen2.5}
\end{figure*}

Figs.~\ref{fig:qwen2.5} show the average normalized brain alignment across the whole brain for both language model families, comparing SLMs, LLMs, and the grouped quantized variants.

\begin{figure*}[!ht]
    \centering
    \includegraphics[width=0.52\linewidth]{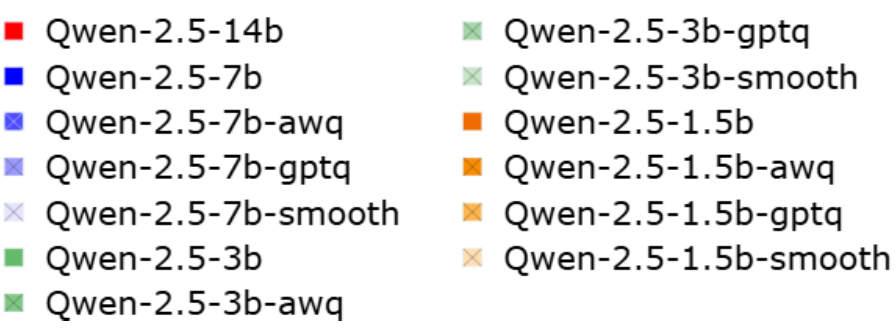}\\
    \includegraphics[width=0.33\linewidth]{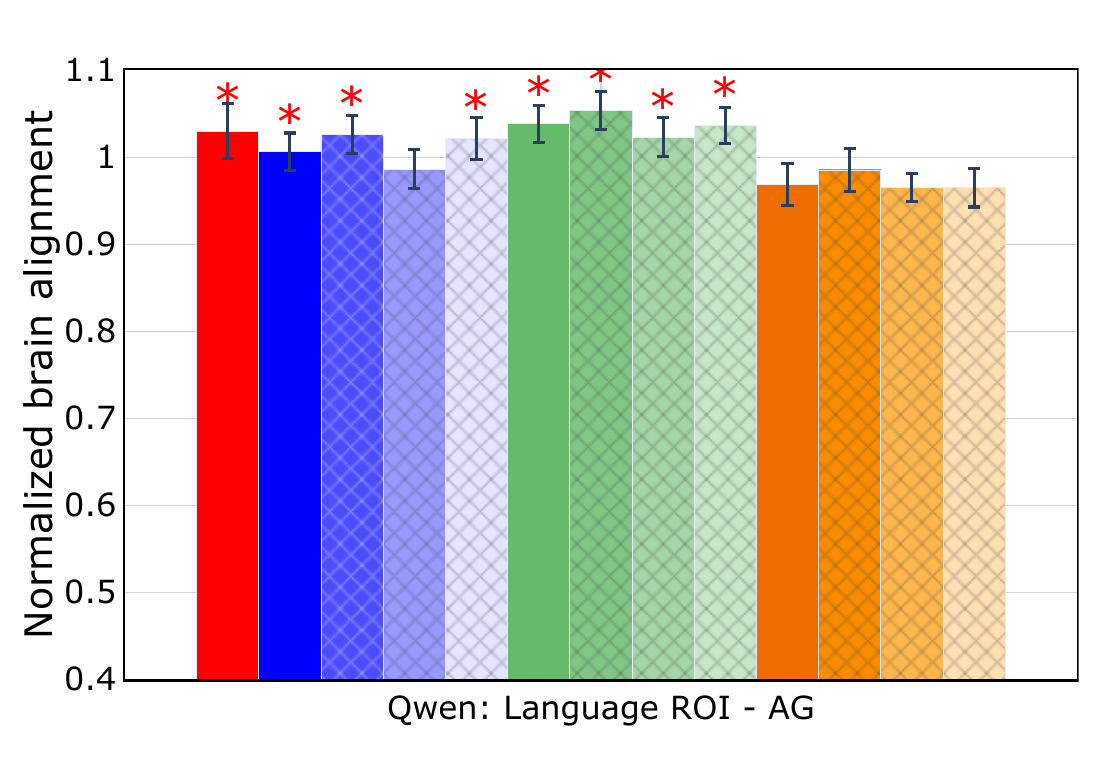}
    \includegraphics[width=0.33\linewidth]{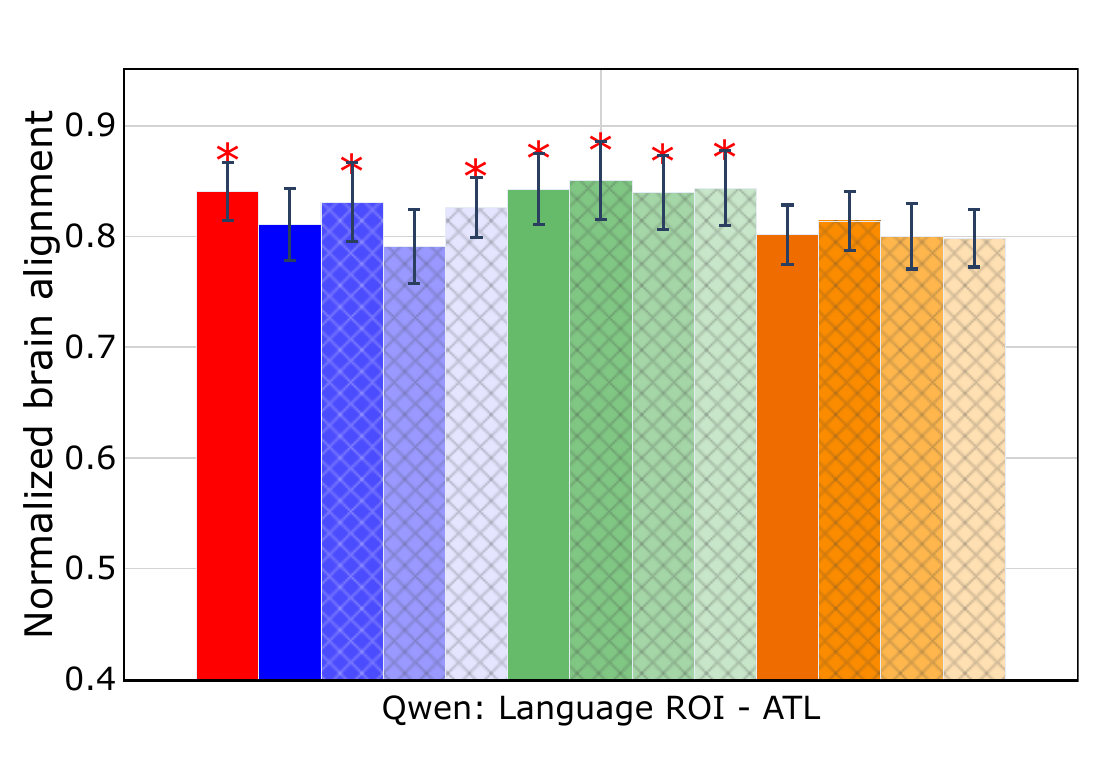}
    \includegraphics[width=0.33\linewidth]{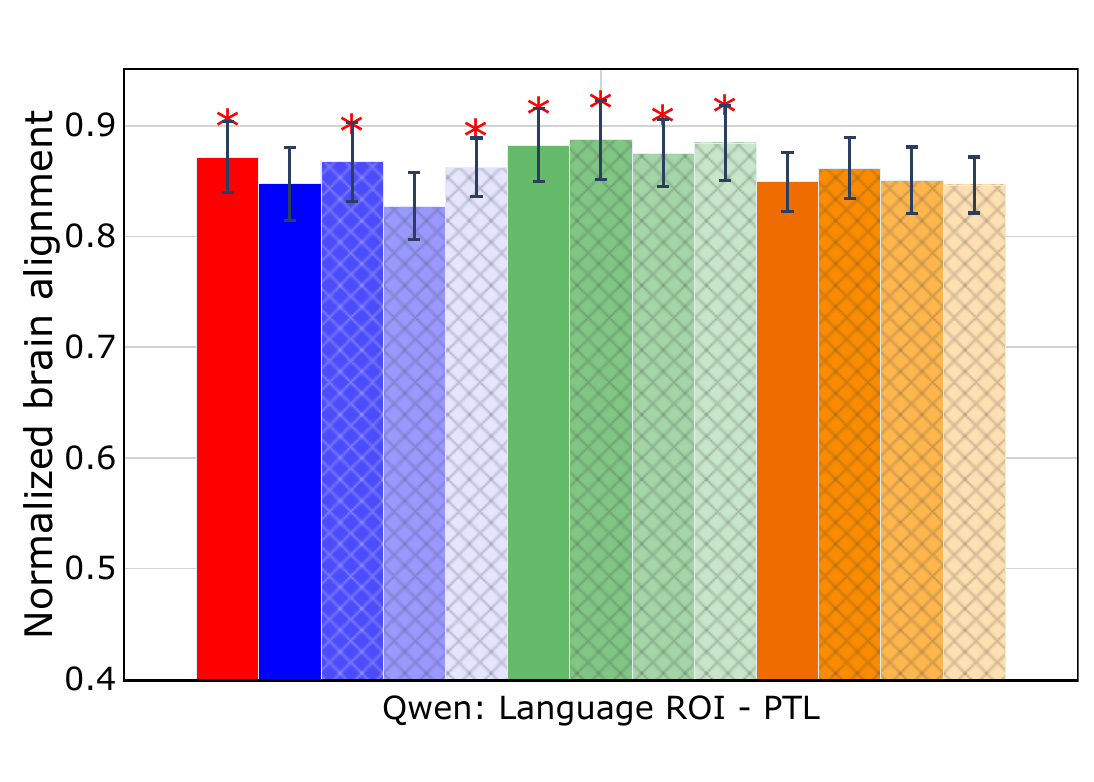}
    \includegraphics[width=0.33\linewidth]{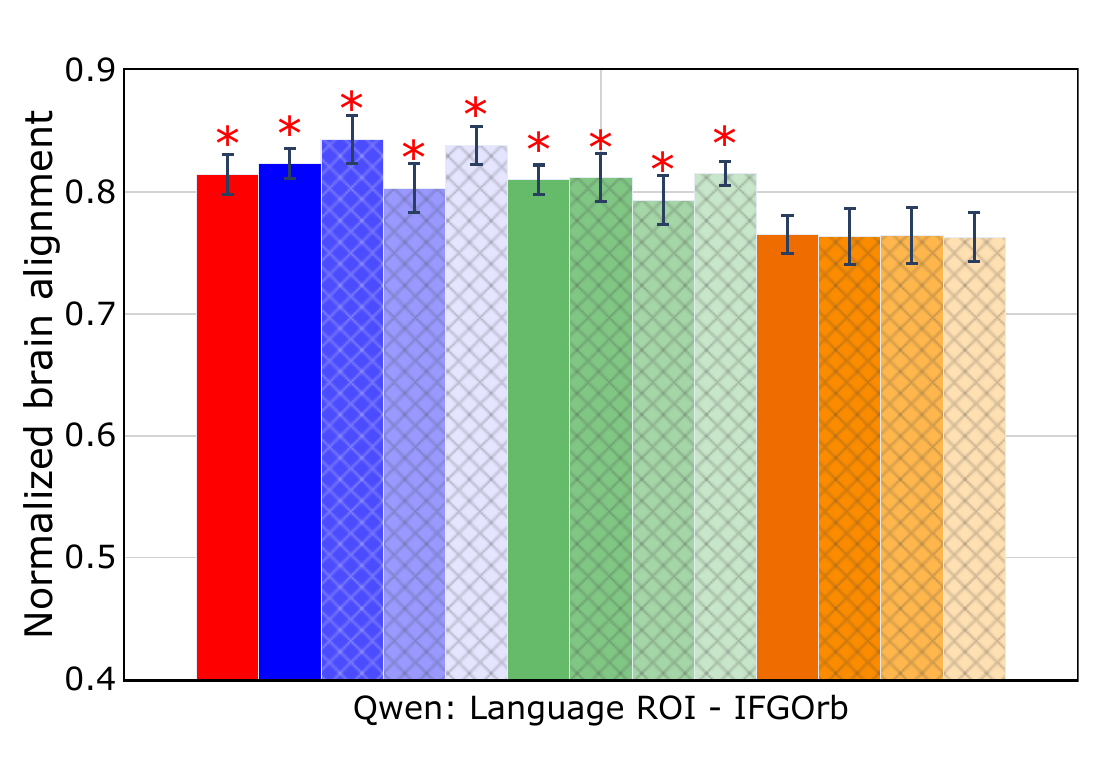}
    \includegraphics[width=0.33\linewidth]{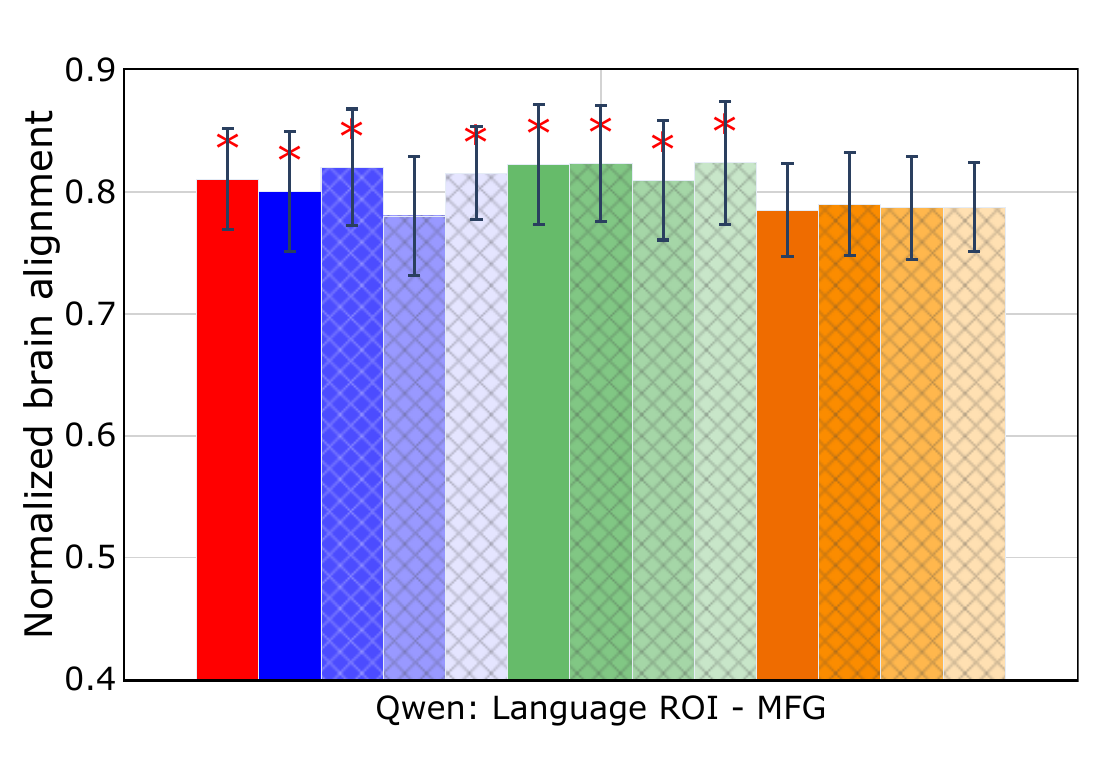}
    \includegraphics[width=0.33\linewidth]{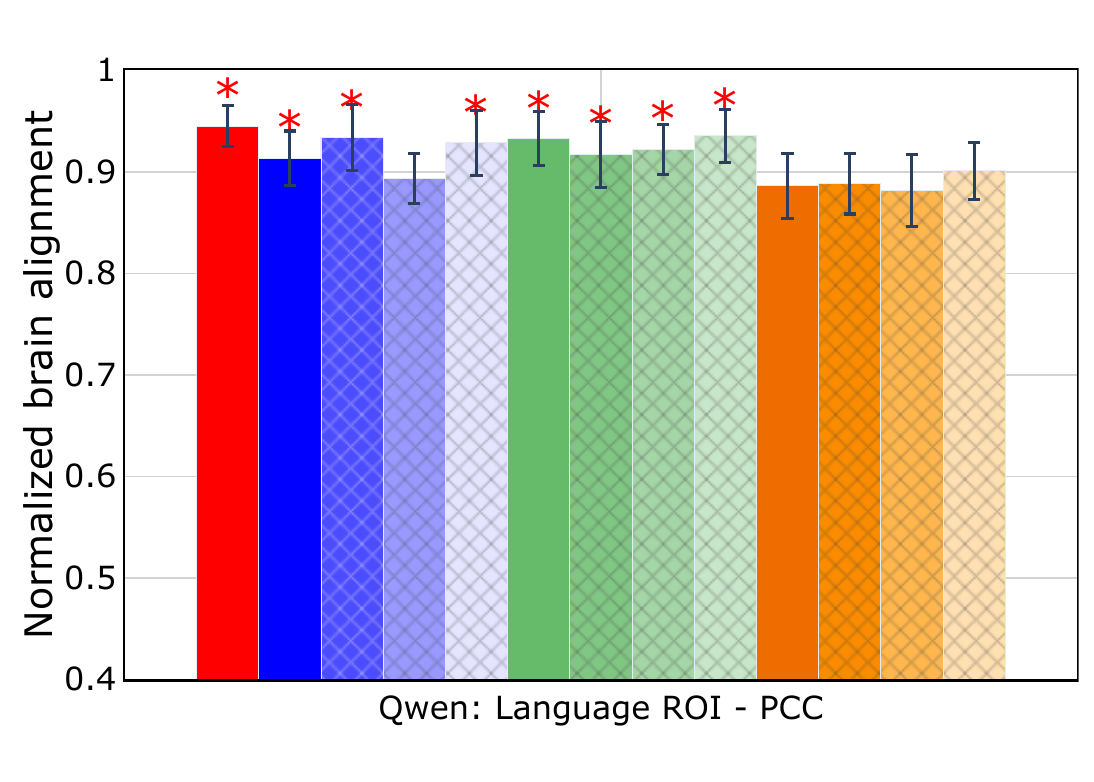}
    \includegraphics[width=0.33\linewidth]{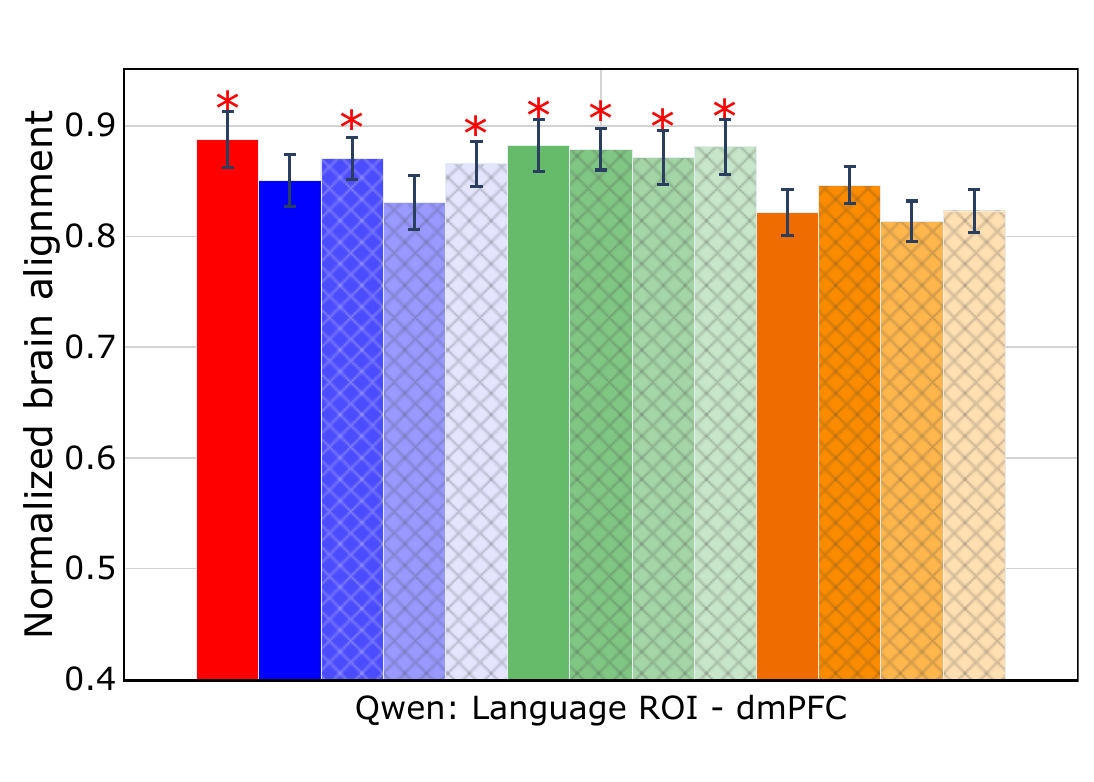}
    \includegraphics[width=0.33\linewidth]{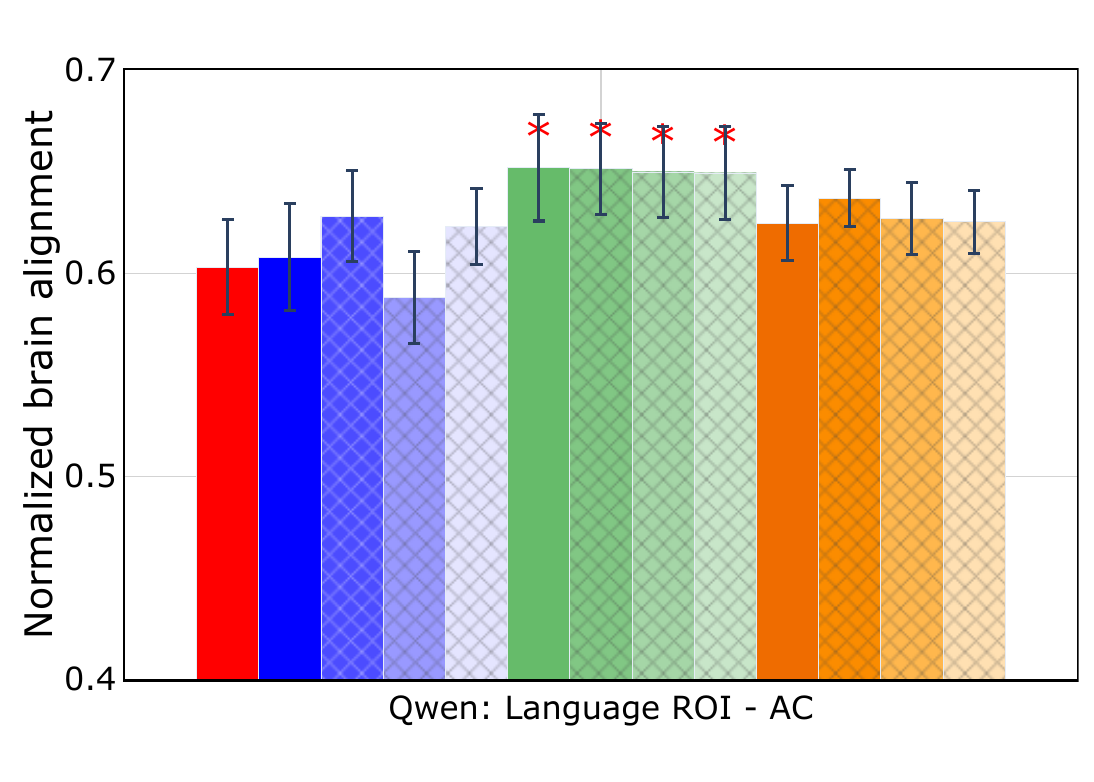}
    \caption{Normalized Predictivity of SLMs, LLMs, and Quantized Language Models for Qwen-2.5 models.}
    \label{fig:qwen_vem_language}
\end{figure*}

\begin{figure*}[!ht]
    \centering
    \includegraphics[width=0.52\linewidth]{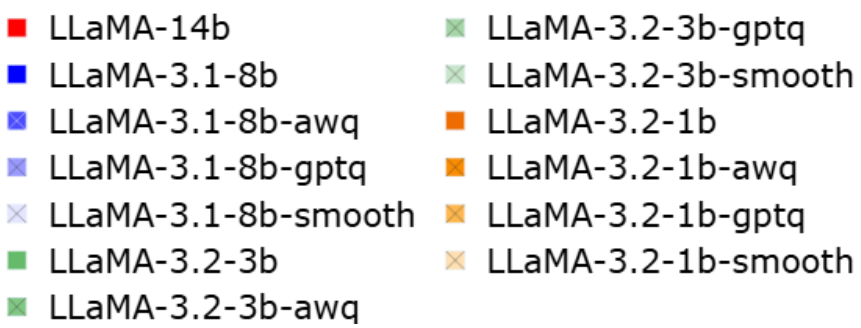} \\
    \includegraphics[width=0.33\linewidth]{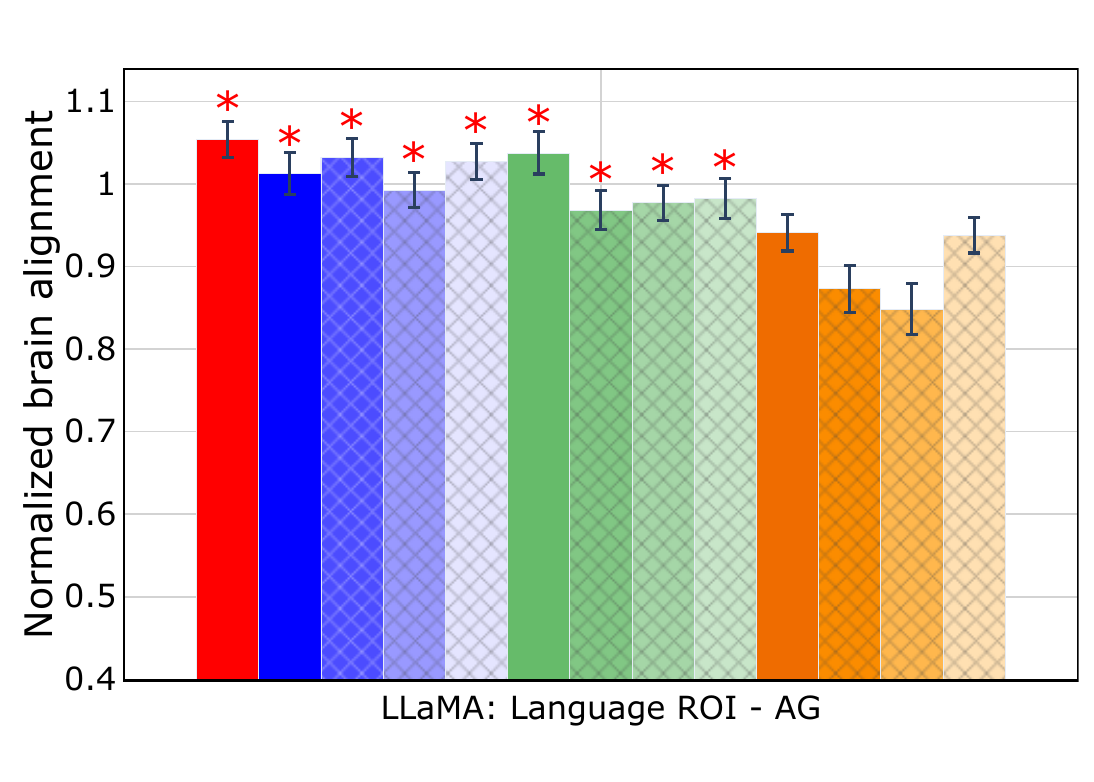}
    \includegraphics[width=0.33\linewidth]{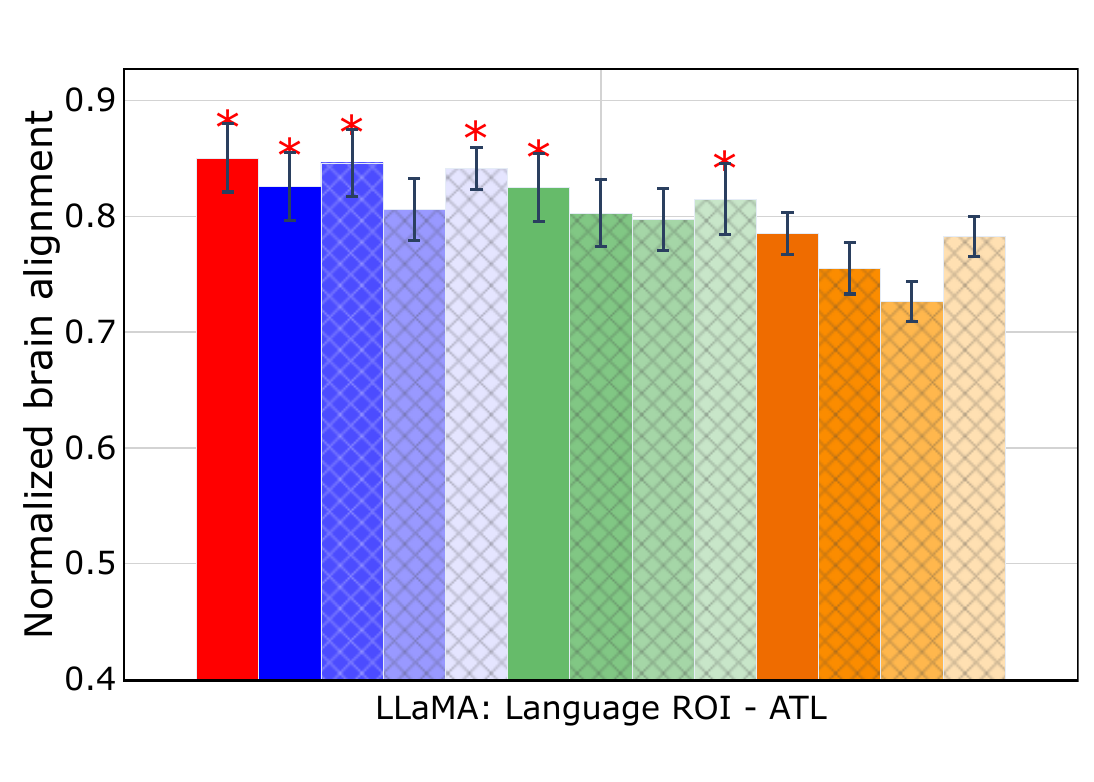}
    \includegraphics[width=0.33\linewidth]{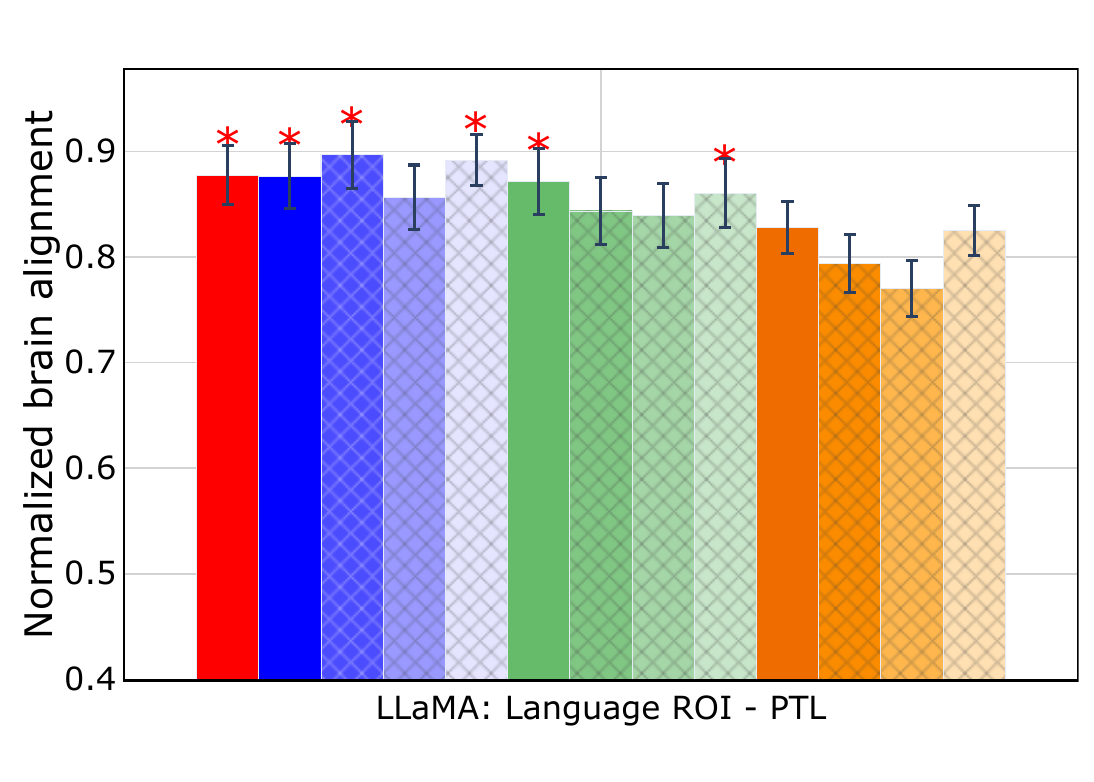}
    \includegraphics[width=0.33\linewidth]{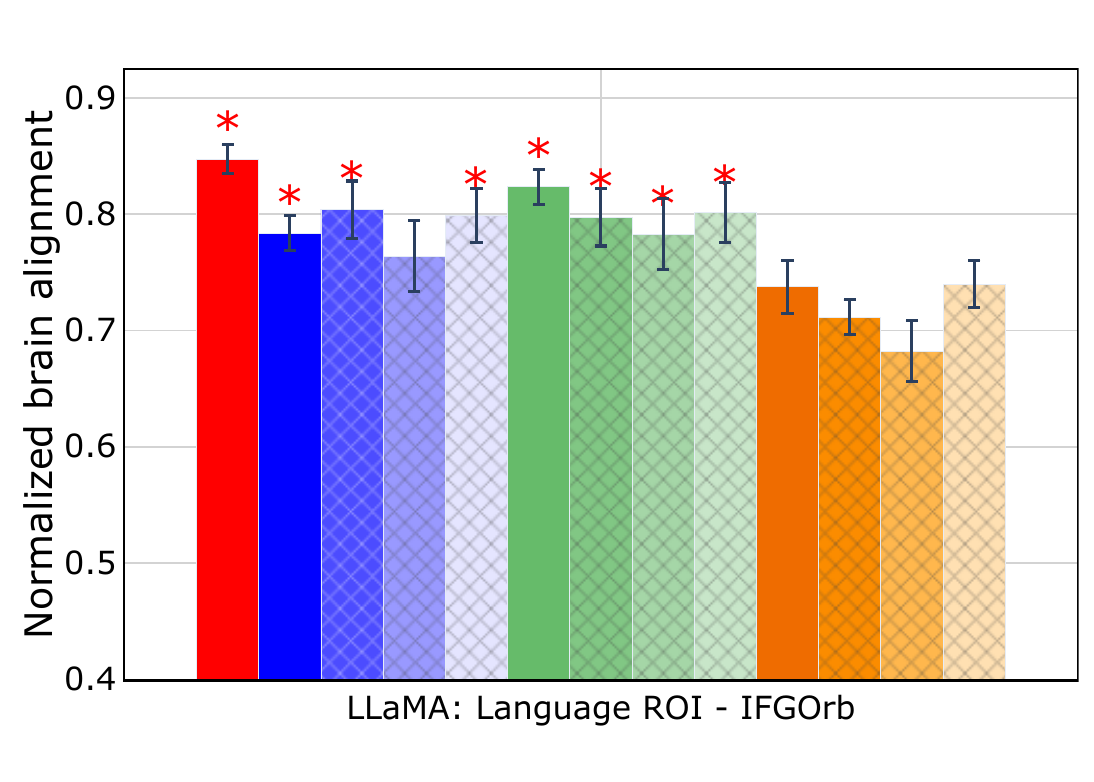}
    \includegraphics[width=0.33\linewidth]{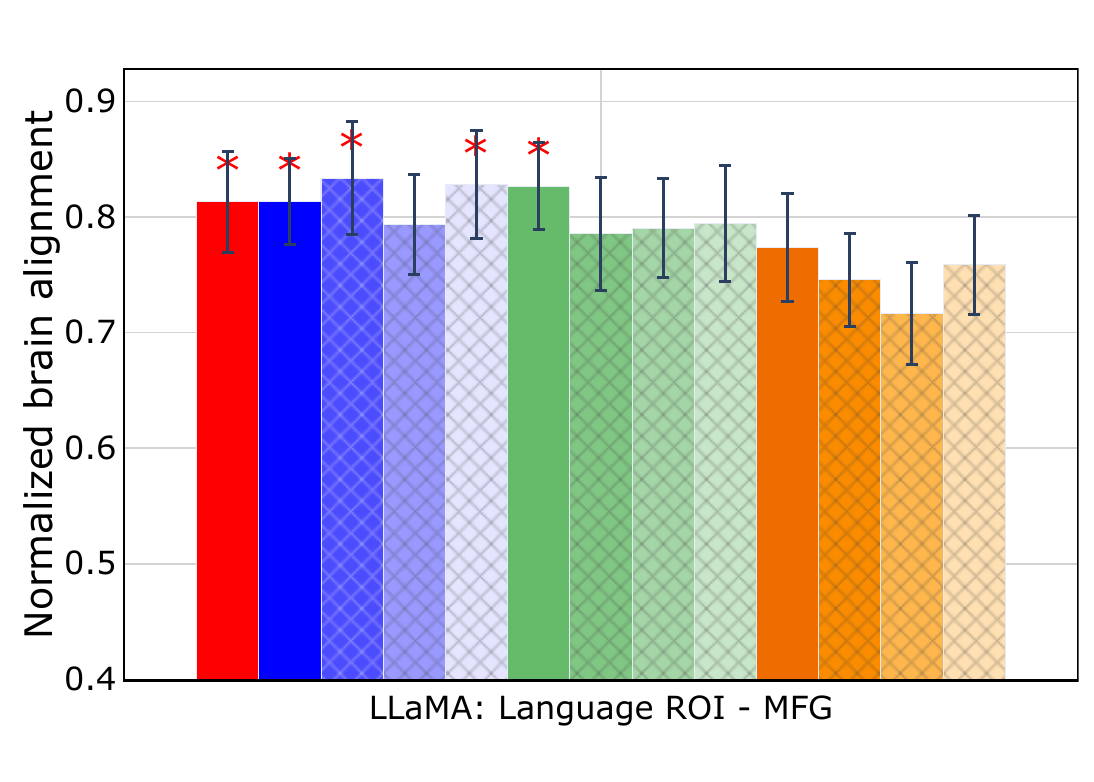}
    \includegraphics[width=0.33\linewidth]{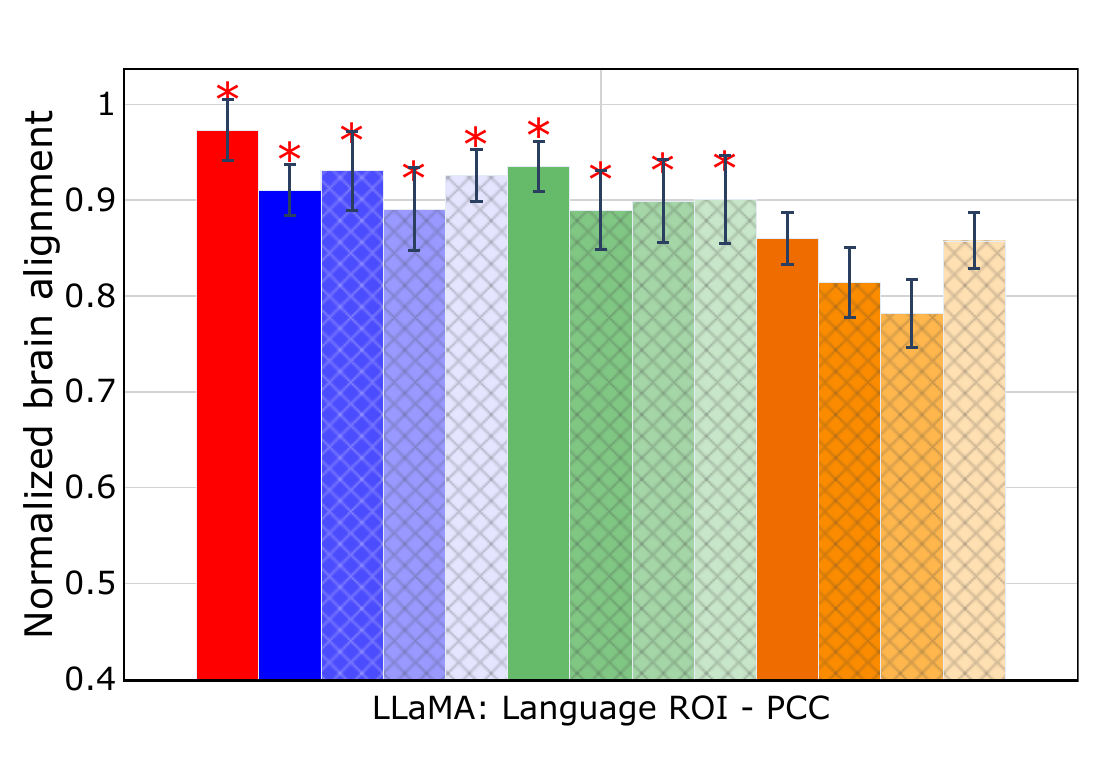}
    \includegraphics[width=0.33\linewidth]{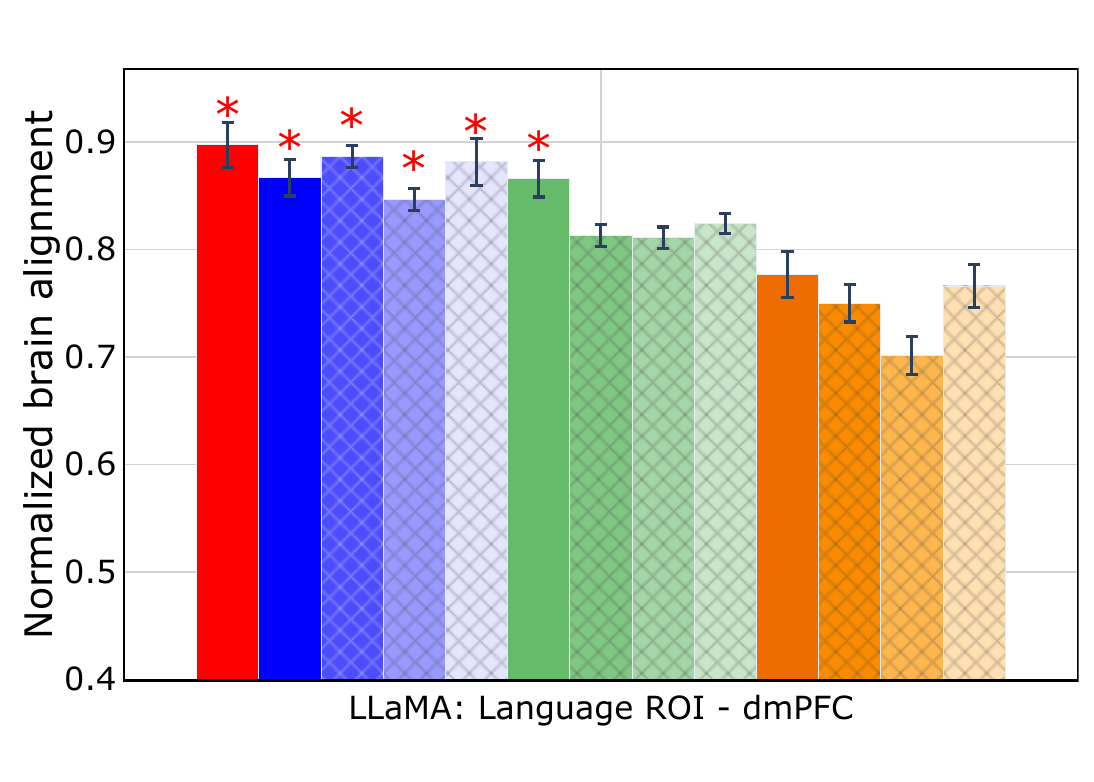}
    \includegraphics[width=0.33\linewidth]{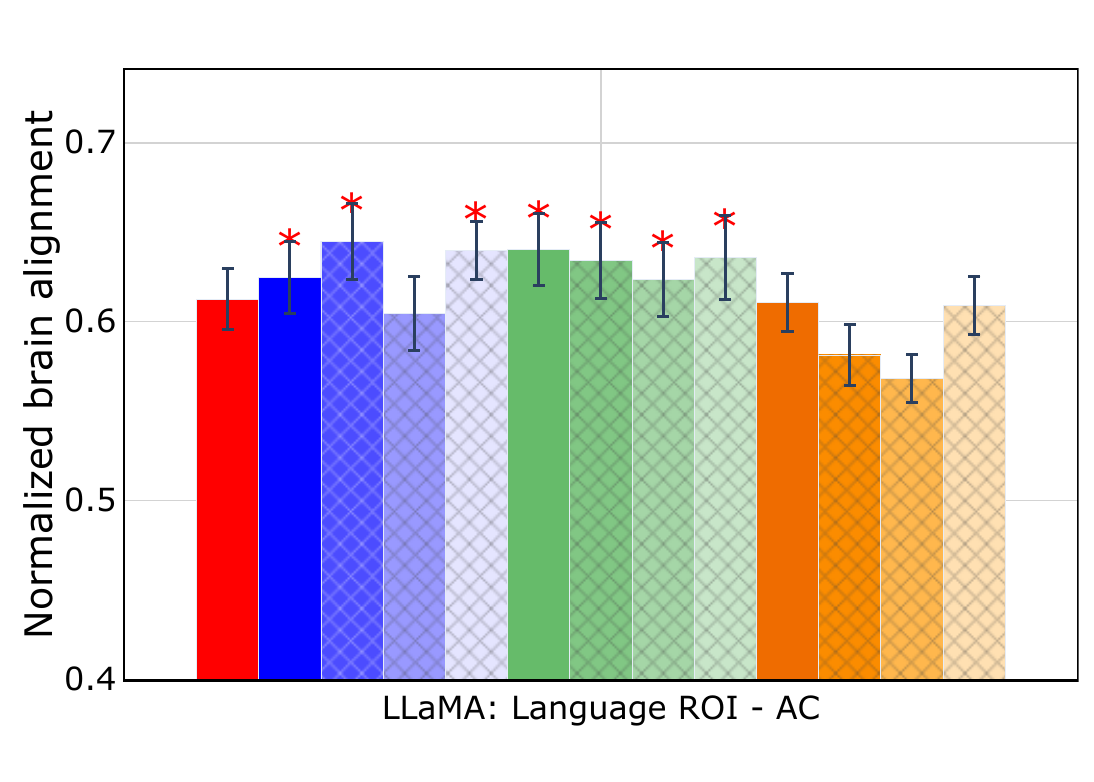}
    \caption{Normalized Predictivity of SLMs, LLMs, and Quantized Language Models for LLaMA-3.2 models.}
    \label{fig:llama_vem_language}
\end{figure*}

Figs.~\ref{fig:qwen_vem_language} and~\ref{fig:llama_vem_language} present the average normalized brain alignment across language-selective ROIs--including AG, ATL, PTL, IFGOrb, MFG, PCC, dmPFC, and AC--for both language model families, comparing SLMs, LLMs, and their quantized variants. 

\begin{figure*}[!ht]
    \centering
    \includegraphics[width=0.52\linewidth]{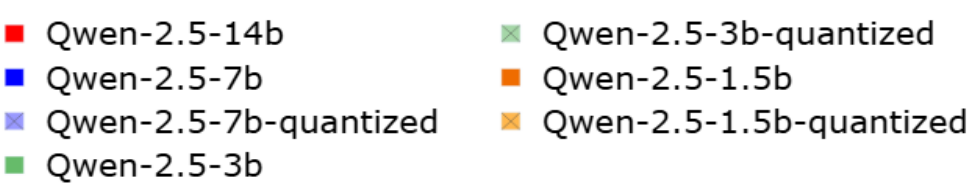}\\
    \includegraphics[width=0.33\linewidth]{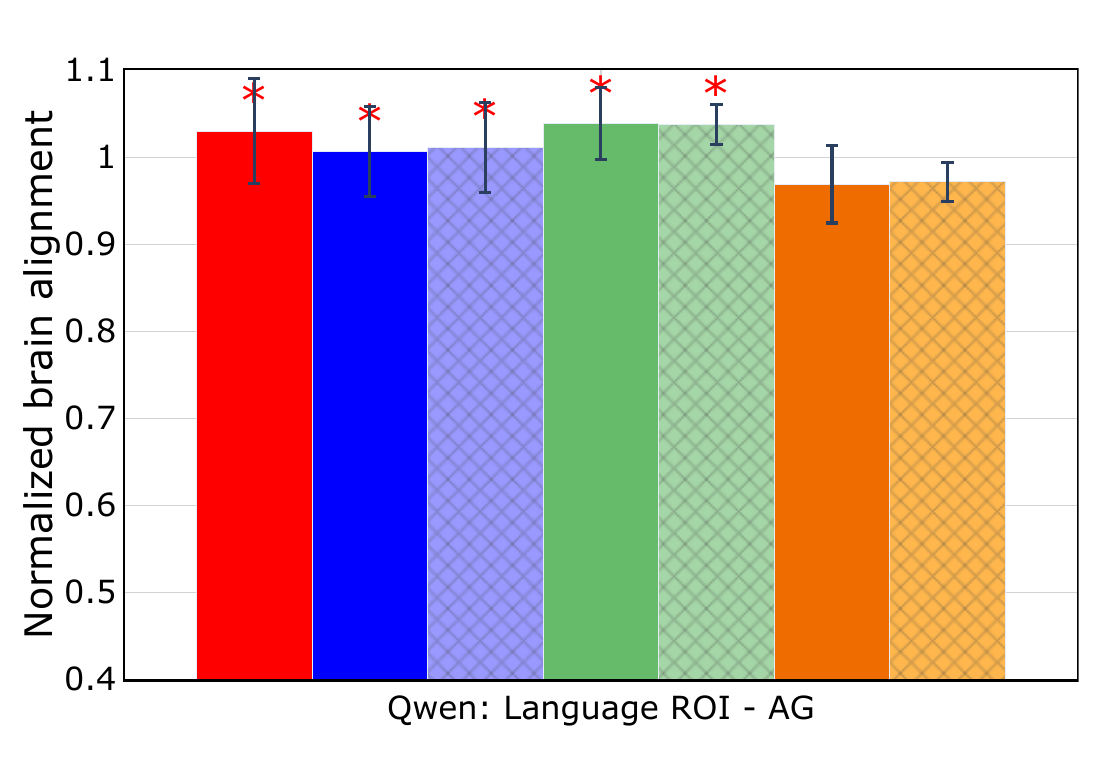}
    \includegraphics[width=0.33\linewidth]{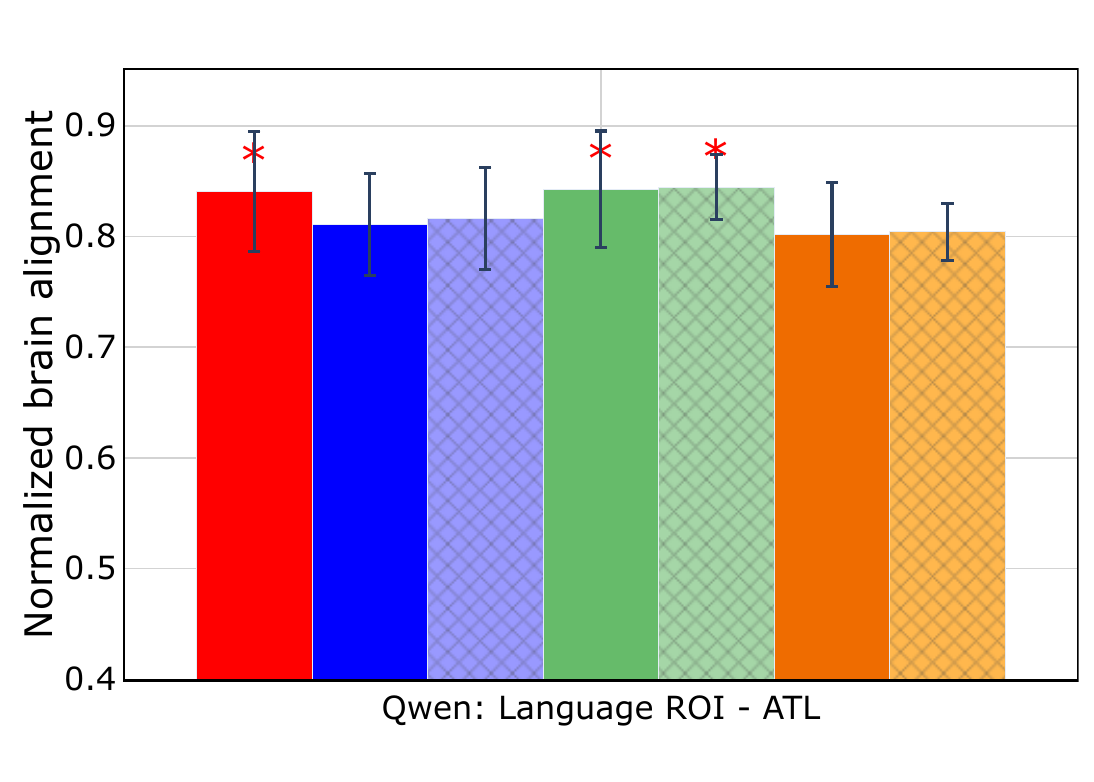}
    \includegraphics[width=0.33\linewidth]{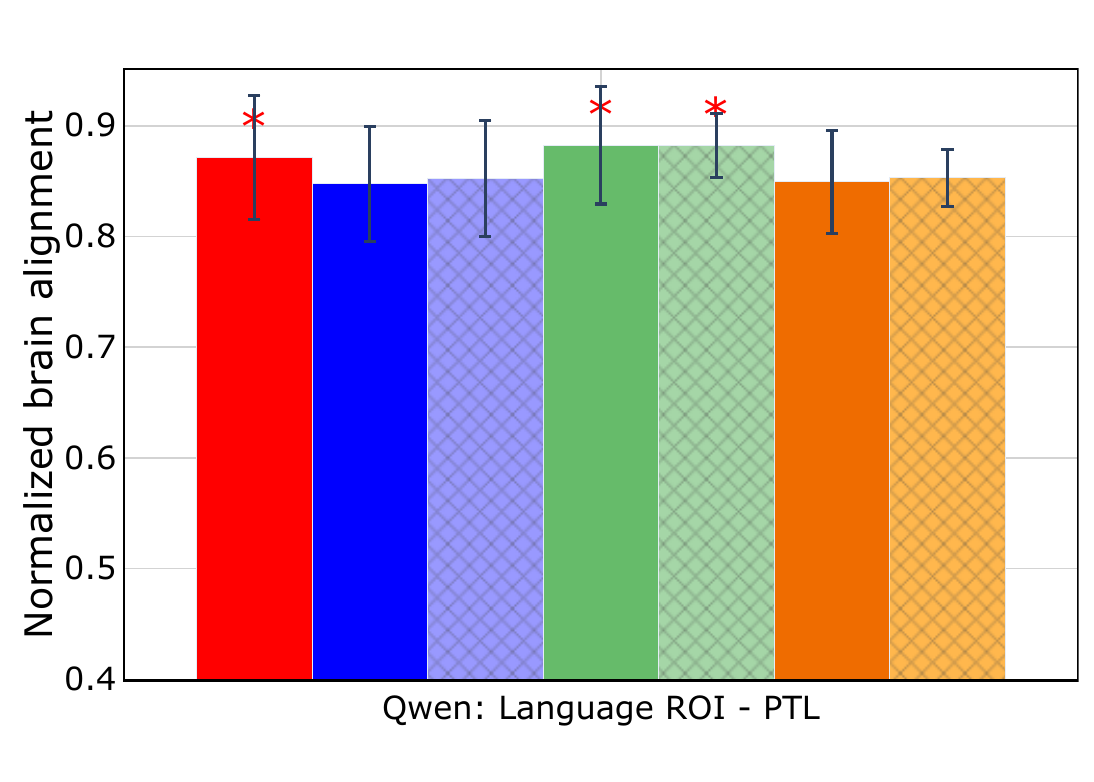}
    \includegraphics[width=0.33\linewidth]{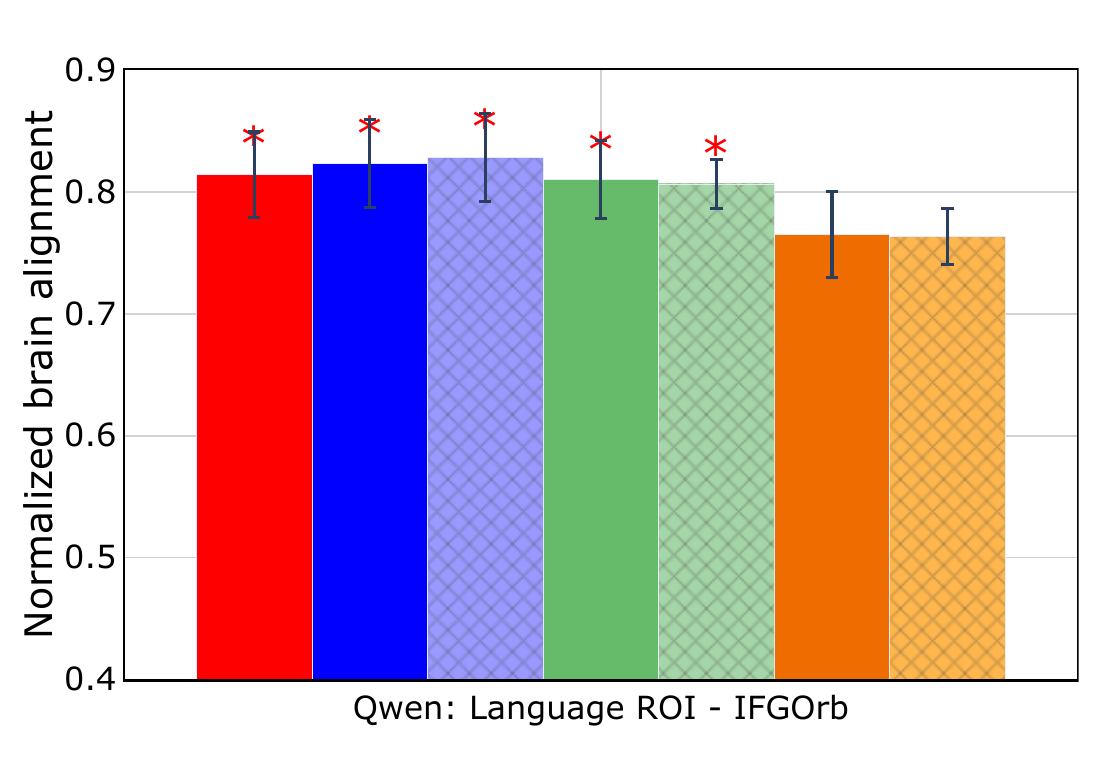}
    \includegraphics[width=0.33\linewidth]{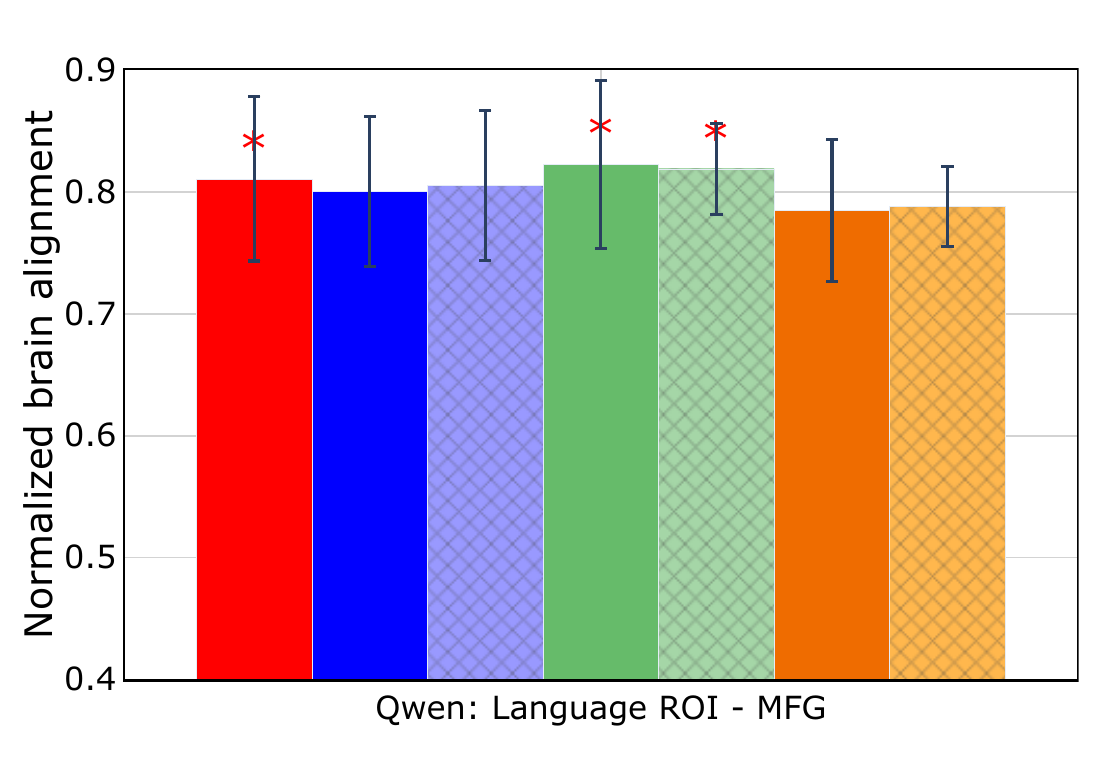}
    \includegraphics[width=0.33\linewidth]{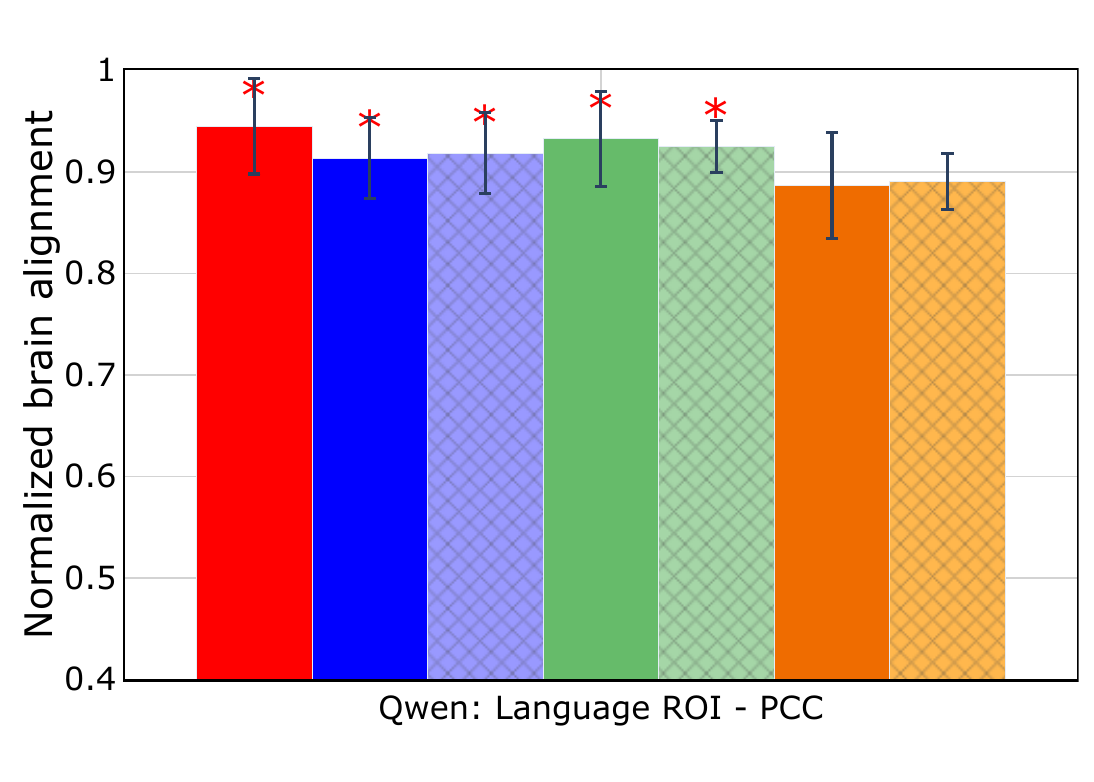}
    \includegraphics[width=0.33\linewidth]{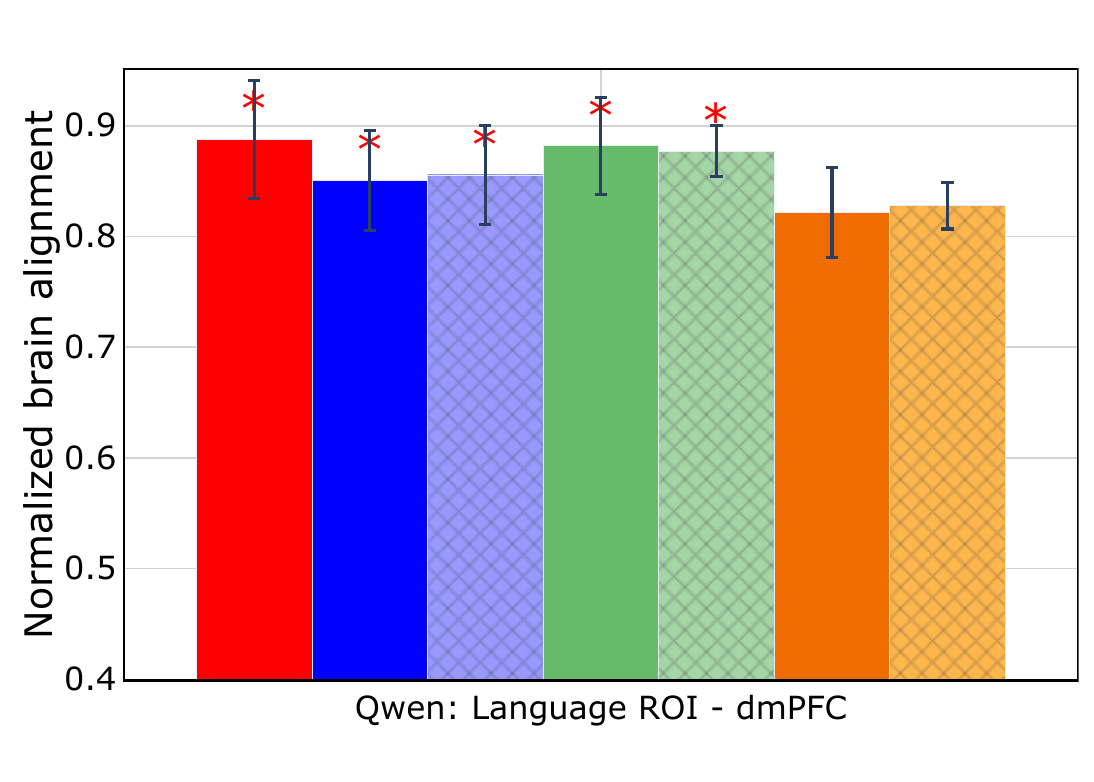}
    \includegraphics[width=0.33\linewidth]{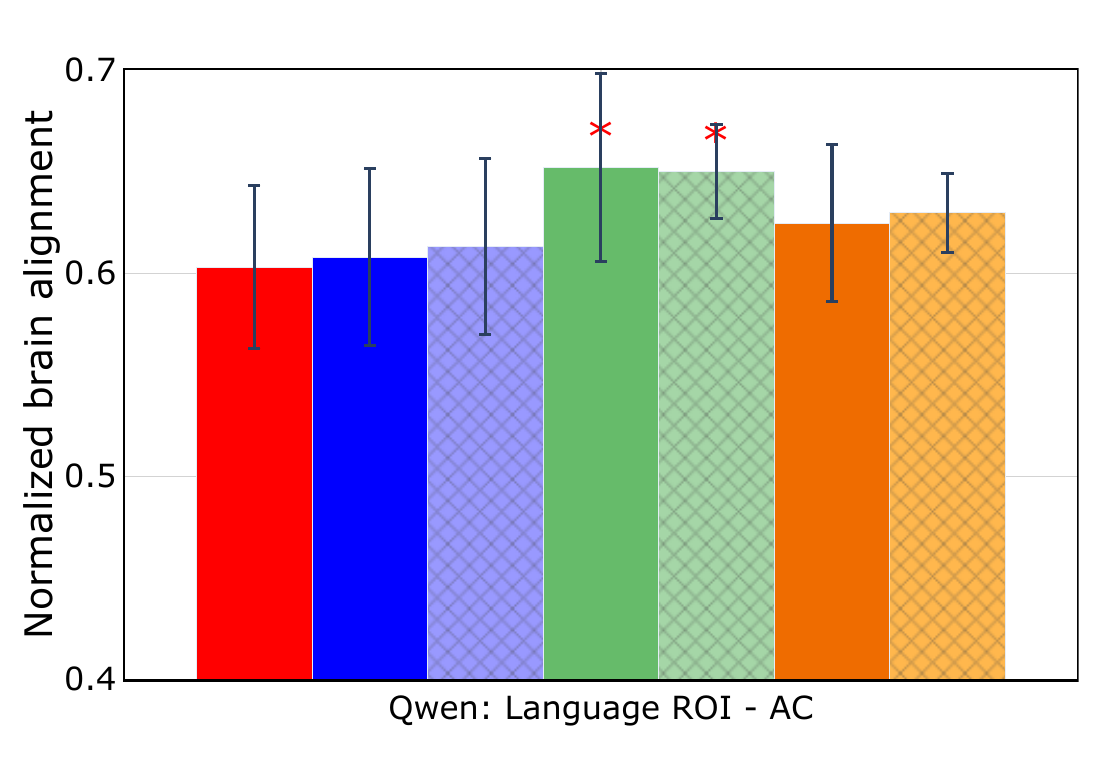}
    \caption{Normalized predictivity of Qwen2.5 SLMs and LLMs, including grouped comparisons of the base and quantized variants.}
    \label{fig:qwen_merged_quantized}
\end{figure*}

\begin{figure*}[!ht]
    \centering
    \includegraphics[width=0.52\linewidth]{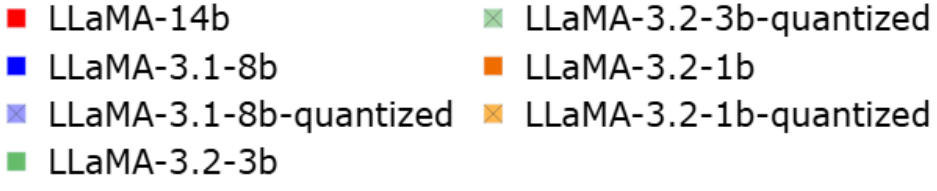} \\
    \includegraphics[width=0.33\linewidth]{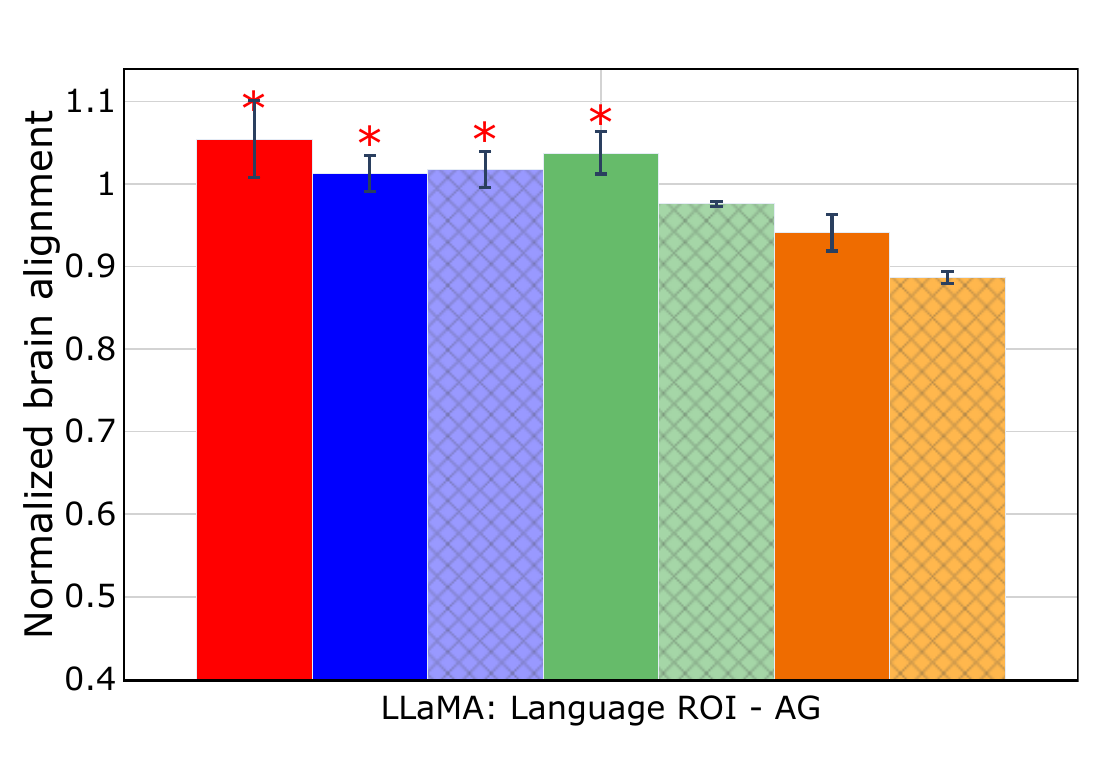}
    \includegraphics[width=0.33\linewidth]{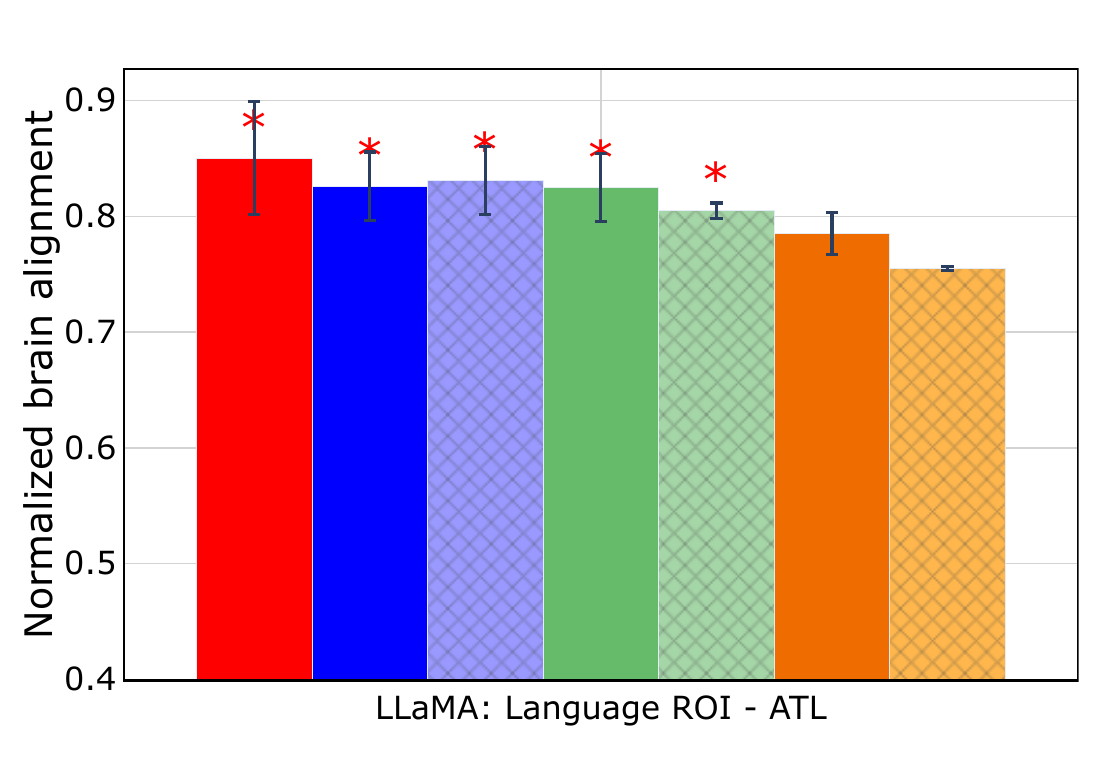}
    \includegraphics[width=0.33\linewidth]{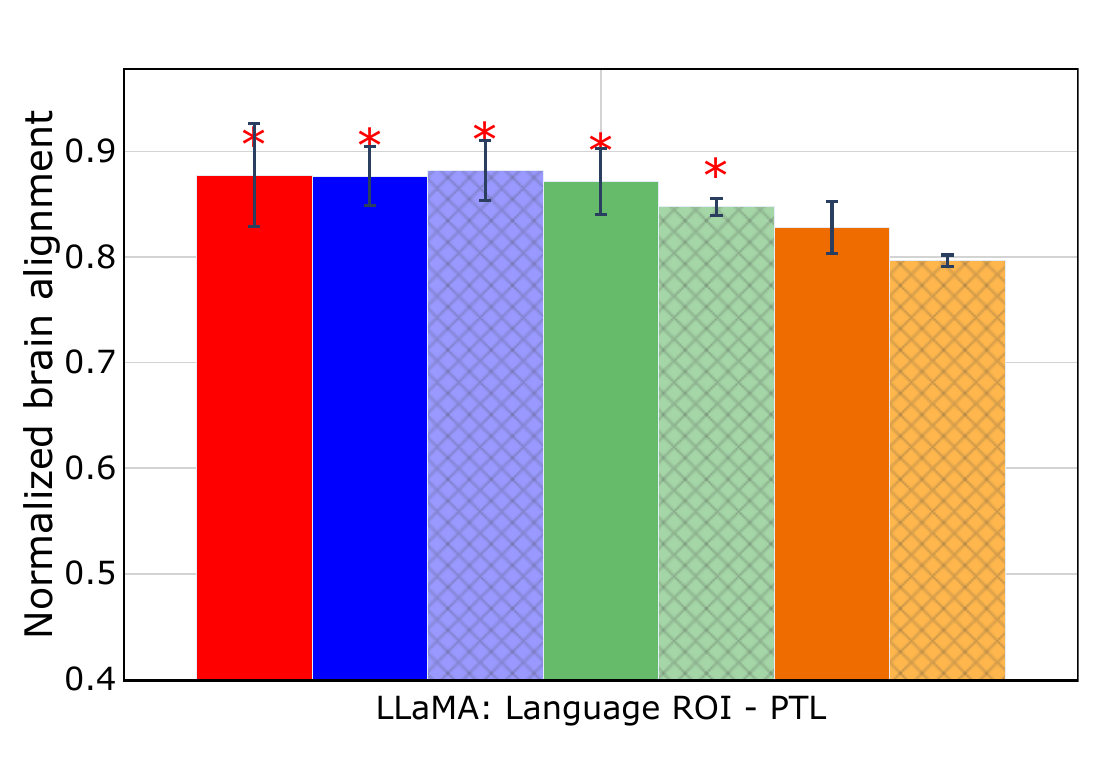}
    \includegraphics[width=0.33\linewidth]{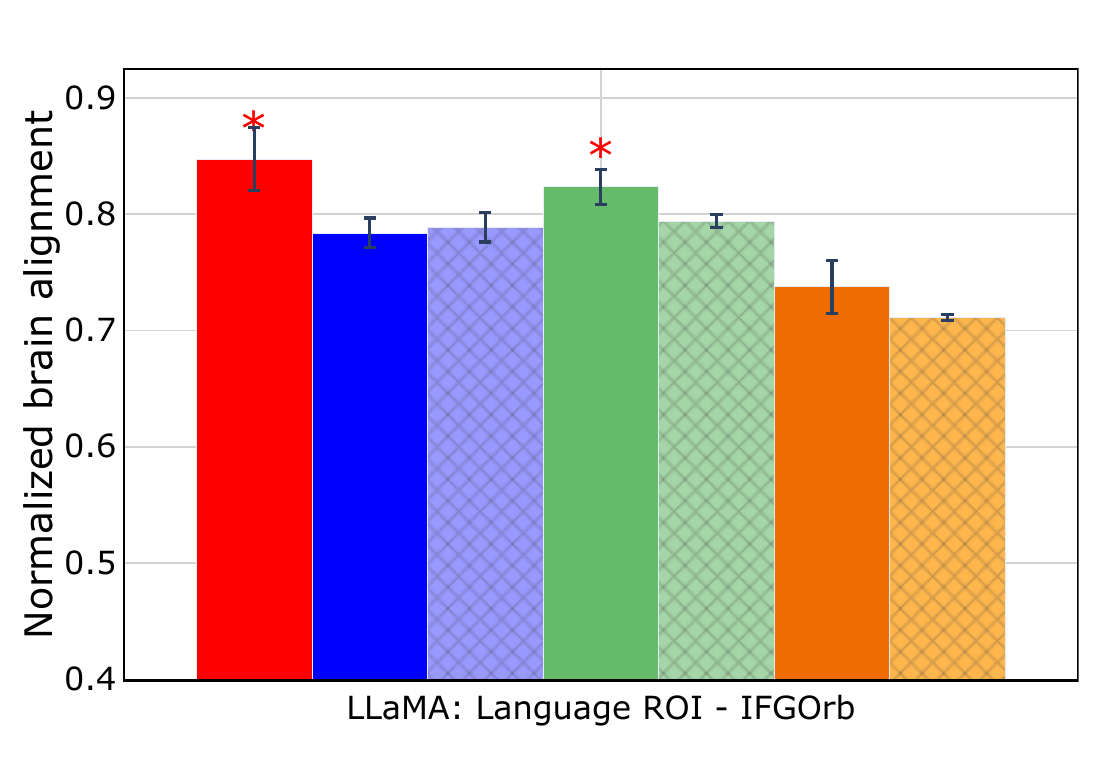}
    \includegraphics[width=0.33\linewidth]{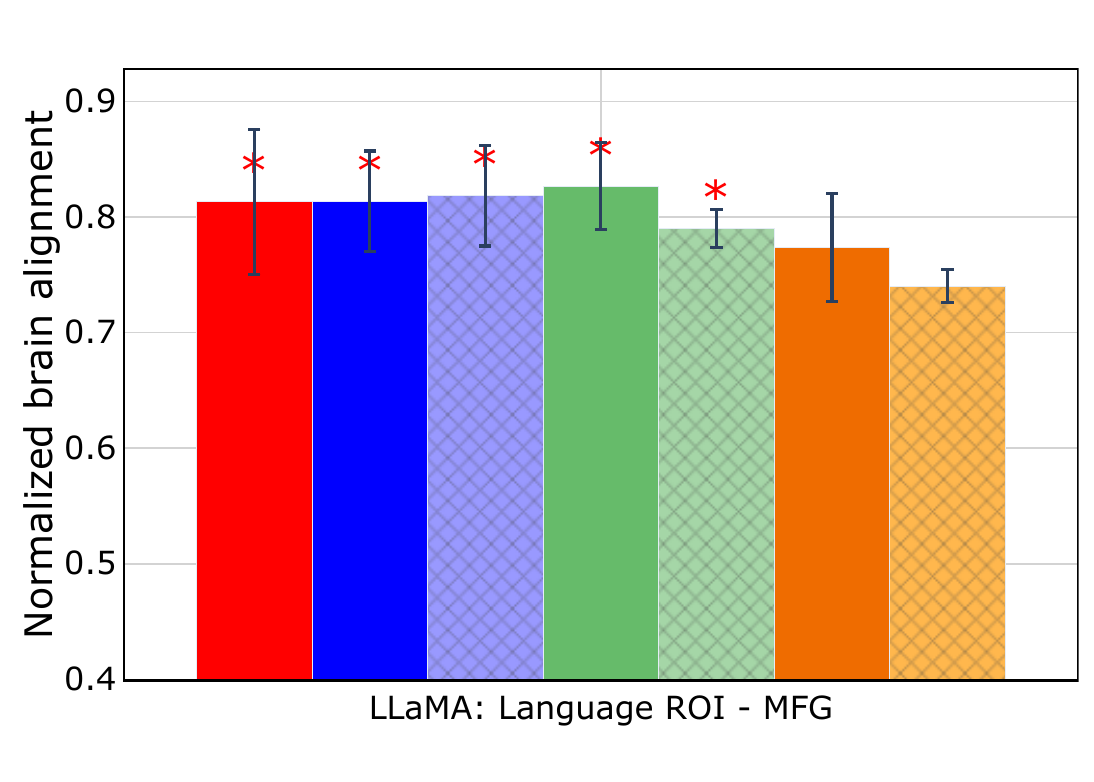}
    \includegraphics[width=0.33\linewidth]{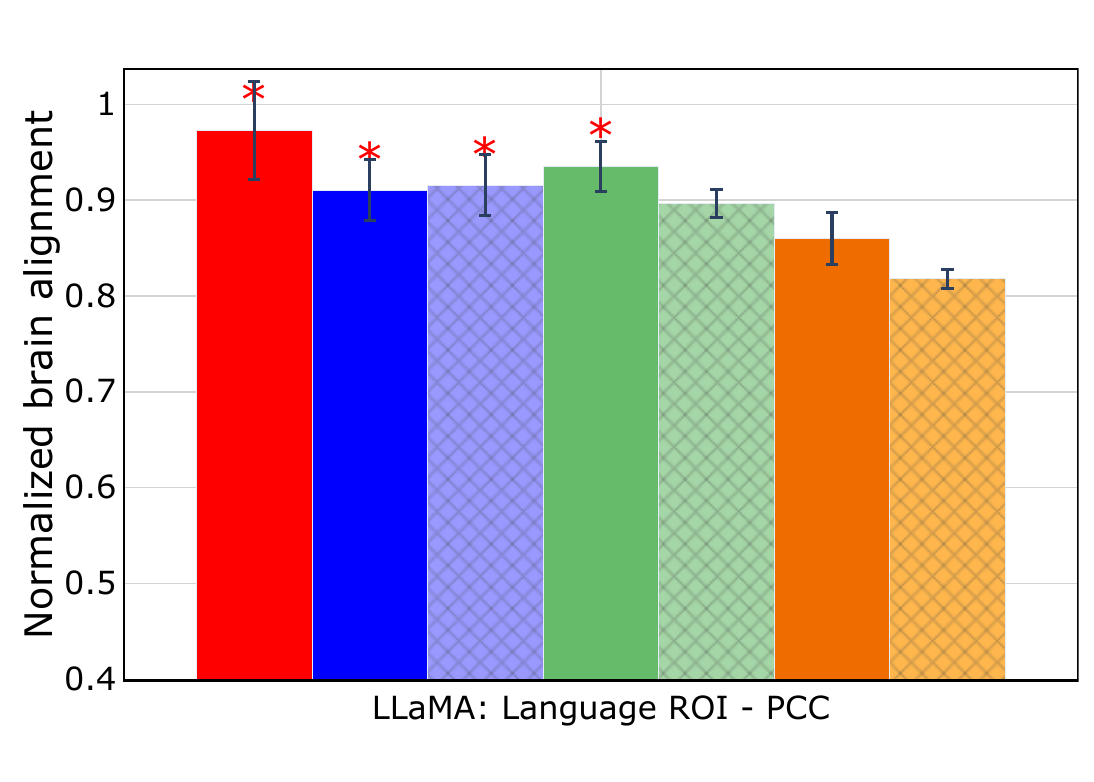}
    \includegraphics[width=0.33\linewidth]{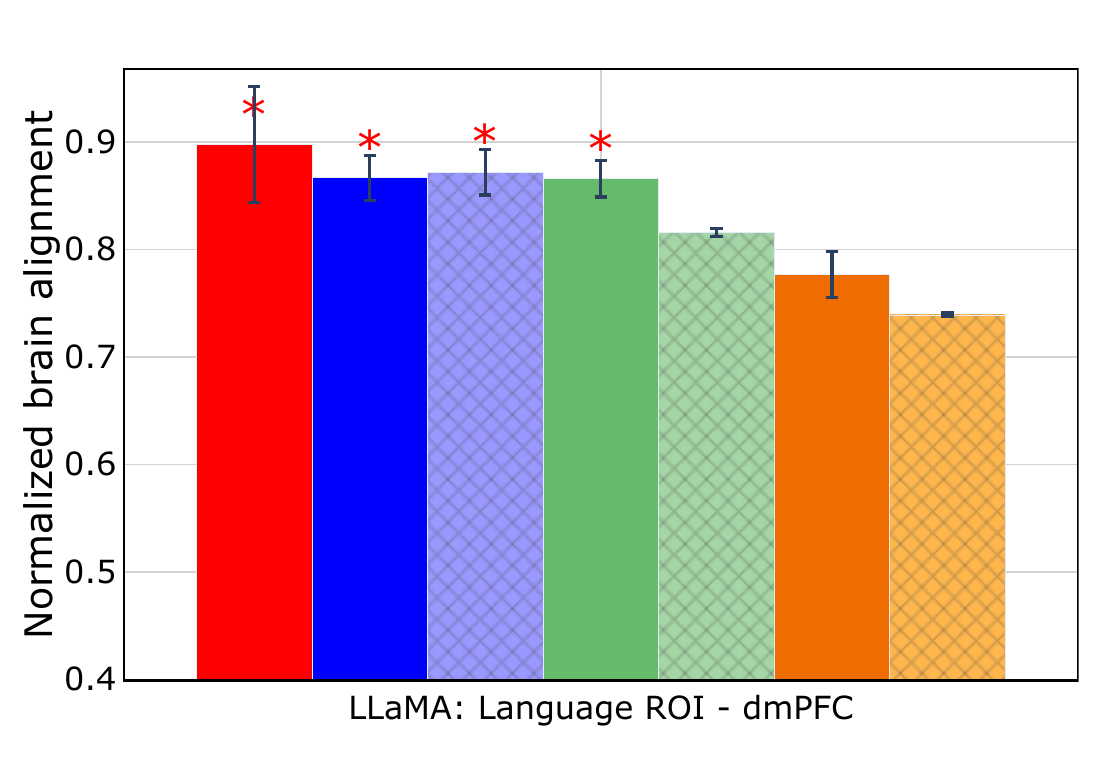}
    \includegraphics[width=0.33\linewidth]{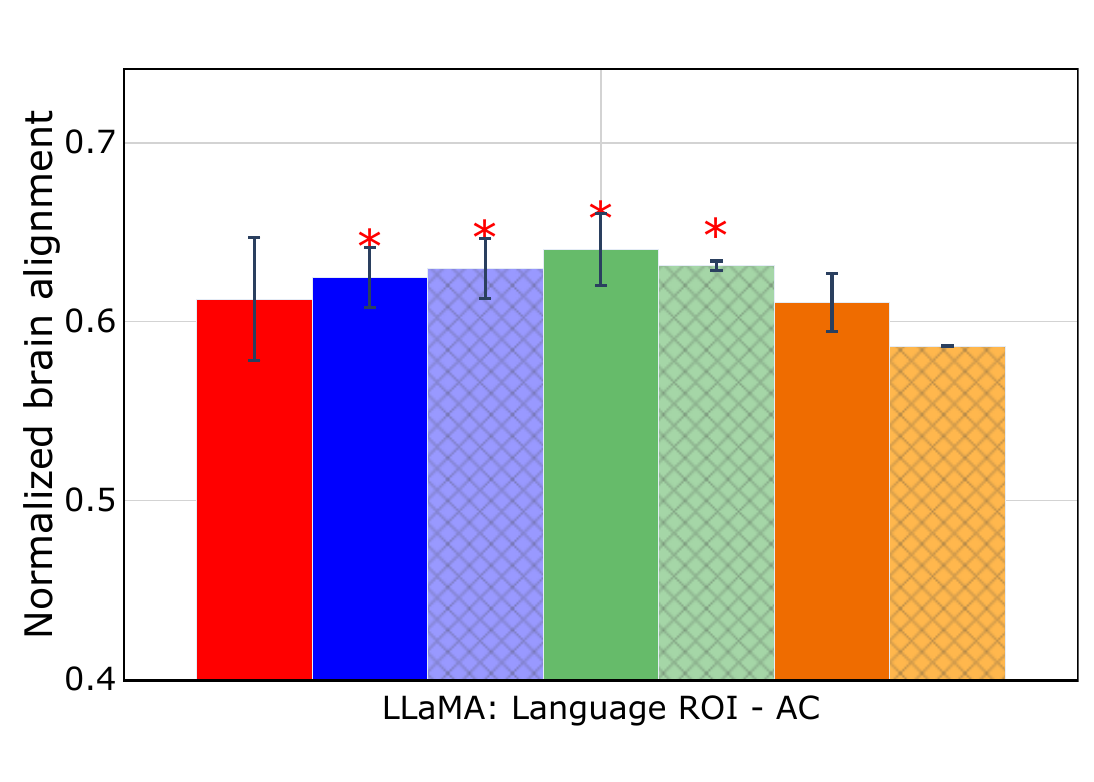}
    \caption{Normalized predictivity of LLaMA3.2 SLMs and LLMs, including grouped comparisons
of the base and quantized variants.}
    \label{fig:llama_merged_quantized}
\end{figure*}


\section{Contrast of estimated model prediction accuracy for various subjects across the two model families}
\label{app:flatMaps}

\begin{figure*}[!ht] 
\centering
\includegraphics[width=0.9\linewidth]{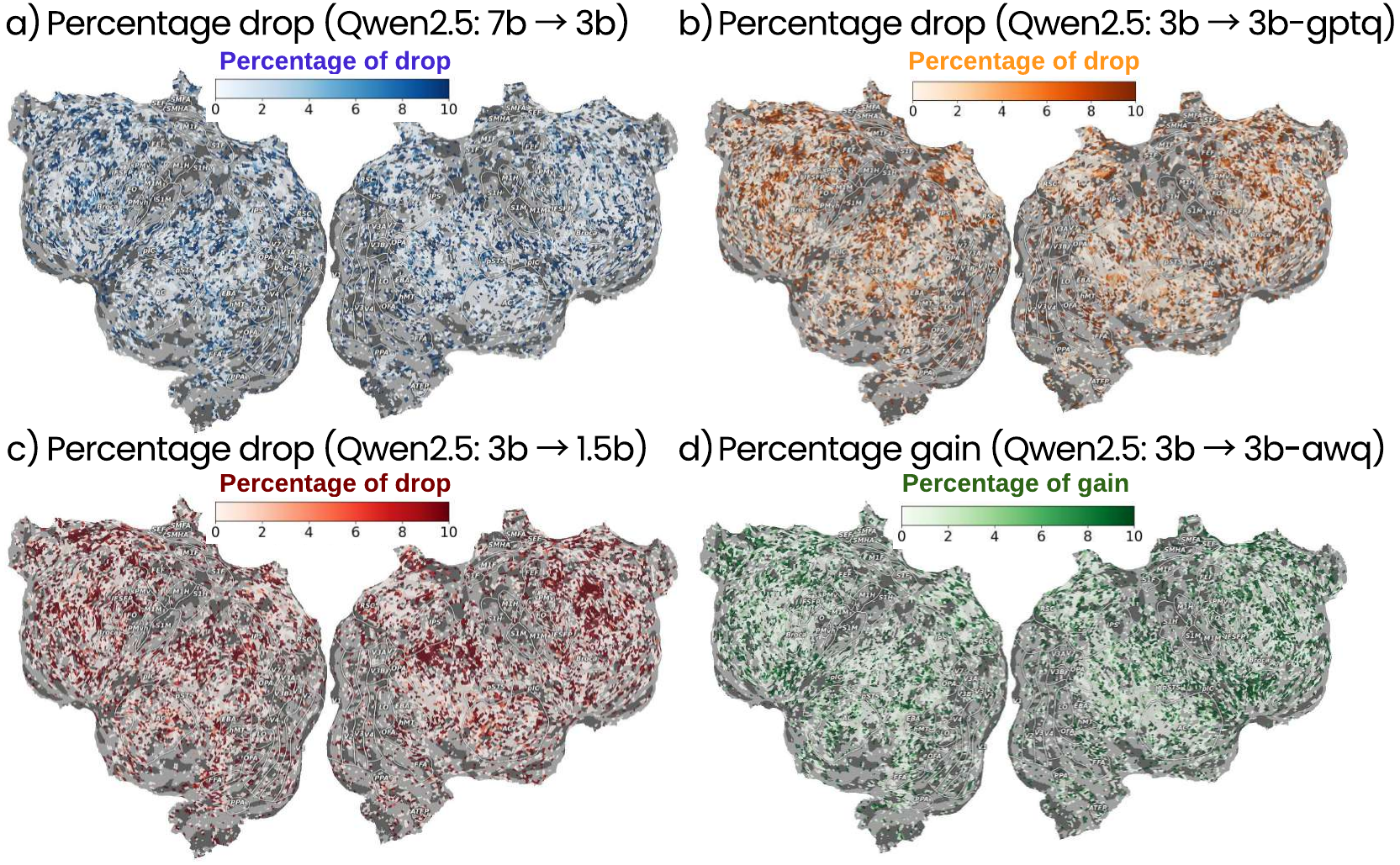}
\caption{Qwen2.5: Percentage change in brain alignment across model scales and quantization methods, shown on the flattened cortical surface of a representative subject (subject-1). Blue, orange, and red voxels indicate regions of information loss ((a) LLMs $\rightarrow$ 3B SLMs, (b) 3B SLMs $\rightarrow$ 3B SLMs GPTQ, (c) 3B SLMs $\rightarrow$ 1.5B SLMs, respectively), (d) while green voxels highlight regions of improvement for 3B SLMs AWQ over 3B SLMs. White voxels denote regions with no change.}
\label{fig:qwen_flatmap_subject01}
\end{figure*}

\begin{figure*}[!ht] 
\centering
\includegraphics[width=0.9\linewidth]{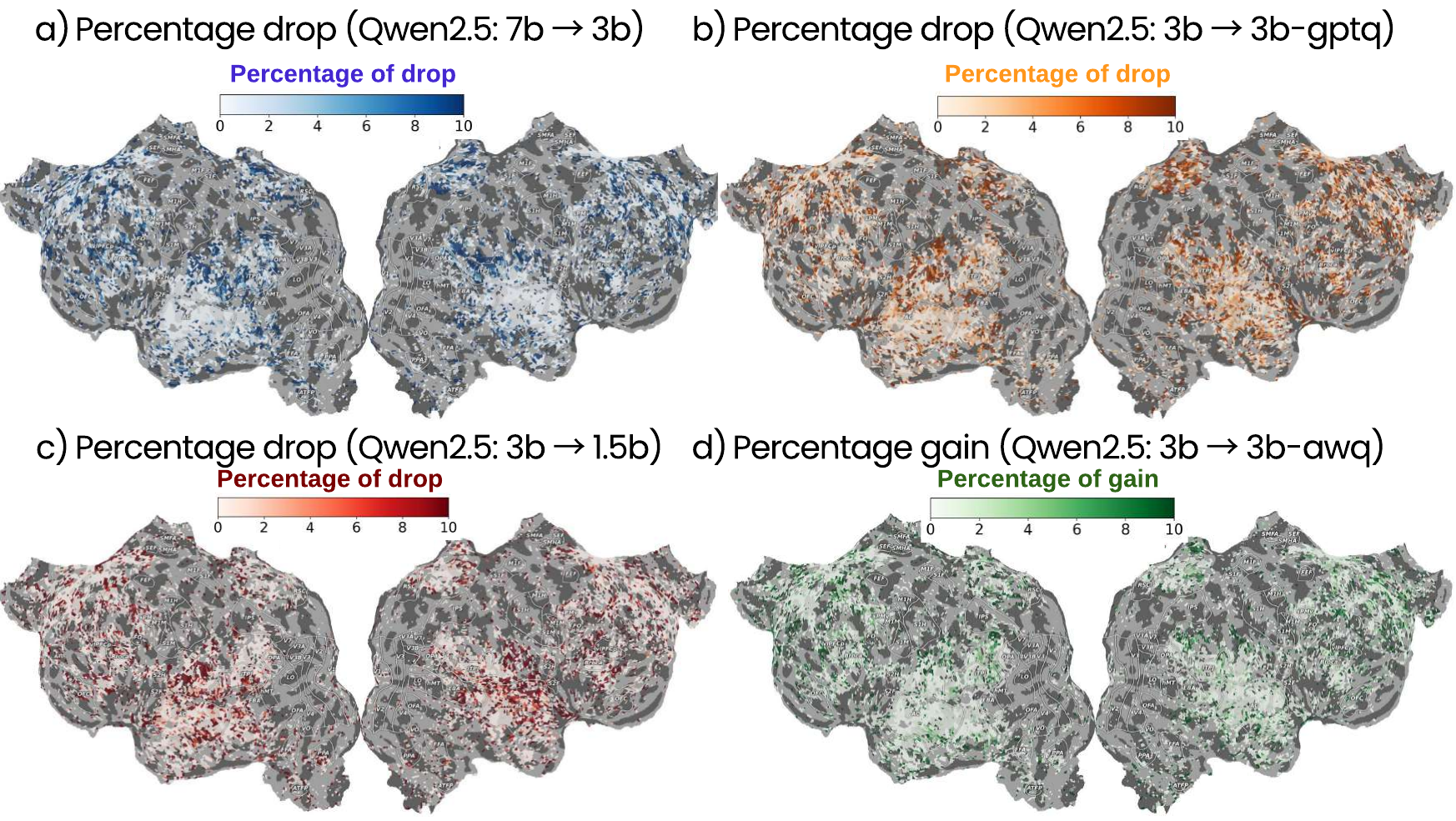}
\caption{Qwen2.5: Percentage change in brain alignment across model scales and quantization methods, shown on the flattened cortical surface of a representative subject (subject-2). Blue, orange, and red voxels indicate regions of information loss ((a) LLMs $\rightarrow$ 3B SLMs, (b) 3B SLMs $\rightarrow$ 3B SLMs GPTQ, (c) 3B SLMs $\rightarrow$ 1.5B SLMs, respectively), (d) while green voxels highlight regions of improvement for 3B SLMs AWQ over 3B SLMs. White voxels denote regions with no change.}
\label{fig:qwen_flatmap_subject02}
\end{figure*}

\begin{figure*}[!ht] 
\centering
\includegraphics[width=0.9\linewidth]{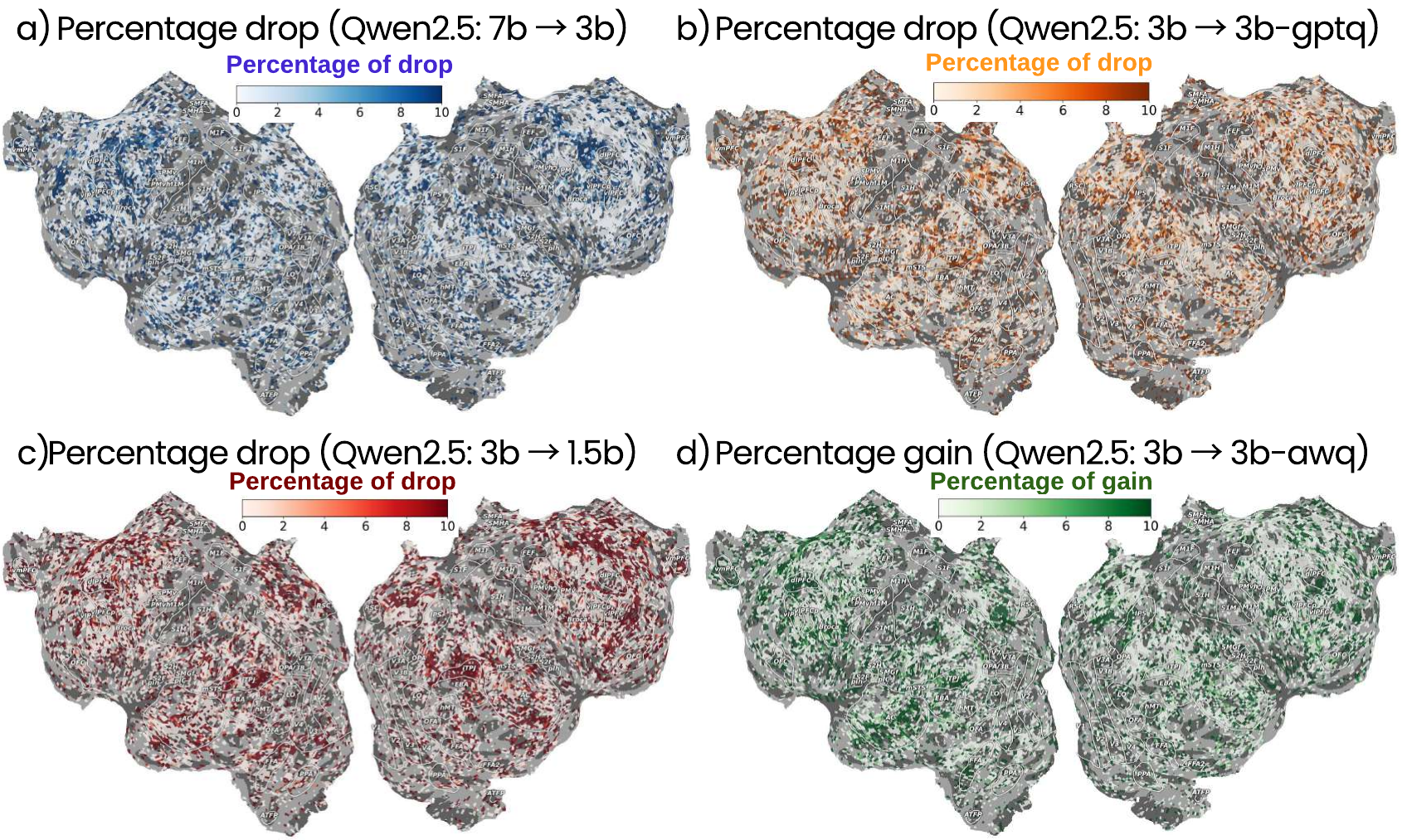}
\caption{Qwen2.5: Percentage change in brain alignment across model scales and quantization methods, shown on the flattened cortical surface of a representative subject (subject-3). Blue, orange, and red voxels indicate regions of information loss ((a) LLMs $\rightarrow$ 3B SLMs, (b) 3B SLMs $\rightarrow$ 3B SLMs GPTQ, (c) 3B SLMs $\rightarrow$ 1.5B SLMs, respectively), (d) while green voxels highlight regions of improvement for 3B SLMs AWQ over 3B SLMs. White voxels denote regions with no change.}
\label{fig:qwen_flatmap_subject03}
\end{figure*}

\begin{figure*}[!ht] 
\centering
\includegraphics[width=0.9\linewidth]{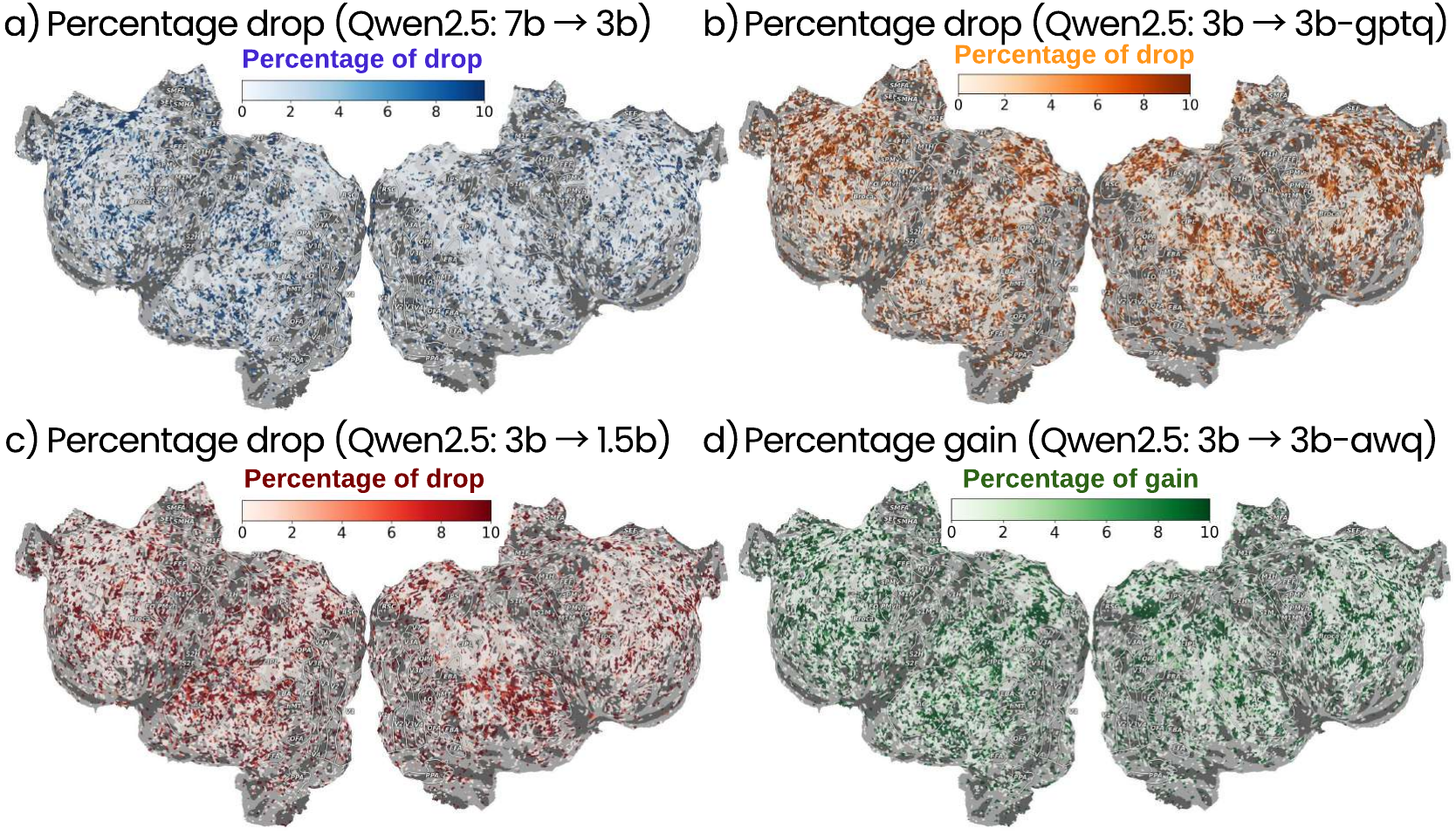}
\caption{Qwen2.5: Percentage change in brain alignment across model scales and quantization methods, shown on the flattened cortical surface of a representative subject (subject-7). Blue, orange, and red voxels indicate regions of information loss ((a) LLMs $\rightarrow$ 3B SLMs, (b) 3B SLMs $\rightarrow$ 3B SLMs GPTQ, (c) 3B SLMs $\rightarrow$ 1.5B SLMs, respectively), (d) while green voxels highlight regions of improvement for 3B SLMs AWQ over 3B SLMs. White voxels denote regions with no change.}
\label{fig:qwen_flatmap_subject07}
\end{figure*}

\begin{figure*}[!ht] 
\centering
\includegraphics[width=0.9\linewidth]{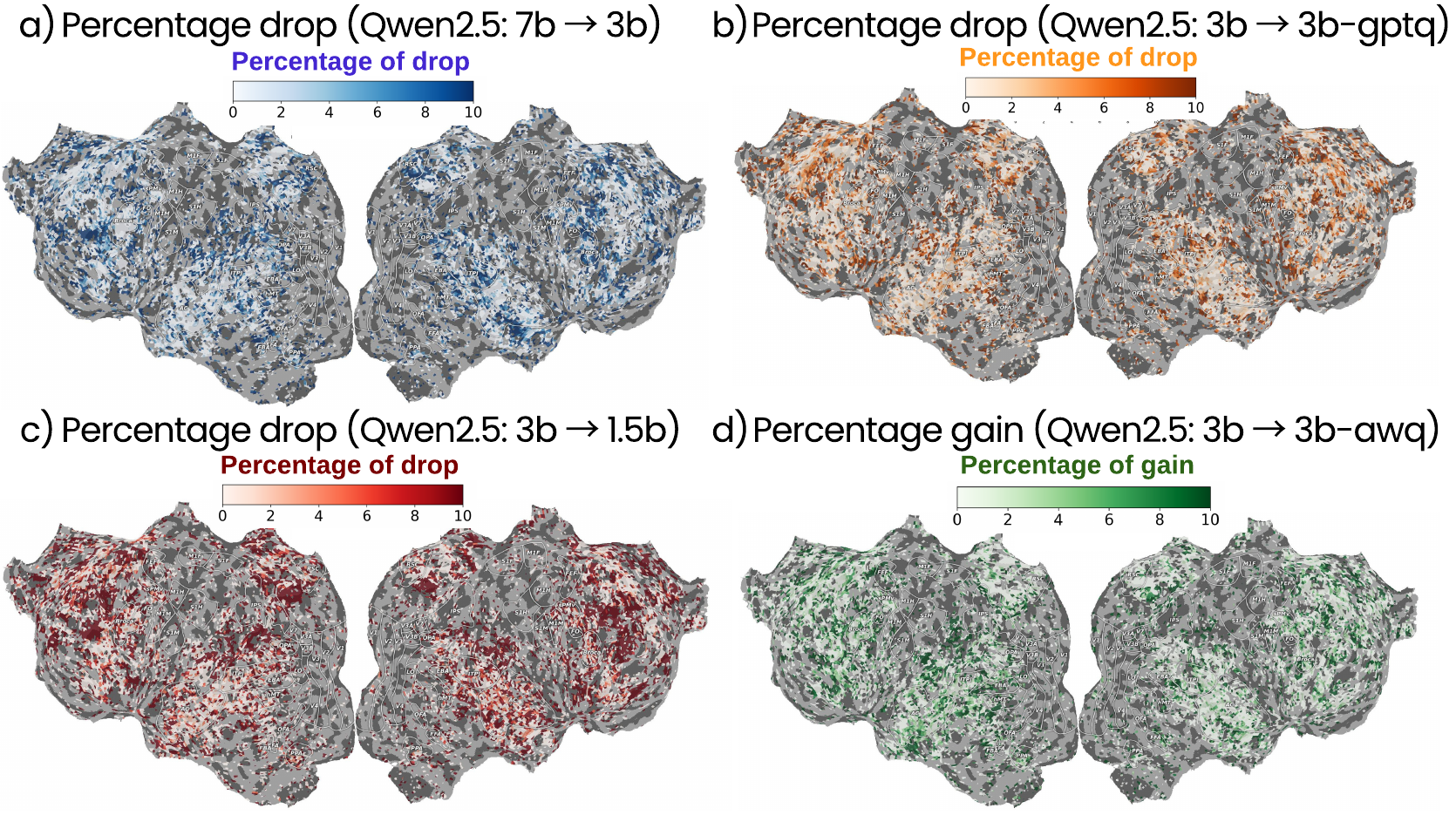}
\caption{Qwen2.5: Percentage change in brain alignment across model scales and quantization methods, shown on the flattened cortical surface of a representative subject (subject-8). Blue, orange, and red voxels indicate regions of information loss ((a) LLMs $\rightarrow$ 3B SLMs, (b) 3B SLMs $\rightarrow$ 3B SLMs GPTQ, (c) 3B SLMs $\rightarrow$ 1.5B SLMs, respectively), (d) while green voxels highlight regions of improvement for 3B SLMs AWQ over 3B SLMs. White voxels denote regions with no change.}
\label{fig:qwen_flatmap_subject08}
\end{figure*}

\begin{figure*}[!ht] 
\centering
\includegraphics[width=0.9\linewidth]{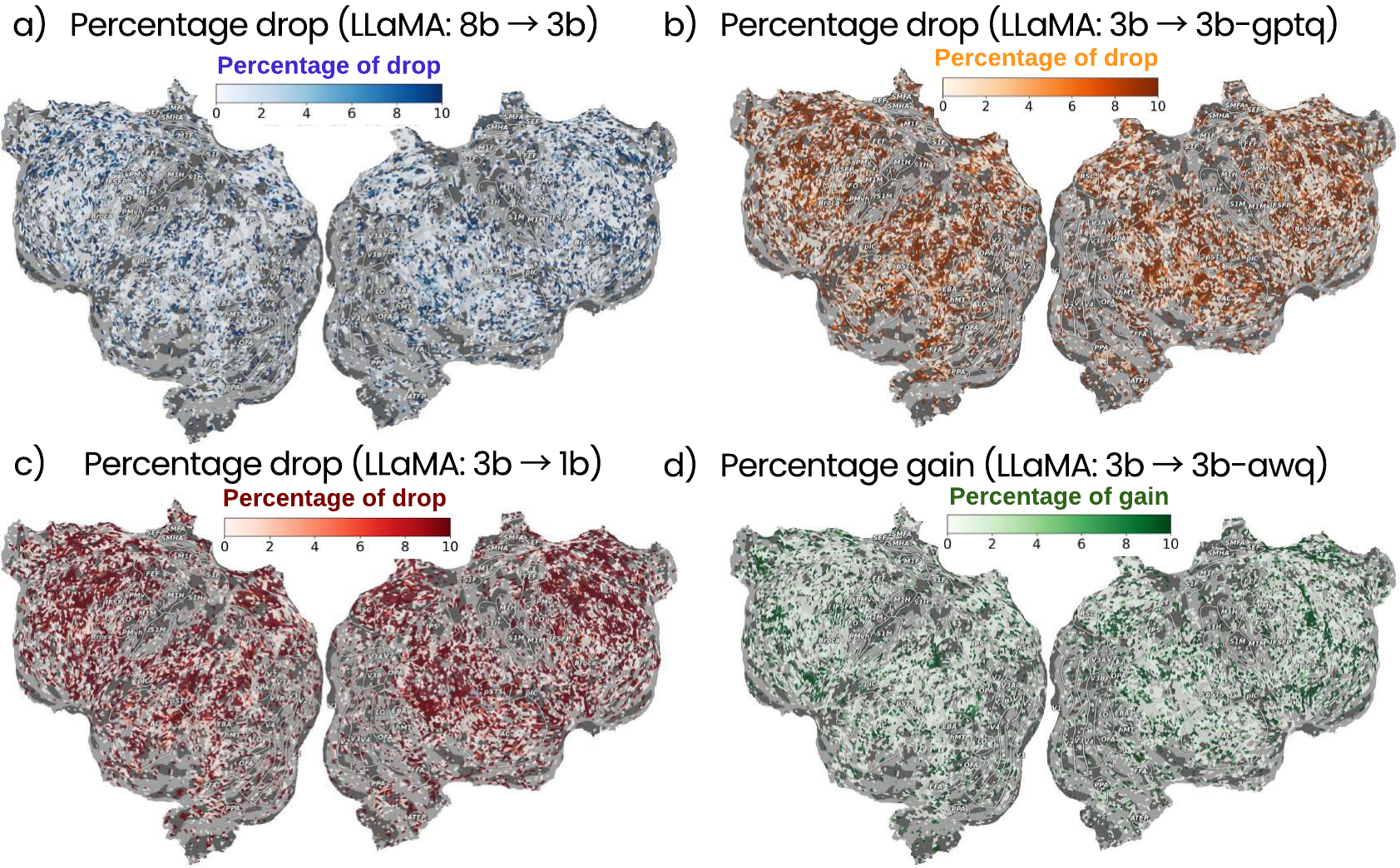}
\caption{LLaMA-3.2: Percentage change in brain alignment across model scales and quantization methods, shown on the flattened cortical surface of a representative subject (subject-1). Blue, orange, and red voxels indicate regions of information loss ((a) LLMs $\rightarrow$ 3B SLMs, (b) 3B SLMs $\rightarrow$ 3B SLMs GPTQ, (c) 3B SLMs $\rightarrow$ 1.5B SLMs, respectively), (d) while green voxels highlight regions of improvement for 3B SLMs AWQ over 3B SLMs. White voxels denote regions with no change.}
\label{fig:llama_flatmap_subject01}
\end{figure*}

\begin{figure*}[!ht] 
\centering
\includegraphics[width=0.9\linewidth]{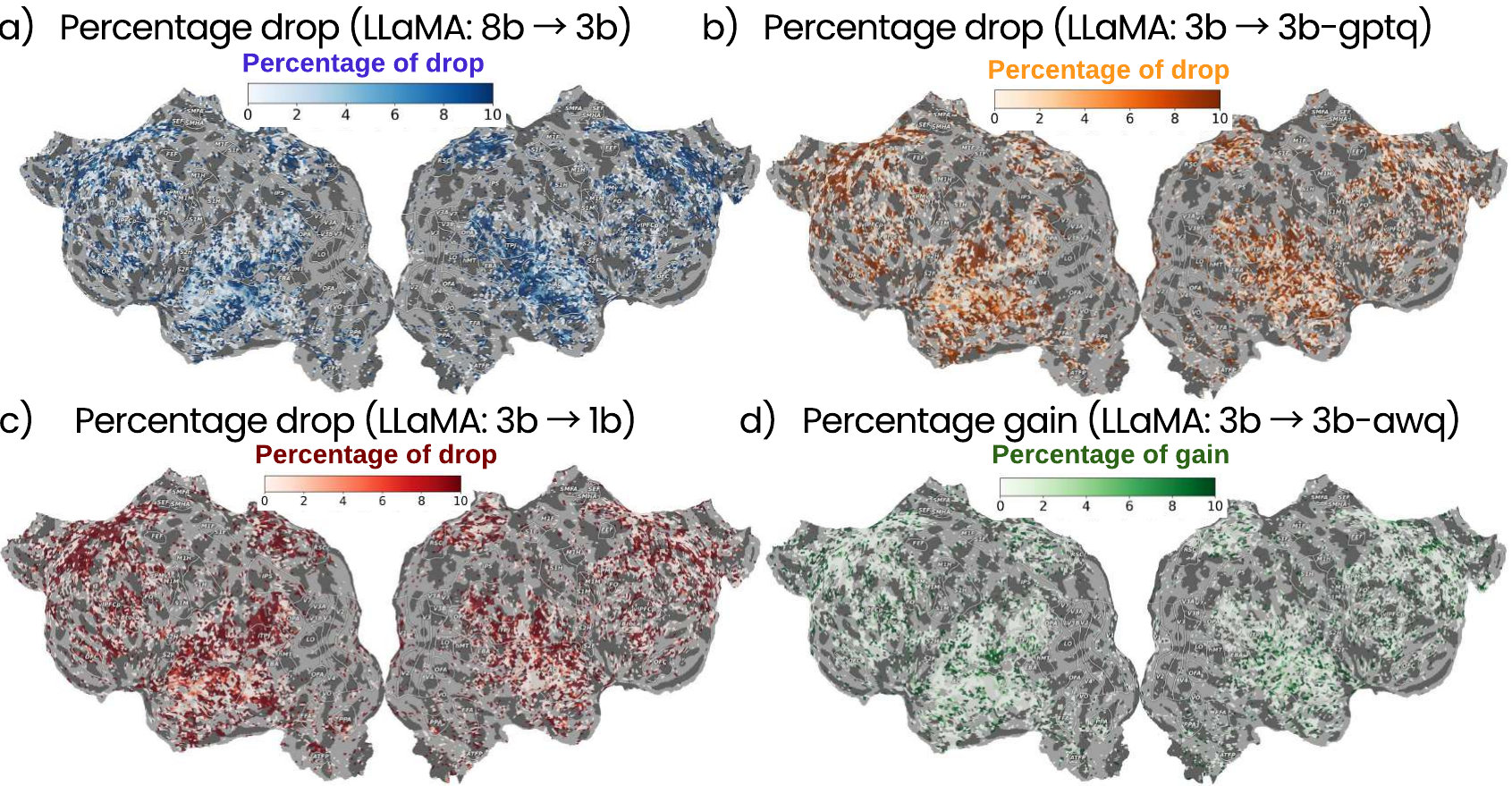}
\caption{LLaMA-3.2: Percentage change in brain alignment across model scales and quantization methods, shown on the flattened cortical surface of a representative subject (subject-2). Blue, orange, and red voxels indicate regions of information loss ((a) LLMs $\rightarrow$ 3B SLMs, (b) 3B SLMs $\rightarrow$ 3B SLMs GPTQ, (c) 3B SLMs $\rightarrow$ 1.5B SLMs, respectively), (d) while green voxels highlight regions of improvement for 3B SLMs AWQ over 3B SLMs. White voxels denote regions with no change.}
\label{fig:llama_flatmap_subject02}
\end{figure*}

\begin{figure*}[!ht] 
\centering
\includegraphics[width=0.9\linewidth]{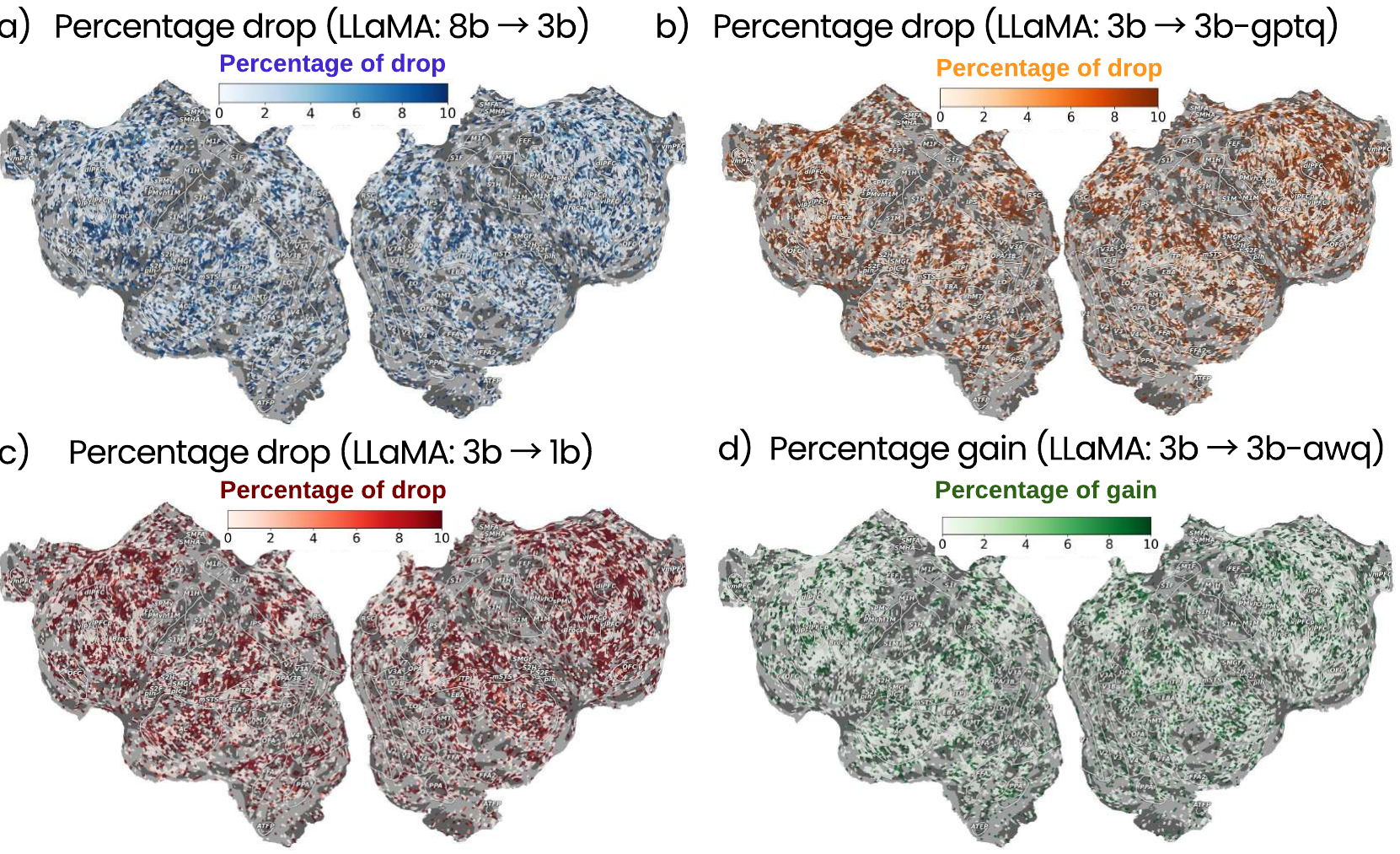}
\caption{LLaMA-3.2: Percentage change in brain alignment across model scales and quantization methods, shown on the flattened cortical surface of a representative subject (subject-3). Blue, orange, and red voxels indicate regions of information loss ((a) LLMs $\rightarrow$ 3B SLMs, (b) 3B SLMs $\rightarrow$ 3B SLMs GPTQ, (c) 3B SLMs $\rightarrow$ 1.5B SLMs, respectively), (d) while green voxels highlight regions of improvement for 3B SLMs AWQ over 3B SLMs. White voxels denote regions with no change.}
\label{fig:llama_flatmap_subject03}
\end{figure*}

\begin{figure*}[!ht] 
\centering
\includegraphics[width=0.9\linewidth]{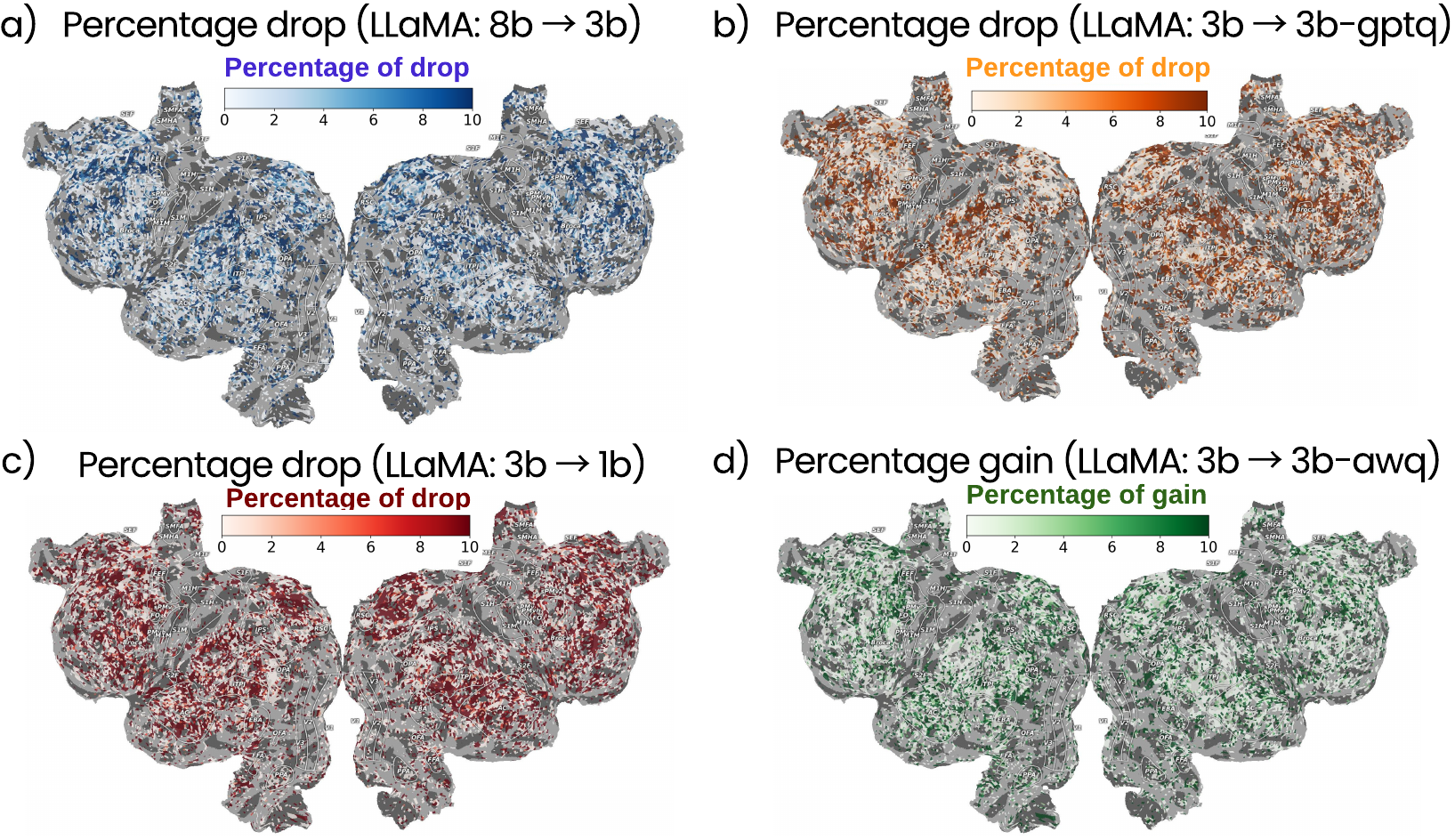}
\caption{LLaMA-3.2: Percentage change in brain alignment across model scales and quantization methods, shown on the flattened cortical surface of a representative subject (subject-7). Blue, orange, and red voxels indicate regions of information loss ((a) LLMs $\rightarrow$ 3B SLMs, (b) 3B SLMs $\rightarrow$ 3B SLMs GPTQ, (c) 3B SLMs $\rightarrow$ 1.5B SLMs, respectively), (d) while green voxels highlight regions of improvement for 3B SLMs AWQ over 3B SLMs. White voxels denote regions with no change.}
\label{fig:llama_flatmap_subject05}
\end{figure*}

\begin{figure*}[!ht] 
\centering
\includegraphics[width=0.9\linewidth]{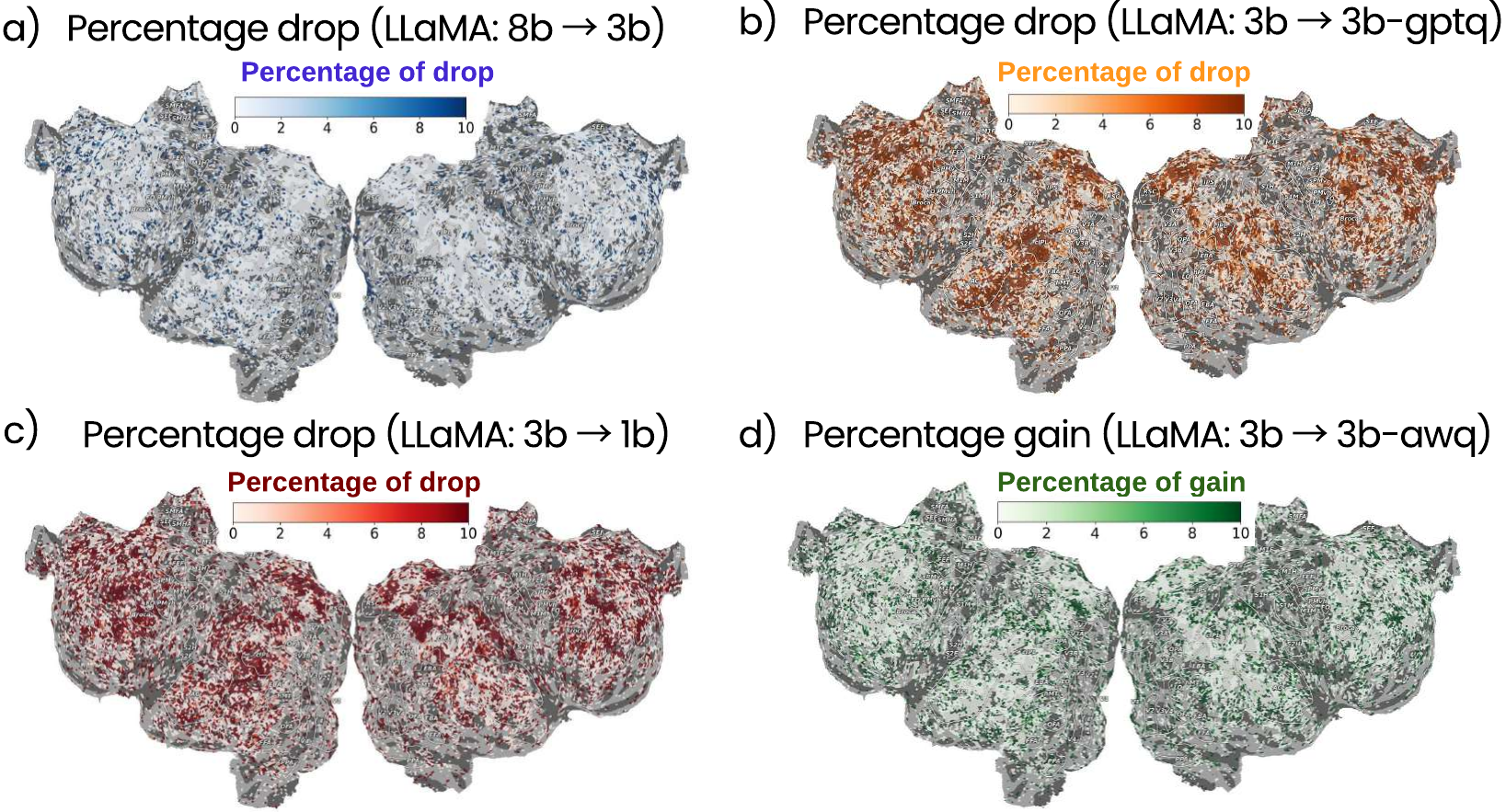}
\caption{LLaMA-3.2: Percentage change in brain alignment across model scales and quantization methods, shown on the flattened cortical surface of a representative subject (subject-8). Blue, orange, and red voxels indicate regions of information loss ((a) LLMs $\rightarrow$ 3B SLMs, (b) 3B SLMs $\rightarrow$ 3B SLMs GPTQ, (c) 3B SLMs $\rightarrow$ 1.5B SLMs, respectively), (d) while green voxels highlight regions of improvement for 3B SLMs AWQ over 3B SLMs. White voxels denote regions with no change.}
\label{fig:llama_flatmap_subject07}
\end{figure*}

\begin{figure*}[!ht] 
\centering
\includegraphics[width=0.9\linewidth]{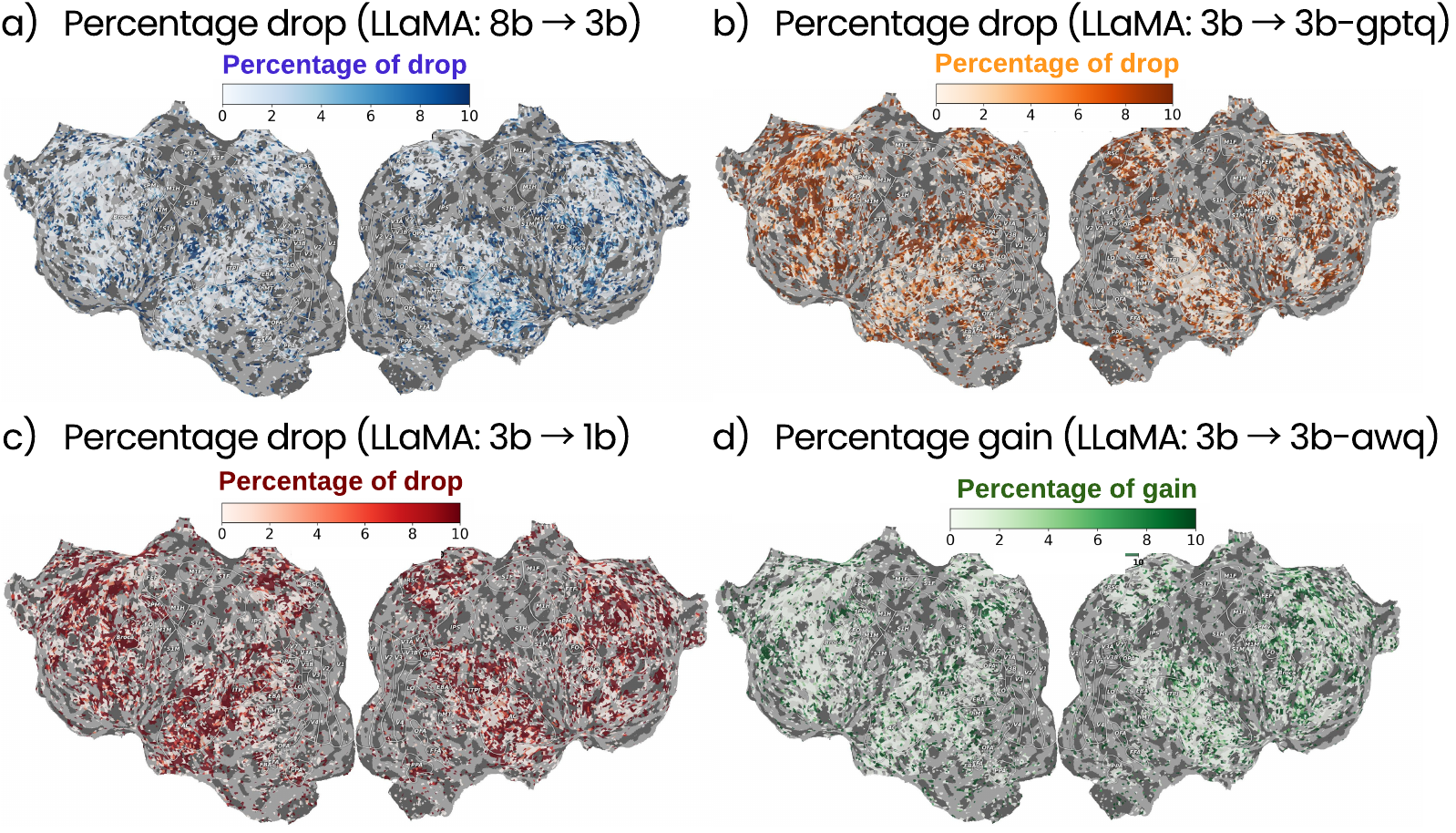}
\caption{LLaMA-3.2: Percentage change in brain alignment across model scales and quantization methods, shown on the flattened cortical surface of a representative subject (subject-8). Blue, orange, and red voxels indicate regions of information loss ((a) LLMs $\rightarrow$ 3B SLMs, (b) 3B SLMs $\rightarrow$ 3B SLMs GPTQ, (c) 3B SLMs $\rightarrow$ 1.5B SLMs, respectively), (d) while green voxels highlight regions of improvement for 3B SLMs AWQ over 3B SLMs. White voxels denote regions with no change.}
\label{fig:llama_flatmap_subject08}
\end{figure*}

Figs.~\ref{fig:qwen_flatmap_subject01},~\ref{fig:qwen_flatmap_subject02},~\ref{fig:qwen_flatmap_subject03},~\ref{fig:qwen_flatmap_subject07}, and~\ref{fig:qwen_flatmap_subject08} show voxel-wise percentage changes in brain alignment across model scales and quantization strategies for Qwen2.5 in the remaining participants. Corresponding results for LLaMA-3.2 are provided in Figs.~\ref{fig:llama_flatmap_subject01},~\ref{fig:llama_flatmap_subject02},~\ref{fig:llama_flatmap_subject03},~\ref{fig:llama_flatmap_subject05},~\ref{fig:llama_flatmap_subject07}, and~\ref{fig:llama_flatmap_subject08}. 

We make the following observations across two families of language models: (i) Scaling down from LLMs to 3B SLMs: reductions in brain alignment are negligible in the bilateral temporal lobe and remain under 5\% in the parietal cortex and IFGorb. Large cortical regions remain white, indicating that 3B SLMs preserve brain-relevant representations comparable to LLMs. The blue-marked voxels are sparse and localized, suggesting only limited information loss. (ii) 3B SLMs to 3B SLMs GPTQ: applying GPTQ leads to widespread orange-marked voxels, reflecting consistent losses across distributed cortical regions. While some areas remain preserved (white voxels), the extent of information loss is greater than that observed with downscaling alone, confirming that GPTQ disproportionately disrupts brain-relevant alignment. (iii) 3B SLMs to 1.5B SLMs: relative to the 3B baseline, we observe extensive red-marked voxels, especially in temporal and language-related regions, indicating larger drops in alignment. This demonstrates the limits of scaling, as ultra-small models fail to capture brain-relevant representations. (iv) 3B SLMs $\rightarrow$ 3B SLMs AWQ: AWQ produces localized green-marked voxels, indicating regions of improved alignment relative to the 3B baseline, particularly in the IFG and AG. Most of the cortex remains unchanged (white), suggesting that AWQ maintains representational fidelity while offering modest regional gains.

\section{Impact of Linguistic Competence for Qwen2.5 and LLaMA-3.2 Models}
\label{app:linguisticCompetencellama}

From Table~\ref{tab:qwen_flashholmes_quant_7b}, we find that discourse competence emerges as the most influential factor. As models scale from smaller to larger versions, discourse probing scores (e.g., bridging, coreference) increase substantially, and these improvements align closely with gains in brain predictivity. By contrast, morphology tasks exert minimal impact: although accuracy improves modestly with scale, alignment scores remain largely unchanged, suggesting morphology is not a primary driver of brain alignment. Syntax and semantics show moderate effects. Both improve steadily with scale, and their contributions parallel alignment increases, though their influence is consistently weaker than discourse. Reasoning tasks (e.g., negation detection, correspondence) exhibit stable performance across model sizes and quantization settings, with alignment remaining relatively robust. This suggests reasoning is less sensitive to scaling but also less explanatory of alignment gains.

Taken together, these findings indicate that discourse-level representations are most critical for capturing brain-relevant information, followed by syntax and semantics, while morphology contributes little. Reasoning, though robust to compression, does not account for the major alignment differences between small and larger models.

\setlength{\tabcolsep}{3pt}
\begin{table*}[!ht]
\centering
\scriptsize
\tabcolsep2pt
\caption{Representative FlashHolmes task scores for Qwen-2.5 (1.5B, 3B, and 7B). Quantized 3B models remain close to full 3B across tasks, while 1.5B models show larger drops, especially in discourse and reasoning. The 7B model achieves the strongest scores overall.}
\label{tab:qwen_flashholmes_quant_7b}
\begin{tabular}{l|l|cccc|cccc|c}
\toprule
\textbf{Category} & \textbf{Task} 
& \multicolumn{4}{c|}{\textbf{Qwen-2.5 1.5B}} 
& \multicolumn{4}{c|}{\textbf{Qwen-2.5 3B}} 
& \multicolumn{1}{c}{\textbf{Qwen-2.5 7B}} \\
 & & FULL & AWQ & GPTQ & Smooth & FULL & AWQ & GPTQ & Smooth & FULL\\
\midrule
Discourse  & Bridging (edge)        & 0.789 & \textbf{0.799} & \textbf{0.798} & \textbf{0.799} & \textbf{0.798} & 0.794 & 0.433 & 0.794 & 0.683 \\
           & Bridging (sentence)    & \textbf{0.800} & \textbf{0.800} & \textbf{0.800} & \textbf{0.800} & \textbf{0.800} & \textbf{0.799} & 0.410 & \textbf{0.799} & 0.667 \\
           & Coreference            & 0.411 & 0.399 & 0.405 & 0.399 & 0.352 & 0.433 & 0.410 & 0.357 & \textbf{0.794} \\
Morphology & Constituent (depth)    & 0.802 & 0.801 & \textbf{0.809} & 0.801 & 0.752 & 0.752 & 0.801 & 0.752 & 0.755 \\
           & Constituent (length)   & 0.827 & 0.826 & \textbf{0.847} & 0.826 & 0.796 & 0.795 & 0.826 & 0.795 & 0.725 \\
Reasoning  & Negation span classify & 0.742 & 0.741 & 0.746 & 0.743 & 0.812 & 0.810 & 0.808 & 0.811 & 0.950 \\
           & Negation correspondence& 0.605 & 0.602 & 0.609 & 0.603 & 0.689 & 0.687 & 0.685 & 0.688 & 0.611 \\
           & SemAntoNeg             & 0.667 & 0.666 & 0.669 & 0.667 & 0.701 & 0.699 & 0.698 & 0.700 & 0.667 \\
Semantics  & Object animacy         & 0.994 & 0.994 & 0.991 & 0.994 & 0.988 & 0.988 & 0.994 & 0.988 & 0.805 \\
           & Object gender          & 0.546 & 0.535 & 0.532 & 0.535 & 0.482 & 0.402 & 0.531 & 0.402 & 0.807 \\
           & Object number          & 0.744 & 0.738 & 0.738 & 0.738 & 0.716 & 0.712 & 0.738 & 0.712 & 0.781 \\
Syntax     & Adjunct island         & 0.704 & 0.643 & 0.689 & 0.643 & 0.678 & 0.677 & 0.643 & 0.676 & 0.515 \\
           & Anaphor gender agr.    & 0.609 & 0.548 & 0.573 & 0.548 & 0.622 & 0.619 & 0.548 & 0.619 & 0.695 \\
           & Anaphor number agr.    & 0.631 & 0.611 & 0.582 & 0.613 & 0.647 & 0.661 & 0.613 & 0.662 & 0.860 \\
\bottomrule
\end{tabular}
\end{table*}

\begin{table*}[!ht]
\centering
\scriptsize
\caption{Representative FlashHolmes task scores for LLaMA-3.2 models (1B, 3B, and 8B). Quantized 3B models remain close to full 3B, while the 8B model achieves the strongest scores overall.}
\label{tab:LLaMA_flashholmes_quant_8b}
\begin{tabular}{l|l|cccc|cccc|c}
\toprule
\textbf{Category} & \textbf{Task} 
& \multicolumn{4}{c|}{\textbf{LLaMA-3.2 1B}} 
& \multicolumn{4}{c|}{\textbf{LLaMA-3.2 3B}} 
& \textbf{LLaMA-3.1 8B} \\
 & & FULL & AWQ & GPTQ & Smooth & FULL & AWQ & GPTQ & Smooth & FULL \\
\midrule
Discourse  & Bridging (edge)        & 0.792 & 0.801 & 0.800 & 0.799 & 0.801 & 0.799 & 0.433 & 0.798 & 0.703 \\
           & Bridging (sentence)    & 0.800 & 0.800 & 0.800 & 0.800 & 0.800 & 0.799 & 0.410 & 0.799 & 0.667 \\
           & Coreference            & 0.392 & 0.401 & 0.405 & 0.400 & 0.417 & 0.433 & 0.410 & 0.411 & 0.776 \\
Morphology & Constituent (depth)    & 0.838 & 0.830 & 0.841 & 0.831 & 0.842 & 0.839 & 0.845 & 0.840 & 0.735 \\
           & Constituent (length)   & 0.877 & 0.875 & 0.890 & 0.876 & 0.897 & 0.896 & 0.895 & 0.894 & 0.705 \\
Reasoning  & Negation span classify & 0.745 & 0.743 & 0.747 & 0.744 & 0.818 & 0.816 & 0.814 & 0.817 & 0.944 \\
           & Negation correspondence& 0.612 & 0.610 & 0.614 & 0.611 & 0.701 & 0.699 & 0.697 & 0.700 & 0.611 \\
           & SemAntoNeg             & 0.672 & 0.670 & 0.673 & 0.671 & 0.709 & 0.707 & 0.706 & 0.708 & 0.667 \\
Semantics  & Object animacy         & 0.981 & 0.981 & 0.980 & 0.981 & 0.989 & 0.989 & 0.988 & 0.989 & 0.882 \\
           & Object gender          & 0.626 & 0.623 & 0.628 & 0.624 & 0.644 & 0.640 & 0.639 & 0.642 & 0.879 \\
           & Object number          & 0.713 & 0.710 & 0.714 & 0.711 & 0.720 & 0.718 & 0.717 & 0.719 & 0.876 \\
Syntax     & Adjunct island         & 0.668 & 0.667 & 0.669 & 0.667 & 0.744 & 0.742 & 0.741 & 0.743 & 0.700 \\
           & Anaphor gender agr.    & 0.678 & 0.675 & 0.680 & 0.676 & 0.739 & 0.738 & 0.736 & 0.737 & 0.650 \\
           & Anaphor number agr.    & 0.660 & 0.659 & 0.662 & 0.660 & 0.671 & 0.670 & 0.669 & 0.671 & 0.845 \\
\bottomrule
\end{tabular}
\end{table*}


\FloatBarrier

\section{Quantitative Analysis across model families}
\label{app:quantitative_analysis}

We quantify scaling differences by performing statistical significance across subjects for the best selective layer per model: Qwen2.5 in Table~\ref{qwen2.5_qunatitative_analysis}, LLaMA in Table~\ref{llama_quantitative_analysis} and DeepSeek~\ref{deepseek_quantitative_analysis}. 
For Qwen2.5 model, the resulting mean best-layer scores are: 1.5B: 0.85+0.09, 3B: 0.92+0.08, 7B: 0.895+0.09, 14B: 0.93+0.10. Paired tests over subjects (n = 9) show that 3B and 14B are statistically indistinguishable (mean difference -0.0004, t(8) = -0.03, p $\approx$ 0.98), while both 3B and 14B significantly outperform the 1.5B model (3B vs. 1.5B: $\Delta$ = 0.07, t(8) = 4.89, p $\approx$ 0.004; 14B vs. 1.5B: $\Delta$ = 0.07, t(8) = 3.16, p $\approx$ 0.025). We also find a modest but significant advantage of 3B and 14B over 7B in best-layer alignment ($\Delta$ $\approx$ 0.04, p $\approx$ 0.02-0.04). These tests support our main claim in this regime: beyond $\sim$3B, increasing model size up to 14B yields at most modest gains in brain alignment, whereas 1B-1.5B models are reliably worse.

For the LLaMA model, as shown in Table~\ref{llama_quantitative_analysis}, we find that (i) 14B $\approx$ 7B: No difference ($\Delta$ = -0.00, p $\approx$ 0.95) - statistically identical (ii) 3B $\approx$ 14B/7B: Slight advantage but not significant (p $>$ 0.05) and (iii) All vs 1B: Highly significant differences (p $<$ 0.001 for 3B and 7B). Overall, the 3B, 7B, and 14B models form a statistically equivalent top tier, all significantly outperforming the 1B model.

Analysis of DeepSeek models (14B, 7B, 3B, 1B parameters), as shown in Table~\ref{deepseek_quantitative_analysis} reveals a clear scaling hierarchy with the 14B and 3B models forming a statistically equivalent top tier. We make the following observations from Table~\ref{deepseek_quantitative_analysis}: (i) 14B $\approx$ 3B: No significant difference (p $\approx$ 0.61), indicating 3B achieves 14B-level performance with ~80\% fewer parameters! (ii) 14B, 3B \texttt{$\gg$} 7B: Highly significant advantages (p $<$ 0.01), (iii) All \texttt{$\gg$} 1B: Very large differences (all p $<$ 0.001).
Overall, across three independent model families (Qwen, LLaMA, DeepSeek) and using best-layer scores with paired tests over nine subjects, we find a consistent pattern: 1B-1.5B models are reliably worse in brain alignment, while 3B models already reach the same level as their 7B-14B counterparts. In Qwen and DeepSeek, 3B and 14B are statistically indistinguishable, whereas both significantly outperform the smallest models; in LLaMA, 3B and 14B again lie in a narrow, non-significantly different range, with 7B closely tracking 14B and clearly above 1B. These results do not overturn global scaling laws, but they do indicate a local plateau in the compressed 1-14B regime: once model capacity reaches $\sim$3B, further scaling yields at most modest gains in brain predictivity, while going below this threshold leads to a robust drop in alignment.


\begin{table}[!ht]
\centering
\scriptsize
\caption{Pairwise differences in LLaMA best-layer scores across models (paired $t$-tests over 9 subjects). $\Delta$ is mean(A--B).}
\label{llama_quantitative_analysis}
\begin{tabular}{|l|c|c|c|}
\hline
Comparison & $\Delta$ (A--B) & $t(8)$ & Sig. (2-sided) \\
\hline
3B -- 14B    &   0.03 &   2.50 & n.s. ($p \approx 0.05$) \\
3B -- 7B     &   0.03 &   1.97 & n.s. ($p \approx 0.11$) \\
3B -- 1B     &   0.10 &   6.89 & $p < 0.001$ \\
14B -- 7B    &  -0.00 &  -0.07 & n.s. ($p \approx 0.95$) \\
14B -- 1B    &   0.07 &   3.12 & $p < 0.05$ \\
7B -- 1B     &   0.07 &  11.43 & $p < 0.001$ \\
\hline
\end{tabular}
\end{table}

\begin{table}[!ht]
\centering
\scriptsize
\caption{Pairwise differences in DeepSeek best-layer scores (paired $t$-tests over 9 subjects). $\Delta$ is mean(A--B).}
\label{deepseek_quantitative_analysis}
\begin{tabular}{|l|c|c|c|}
\hline
Comparison & $\Delta$ (A--B) & $t(8)$ & Sig. (2-sided) \\
\hline
14B -- 3B    &  -0.01 &  -0.54 & n.s. ($p \approx 0.61$) \\
14B -- 7B    &   0.07 &   4.84 & $p < 0.01$ \\
14B -- 1B    &   0.19 &  11.18 & $p < 0.001$ \\
3B -- 7B     &   0.08 &   4.18 & $p < 0.01$ \\
3B -- 1B     &   0.19 &  10.85 & $p < 0.001$ \\
7B -- 1B     &   0.11 &   9.28 & $p < 0.001$ \\
\hline
\end{tabular}
\end{table}

\section{Statistical Validation of Quantization Effects}
\label{statistical_variability}

\noindent\textbf{Quantization Effects - Qwen2.5.}
    We now provide formal statistical tests and variability measures for the quantization comparisons. For each Qwen2.5 model (1.5B, 3B, 7B), and for each quantization method (Full (FP16), AWQ, GPTQ, SmoothQuant), we compute best-layer brain alignment per subject and run paired $t$-tests across subjects between methods (Table~\ref{statistical_results_qwen2.5}). Negative $\Delta$ in rows of the form “FP16–AWQ” indicates that AWQ outperforms FP16. For Qwen2.5–7B (Table~\ref{statistical_results_qwen2.5}, left), AWQ and SmoothQuant are significantly better than both FP16 and GPTQ (e.g., FP16–AWQ: $\Delta = -0.020$, $t(8) = -6.10$, $p < 0.001$; AWQ–GPTQ: $\Delta = 0.040$, $t(8) = 7.10$, $p < 0.001$), while GPTQ is significantly worse than FP16. For Qwen2.5–3B (Table~\ref{statistical_results_qwen2.5}, right), none of the quantized variants differ significantly from FP16, but AWQ and SmoothQuant significantly outperform GPTQ, suggesting that well-designed quantization preserves alignment whereas GPTQ exhibits a modest degradation. For Qwen2.5–1.5B (Table~\ref{statistical_results_qwen2.5}, bottom), AWQ is significantly better than FP16 ($\Delta = -0.024$, $t(8) = -4.04$, $p < 0.01$), whereas GPTQ and SmoothQuant are statistically indistinguishable from FP16, and differences among the three quantized variants do not reach significance after correction.

\begin{table*}[!ht]
\centering
\scriptsize
\caption{Pairwise comparisons of brain-alignment differences across quantization methods for Qwen2.5 models. Each Table reports mean differences ($\Delta$), $t$-statistics, and two-sided significance tests for 7B (left), 3B (right), and 1.5B (bottom).}
\label{statistical_results_qwen2.5}
\begin{minipage}{0.49\linewidth}
\centering
\begin{tabular}{|l|c|c|c|}
\hline
Comparison (A–B) & $\Delta$ & $t(8)$ & Sig. \\
\hline
Qwen2.5-7B–AWQ         & -0.020 & -6.10 & $p<0.001$ \\
Qwen2.5-7B–GPTQ        &  0.020 &  6.20 & $p<0.001$ \\
Qwen2.5-7B–SmoothQuant & -0.005 & -3.50 & $p<0.016$ \\
AWQ–GPTQ         &  0.040 &  7.10 & $p<0.001$ \\
AWQ–SmoothQuant  &  0.015 &  4.20 & $p<0.008$ \\
GPTQ–SmoothQuant & -0.025 & -4.90 & $p<0.004$ \\
\hline
\end{tabular}
\end{minipage}
\hfill
\begin{minipage}{0.49\linewidth}
\centering
\begin{tabular}{|l|c|c|c|}
\hline
Comparison (A–B) & $\Delta$ & $t(8)$ & Sig. \\
\hline
Qwen2.5-3B–AWQ         & -0.010 & -1.23 & n.s. ($p\approx0.28$) \\
Qwen2.5-3B–GPTQ        &  0.014 &  2.24 & n.s. ($p\approx0.08$) \\
Qwen2.5-3B–SmoothQuant & -0.007 & -1.55 & n.s. ($p\approx0.18$) \\
AWQ–GPTQ         &  0.024 &  3.10 & $p<0.05$ \\
AWQ–SmoothQuant  &  0.003 &  0.34 & n.s. ($p\approx0.75$) \\
GPTQ–SmoothQuant & -0.020 & -3.12 & $p<0.05$ \\
\hline
\end{tabular}
\end{minipage}
\hfill
\begin{minipage}{0.49\linewidth}
\centering
\begin{tabular}{|l|c|c|c|}
\hline
Comparison (A–B) & $\Delta$ & $t(8)$ & Sig. \\
\hline
Qwen2.5-1.5B–AWQ         & -0.024 & -4.04 & $p<0.01$ \\
Qwen2.5-1.5B–GPTQ        &  0.002 &  0.18 & n.s. ($p\approx0.86$) \\
Qwen2.5-1.5B–SmoothQuant & -0.004 & -0.71 & n.s. ($p\approx0.51$) \\
AWQ–GPTQ         &  0.026 &  2.03 & n.s. ($p\approx0.10$) \\
AWQ–SmoothQuant  &  0.020 &  2.43 & n.s. ($p\approx0.06$) \\
GPTQ–SmoothQuant & -0.006 & -0.42 & n.s. ($p\approx0.69$) \\
\hline
\end{tabular}
\end{minipage}
\end{table*}

We also summarize quantization performance at the level of mean $\pm$ standard deviation across subjects in Table~\ref{qwen2.5_summary_quantization}. Across all Qwen sizes, AWQ and SmoothQuant closely or slightly exceed full models (FP16) in mean best-layer alignment (differences on the order of 0.01–0.02, within one standard deviation), whereas GPTQ tends to be lower than FP16, especially for 7B and 3B. Together, Table 8 and Table 9 show that (i) some apparent improvements in the figures are within noise and are now explicitly reported as non-significant, and (ii) the main qualitative pattern is statistically supported: well-designed quantization (AWQ/SmoothQuant) preserves brain alignment at near-full-precision levels, while GPTQ produce a modest but reliable degradation.

\begin{table}[!ht]
\centering
\scriptsize
\caption{Quantization method performance across Qwen models (mean $\pm$ std over 9 subjects).}
\label{qwen2.5_summary_quantization}
\begin{tabular}{lcccc}
\toprule
Model & Full precision (FP16) & AWQ & GPTQ & SmoothQuant \\
\midrule
Qwen-7B    & $0.886 \pm 0.092$ & $0.906 \pm 0.092$ & $0.866 \pm 0.092$ & $0.891 \pm 0.092$ \\
Qwen-3B    & $0.923 \pm 0.080$ & $0.933 \pm 0.085$ & $0.910 \pm 0.091$ & $0.930 \pm 0.085$ \\
Qwen-1.5B  & $0.850 \pm 0.087$ & $0.874 \pm 0.099$ & $0.848 \pm 0.088$ & $0.854 \pm 0.084$ \\
\bottomrule
\end{tabular}
\end{table}

\begin{figure*}[!ht]
    \centering
    \includegraphics[width=\linewidth]{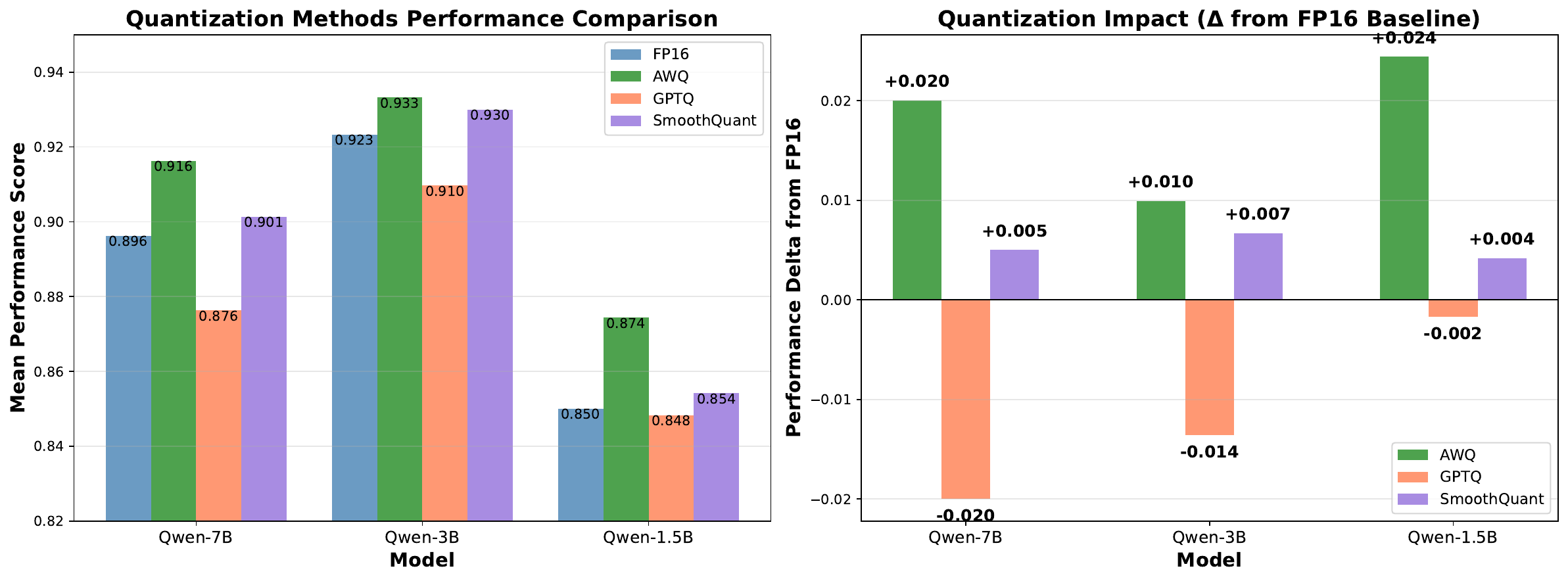}
    \caption{Qwen2.5 Quantization Analysis: (left) Quantization methods comparison, (right) Quantization impact}
    \label{fig:qwen_quantization_summary}
\end{figure*}

\noindent\textbf{Quantization effects in LLaMA-3.}
We performed the same best-layer, paired $t$-test analysis for LLaMA-3 models (1B, 3B, 8B). For LLaMA-3-8B, all pairwise differences between FP16, AWQ, GPTQ, and SmoothQuant are highly significant (Table~\ref{statistical_results_llama3}, left). Negative $\Delta$ in rows of the form “FP16–AWQ” indicates AWQ $>$ FP16; specifically, FP16–AWQ is negative ($\Delta$ = -0.010, $p < 0.001$), while FP16–GPTQ is positive ($\Delta$ = 0.020, $p < 0.001$) and AWQ-GPTQ is strongly positive ($\Delta$ = 0.030, $p < 0.001$). This implies the ordering AWQ $>$ FP16 $ge$ SmoothQuant $>$ GPTQ for 8B. For LLaMA-3-3B (Table~\ref{statistical_results_llama3}, right), GPTQ is significantly worse than FP16 ($\Delta$ = 0.059, $p < 0.05$), while AWQ and SmoothQuant are not significantly different from FP16. SmoothQuant significantly outperforms both AWQ and GPTQ (AWQ–SmoothQuant: $\Delta$ = -0.011, $p < 0.05$; GPTQ–SmoothQuant: $\Delta$ = -0.023, $p < 0.01$), indicating that GPTQ is the main outlier, with AWQ/SmoothQuant and FP16 forming a higher-performing cluster. For LLaMA-3-1B (Table~\ref{statistical_results_llama3}, bottom), GPTQ again shows a clear degradation: FP16–GPTQ is positive and significant ($\Delta$ = 0.076, $p < 0.01$), and GPTQ–SmoothQuant is strongly negative ($\Delta$ = -0.074, $p < 0.01$). In contrast, AWQ and SmoothQuant are statistically indistinguishable from FP16 (all $p > 0.1$), and differences among the three quantized variants other than GPTQ do not reach significance.

Table~\ref{llama_summary_across_models} summarizes quantization performance for LLaMA-3 models (8B, 3B, 1B). For LLaMA-3-8B, AWQ achieves the highest mean alignment (0.911), followed by FP16 and SmoothQuant, with GPTQ lowest (0.881). For LLaMA-3-3B, FP16 has the highest mean (0.929), while all three quantized variants are somewhat lower, with SmoothQuant $>$ AWQ $>$ GPTQ. For LLaMA-3-1B, FP16 and SmoothQuant are nearly identical and clearly above AWQ and GPTQ, with GPTQ again lowest. 
Overall, the LLaMA results align with our Qwen analyses: GPTQ consistently yields lower brain alignment than FP16, whereas AWQ and SmoothQuant generally preserve full-precision performance, sometimes even slightly improving upon it. This validates our conclusion that carefully designed quantization (AWQ/SmoothQuant) can maintain brain alignment at near full-precision levels, while some schemes (GPTQ) introduce a modest but reliable degradation.

\begin{table*}[!ht]
\centering
\scriptsize
\caption{Pairwise comparisons of brain-alignment differences across quantization methods for LLaMA-3 models. Each Table reports mean differences ($\Delta$), $t$-statistics, and two-sided significance tests for 8B (left), 3B (right), and 1B (bottom).}
\label{statistical_results_llama3}
\begin{minipage}{0.42\linewidth}
\centering
\begin{tabular}{|l|c|c|c|}
\hline
Comparison (A–B) & $\Delta$ & $t(8)$ & Sig. \\
\hline
LLaMA-3-8B -- AWQ               &  -0.010 &   -inf & $p < 0.001$ \\
LLaMA-3-8B -- GPTQ              &   0.020 &    inf & $p < 0.001$ \\
LLaMA-3-8B -- SmoothQuant       &   0.005 & 213621227803258.03 & $p < 0.001$ \\
AWQ -- GPTQ               &   0.030 &    inf & $p < 0.001$ \\
AWQ -- SmoothQuant        &   0.015 & 640834349907832.12 & $p < 0.001$ \\
GPTQ -- SmoothQuant       &  -0.015 & -640834349907835.25 & $p < 0.001$ \\
\hline
\end{tabular}
\end{minipage}
\hfill
\begin{minipage}{0.42\linewidth}
\centering
\begin{tabular}{|l|c|c|c|}
\hline
Comparison (A–B) & $\Delta$ & $t(8)$ & Sig. \\
\hline
LLaMA-3-3B -- AWQ               &   0.047 &   1.85 & n.s. ($p \approx 0.12$) \\
LLaMA-3-3B -- GPTQ              &   0.059 &   2.57 & $p < 0.05$ \\
LLaMA-3-3B -- SmoothQuant       &   0.036 &   1.47 & n.s. ($p \approx 0.20$) \\
AWQ -- GPTQ               &   0.012 &   1.87 & n.s. ($p \approx 0.12$) \\
AWQ -- SmoothQuant        &  -0.011 &  -3.28 & $p < 0.05$ \\
GPTQ -- SmoothQuant       &  -0.023 &  -4.08 & $p < 0.01$ \\
\hline
\end{tabular}
\end{minipage}
\hfill
\begin{minipage}{0.49\linewidth}
\centering
\begin{tabular}{|l|c|c|c|}
\hline
Comparison (A–B) & $\Delta$ & $t(8)$ & Sig. \\
\hline
LLaMA-3-1B -- AWQ               &   0.035 &   1.88 & n.s. ($p \approx 0.12$) \\
LLaMA-3-1B -- GPTQ              &   0.076 &   6.64 & $p < 0.01$ \\
LLaMA-3-1B -- SmoothQuant       &   0.002 &   0.56 & n.s. ($p \approx 0.60$) \\
AWQ -- GPTQ               &   0.041 &   3.20 & $p < 0.05$ \\
AWQ -- SmoothQuant        &  -0.032 &  -1.78 & n.s. ($p \approx 0.14$) \\
GPTQ -- SmoothQuant       &  -0.074 &  -5.61 & $p < 0.01$ \\
\hline
\end{tabular}
\end{minipage}
\end{table*}

\begin{table}[h]
\centering
\scriptsize
\caption{Quantization method performance across LLaMA models (mean $\pm$ std over 9 subjects).}
\label{llama_summary_across_models}
\begin{tabular}{|l|c|c|c|c|}
\hline
Model & FP16 & AWQ & GPTQ & SmoothQuant \\
\hline
LLaMA-3-8B      & $0.901 \pm 0.094$ & $0.911 \pm 0.094$ & $0.881 \pm 0.094$ & $0.896 \pm 0.094$ \\
LLaMA-3-3B      & $0.929 \pm 0.071$ & $0.882 \pm 0.102$ & $0.870 \pm 0.092$ & $0.893 \pm 0.102$ \\
LLaMA-3-1B      & $0.830 \pm 0.091$ & $0.795 \pm 0.133$ & $0.754 \pm 0.115$ & $0.828 \pm 0.092$ \\
\hline
\end{tabular}
\end{table}

\begin{figure*}[!ht]
    \centering
    \includegraphics[width=\linewidth]{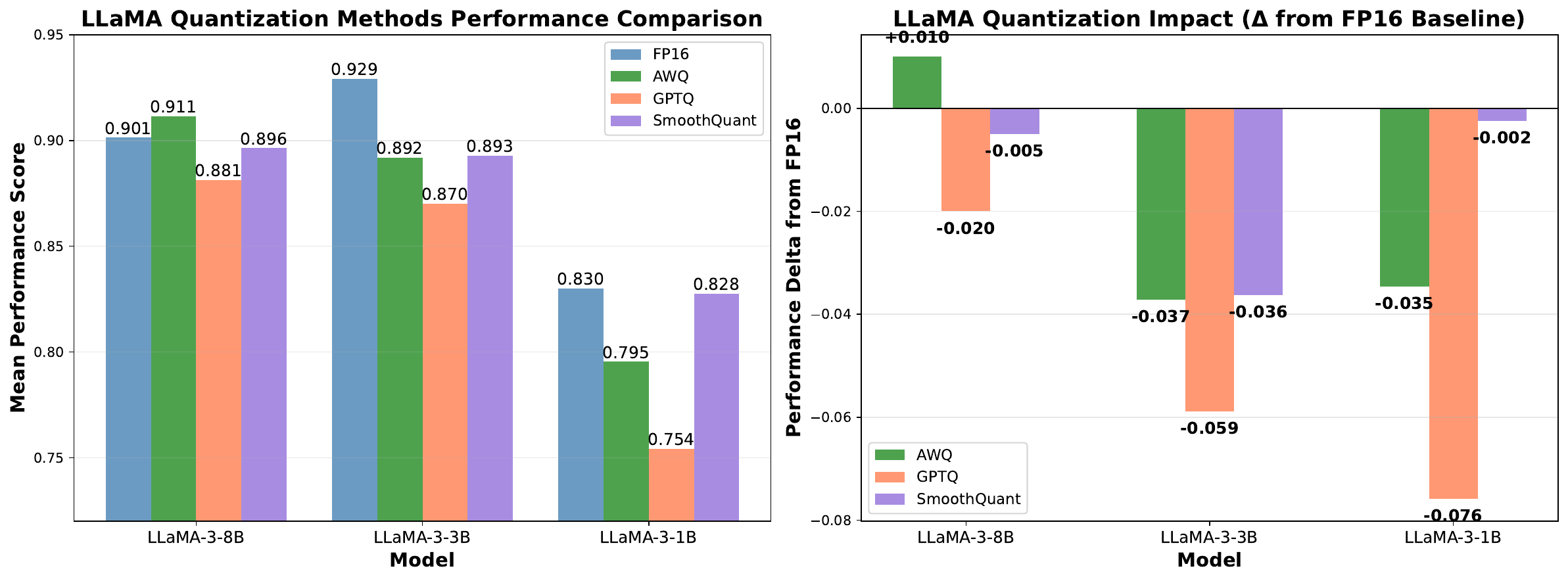}
    \caption{LLaMA-3 Quantization Analysis: (left) Quantization methods comparison, (right) Quantization impact}
    \label{fig:llama_quantization_summary}
\end{figure*}

\section{ROI-Specific Analysis, Best Layer Selection and Subject Variability.}
\label{app:roi_analysis_layer_subject_variablity}

In our analyses, we extract activations from every transformer layer, and fit a separate voxel-wise encoding model for each layer. For each model, we then compute brain alignment layer-by-layer across the language ROIs and identify the best layer as the one with the highest mean normalized predictivity. The main size/quantization comparisons are reported using this model-specific best layer (see Tables~\ref{tab:qwen2.5_optimal_layer_rois} and~\ref{tab:llama3_optimal_layer_rois}).

From Table~\ref{tab:qwen2.5_optimal_layer_rois}, across language ROIs we find that the best layers are highly consistent within a given model: the same (or adjacent) layer tends to be optimal across ROIs, so we treat the best layer as a model-level property when summarizing results. Overall, across models we observe the familiar pattern that middle-to-late layers yield the strongest brain alignment, with early layers performing clearly worse.

Regarding quantization, we also examined whether AWQ, GPTQ, or SmoothQuant systematically shift the optimal layer. We did not observe any systematic change: for a given architecture, the best layer under quantization is typically the same as in FP16 or within $\pm$1–2 layers in the same middle/late portion of the network. In other words, quantization affects the magnitude of brain alignment (as analyzed in Tables~\ref{qwen2.5_summary_quantization} and~\ref{llama_summary_across_models}), but not the qualitative position of the brain-optimal layers. We now clarify this procedure and these observations in the main text and refer explicitly to the layer-wise summaries in Tables~\ref{tab:qwen2.5_optimal_layer_rois} and~\ref{tab:llama3_optimal_layer_rois}.

The best layers are consistent in each model across ROIs; so, we consider best layers specific to each model while reporting the results. Overall, across models, middle to late layers show better brain alignment. Mostly 3b model is the best in terms of brain alignment across ROIs.

\begin{table*}[t]
\centering
\scriptsize
\caption{Qwen2.5 Cross-Model Size Comparisons}
\label{qwen2.5_cross_model_comparisons}
\begin{tabular}{|l|c|c|c|c|c|c|c|}
\hline
Comparison & Base Mean & Comp Mean & Difference & Effect Size & \% Sig & Median p & Interpretation \\
\hline
7B vs 3B & 0.7208 & 0.6996 & -0.0213 & 0.596 & 0.0\% & 0.549 & Not significant \\
14B vs 3B & 0.7208 & 0.6912 & -0.0296 & 1.184 & 27.8\% & 0.111 & Trend (p $<$ 0.15) \\
14B vs 7B & 0.6996 & 0.6912 & -0.0083 & 0.874 & 7.8\% & 0.356 & Not significant \\
\hline
\end{tabular}
\end{table*}

\begin{table*}[!ht]
\centering
\scriptsize
\caption{ROI-Specific Layer Performance Summary}
\label{tab:qwen2.5_optimal_layer_rois}
\begin{tabular}{|l| c| c| c| c| c| c|}
\hline
ROI & Model (Qwen2.5) & Overall Mean$\pm$SD & Best Layer & Best Layer Mean$\pm$SD & Worst Layer & Layer Range \\
\hline
AG & 1.5B & 0.8341$\pm$0.1546 & L14 & 0.9466$\pm$0.1099 & L1 & 0.5478--0.9466 \\
AG & 3B & 0.8558$\pm$0.1823 & L22 & 1.0091$\pm$0.1157 & L1 & 0.5418--1.0091 \\
AG & 7B & 0.8417$\pm$0.2119 & L15 & 0.9818$\pm$0.1390 & L1 & 0.4535--0.9818 \\
AG & 14B & 0.8178$\pm$0.2085 & L24 & 1.0143$\pm$0.1431 & L1 & 0.4913--1.0143 \\ \hline
ATL & 1.5B & 0.7161$\pm$0.1248 & L14 & 0.7904$\pm$0.1302 & L1 & 0.5296--0.7904 \\
ATL & 3B & 0.7251$\pm$0.1422 & L21 & 0.8304$\pm$0.1309 & L1 & 0.5108--0.8304 \\
ATL & 7B & 0.7016$\pm$0.1611 & L16 & 0.7891$\pm$0.1072 & L1 & 0.4620--0.7891 \\
ATL & 14B & 0.6974$\pm$0.1579 & L25 & 0.8362$\pm$0.1333 & L1 & 0.4630--0.8362 \\ \hline
PTL & 1.5B & 0.7697$\pm$0.1300 & L15 & 0.8335$\pm$0.1078 & L1 & 0.5944--0.8335 \\
PTL & 3B & 0.7763$\pm$0.1394 & L21 & 0.8725$\pm$0.1313 & L1 & 0.5843--0.8725 \\
PTL & 7B & 0.7443$\pm$0.1616 & L15 & 0.8353$\pm$0.1231 & L3 & 0.5166--0.8353 \\
PTL & 14B & 0.7471$\pm$0.1513 & L25 & 0.8592$\pm$0.1386 & L1 & 0.5353--0.8592 \\ \hline
IFG & 1.5B & 0.7726$\pm$0.1716 & L14 & 0.8730$\pm$0.1480 & L1 & 0.5366--0.8730 \\
IFG & 3B & 0.7801$\pm$0.1872 & L21 & 0.9309$\pm$0.1377 & L1 & 0.5058--0.9309 \\
IFG & 7B & 0.7665$\pm$0.1996 & L15 & 0.9045$\pm$0.1305 & L3 & 0.4263--0.9045 \\
IFG & 14B & 0.7639$\pm$0.2079 & L25 & 0.9516$\pm$0.1027 & L1 & 0.4938--0.9516 \\ \hline
MFG & 1.5B & 0.6929$\pm$0.1494 & L15 & 0.7618$\pm$0.1540 & L1 & 0.5068--0.7618 \\
MFG & 3B & 0.6984$\pm$0.1732 & L21 & 0.8006$\pm$0.1746 & L1 & 0.5213--0.8006 \\
MFG & 7B & 0.6716$\pm$0.1883 & L15 & 0.7732$\pm$0.1553 & L1 & 0.4283--0.7732 \\
MFG & 14B & 0.6689$\pm$0.1642 & L24 & 0.7921$\pm$0.1506 & L1 & 0.4682--0.7921 \\ \hline
IFGOrb & 1.5B & 0.6193$\pm$0.1589 & L14 & 0.7249$\pm$0.1007 & L1 & 0.4160--0.7249 \\
IFGOrb & 3B & 0.6401$\pm$0.1712 & L22 & 0.7647$\pm$0.0866 & L1 & 0.4072--0.7647 \\
IFGOrb & 7B & 0.6403$\pm$0.1966 & L15 & 0.7765$\pm$0.1080 & L1 & 0.2891--0.7765 \\
IFGOrb & 14B & 0.6159$\pm$0.1878 & L25 & 0.7803$\pm$0.0772 & L3 & 0.3628--0.7803 \\ \hline
PCC & 1.5B & 0.7638$\pm$0.1732 & L15 & 0.8618$\pm$0.1261 & L1 & 0.4904--0.8618 \\
PCC & 3B & 0.7651$\pm$0.1878 & L22 & 0.9093$\pm$0.1203 & L1 & 0.4654--0.9093 \\
PCC & 7B & 0.7509$\pm$0.2118 & L15 & 0.8904$\pm$0.0938 & L1 & 0.3571--0.8904 \\
PCC & 14B & 0.7235$\pm$0.2095 & L24 & 0.9367$\pm$0.1163 & L1 & 0.4205--0.9367 \\ \hline
dmPFC & 1.5B & 0.6884$\pm$0.1392 & L14 & 0.8089$\pm$0.0991 & L1 & 0.4443--0.8089 \\
dmPFC & 3B & 0.6964$\pm$0.1726 & L21 & 0.8685$\pm$0.1049 & L1 & 0.4192--0.8685 \\
dmPFC & 7B & 0.6888$\pm$0.1825 & L15 & 0.8226$\pm$0.1018 & L1 & 0.3628--0.8226 \\
dmPFC & 14B & 0.6692$\pm$0.1913 & L25 & 0.8859$\pm$0.1320 & L1 & 0.3712--0.8859 \\ \hline
AC & 1.5B & 0.5587$\pm$0.0906 & L15 & 0.5963$\pm$0.0762 & L1 & 0.4727--0.5963 \\
AC & 3B & 0.5634$\pm$0.1056 & L21 & 0.6303$\pm$0.0890 & L1 & 0.4668--0.6303 \\
AC & 7B & 0.5204$\pm$0.1243 & L15 & 0.5867$\pm$0.1017 & L3 & 0.3757--0.5867 \\
AC & 14B & 0.5241$\pm$0.0909 & L24 & 0.5831$\pm$0.1042 & L1 & 0.4095--0.5831 \\
\hline
\end{tabular}
\end{table*}

\begin{table}[!ht]
\centering
\scriptsize
\caption{Qwen2.5: Subject Variability at Optimal Layers}
\label{tab:qwen2.5_subject_variability}
\begin{tabular}{|l| c| c| c| c| c| c|}
\hline
\textbf{Model} & \textbf{Layer} & \textbf{Mean $\pm$ SD} & \textbf{SEM} & \textbf{95\% CI} & \textbf{CV (\%)} & \textbf{Variability} \\
\hline
1.5B  & L14 & $0.7956 \pm 0.0892$ & 0.0364 & [0.724, 0.867] & 11.21\% & LOW \\
3B    & L22 & $0.8309 \pm 0.0835$ & 0.0341 & [0.764, 0.898] & 10.05\% & LOW \\
7B    & L15 & $0.8171 \pm 0.0852$ & 0.0348 & [0.749, 0.885] & 10.43\% & LOW \\
14B   & L24 & $0.8411 \pm 0.0883$ & 0.0361 & [0.770, 0.912] & 10.50\% & LOW \\
\hline
\end{tabular}
\end{table}


\begin{table}[!ht]
\centering
\scriptsize
\caption{Cross-Model Comparisons for Llama-3}
\label{llama3_cross_model_comparison}
\begin{tabular}{|l| c| c| c| c| c| c| l|}
\hline
\textbf{Comparison} & \textbf{Base Mean} & \textbf{Comp Mean} & \textbf{Difference} &
\textbf{Effect Size} & \textbf{\% Sig} & \textbf{Median $p$} & \textbf{Interpretation} \\
\hline
8B vs 3B   & 0.7409 & 0.7300 & -0.0110 & 0.425 & 5.2\%  & 0.417 & Not significant \\
14B vs 3B  & 0.7409 & 0.7494 & +0.0084 & 0.811 & 27.4\% & 0.227 & Moderate effect \\
14B vs 8B  & 0.7285 & 0.7517 & +0.0232 & 0.577 & 14.6\% & 0.320 & Not significant \\
\hline
\end{tabular}
\end{table}

\begin{table*}[!ht]
\centering
\scriptsize
\caption{LLaMA3: ROI-Specific Layer Performance Summary}
\label{tab:llama3_optimal_layer_rois}
\begin{tabular}{|l| c| c| c| c| c| c| c| c|}
\toprule
\textbf{ROI} & \textbf{Model (LLaMA-3)} & \textbf{Optimal Layer} & \textbf{Optimal Value} & \textbf{SD} & \textbf{CI Low} & \textbf{CI High} & \textbf{High-Perf Range} & \textbf{Total Layers} \\
\hline
AG       & 1B  & 9  & 0.9189 & 0.0867 & 0.8495 & 0.9882 & 7--14  & 16 \\
ATL      & 1B  & 9  & 0.7751 & 0.0789 & 0.7119 & 0.8382 & 5--14  & 16 \\
PTL      & 1B  & 8  & 0.8149 & 0.1134 & 0.7242 & 0.9056 & 5--14  & 16 \\
IFG      & 1B  & 8  & 0.8312 & 0.1209 & 0.7345 & 0.9279 & 5--15  & 16 \\
MFG      & 1B  & 9  & 0.7477 & 0.1623 & 0.6178 & 0.8775 & 7--15  & 16 \\
IFGOrb   & 1B  & 7  & 0.6937 & 0.0868 & 0.6242 & 0.7631 & 5--15  & 16 \\
PCC      & 1B  & 7  & 0.8364 & 0.1109 & 0.7476 & 0.9251 & 5--14  & 16 \\
dmPFC    & 1B  & 9  & 0.7591 & 0.0995 & 0.6795 & 0.8387 & 7--11  & 16 \\
EarlyAud & 1B  & 8  & 0.6010 & 0.0802 & 0.5368 & 0.6651 & 5--9   & 16 \\
\hline
AG       & 3B  & 12 & 1.0206 & 0.1073 & 0.9347 & 1.1065 & 9--19  & 28 \\
ATL      & 3B  & 12 & 0.8150 & 0.1111 & 0.7261 & 0.9039 & 9--21  & 28 \\
PTL      & 3B  & 13 & 0.8654 & 0.1246 & 0.7657 & 0.9651 & 9--21  & 28 \\
IFG      & 3B  & 13 & 0.9255 & 0.1244 & 0.8259 & 1.0250 & 9--22  & 28 \\
MFG      & 3B  & 13 & 0.8127 & 0.1439 & 0.6975 & 0.9278 & 11--22 & 28 \\
IFGOrb   & 3B  & 12 & 0.8030 & 0.0739 & 0.7439 & 0.8622 & 11--17 & 28 \\
PCC      & 3B  & 14 & 0.9151 & 0.1105 & 0.8267 & 1.0036 & 9--22  & 28 \\
dmPFC    & 3B  & 12 & 0.8464 & 0.0846 & 0.7787 & 0.9141 & 11--17 & 28 \\
EarlyAud & 3B  & 13 & 0.6271 & 0.0905 & 0.5547 & 0.6995 & 12--17 & 28 \\
\hline
AG       & 7B  & 14 & 0.9802 & 0.1095 & 0.8927 & 1.0678 & 7--29  & 32 \\
ATL      & 7B  & 14 & 0.8098 & 0.1198 & 0.7139 & 0.9056 & 7--25  & 32 \\
PTL      & 7B  & 14 & 0.8644 & 0.1082 & 0.7778 & 0.9509 & 7--25  & 32 \\
IFG      & 7B  & 14 & 0.8974 & 0.1270 & 0.7958 & 0.9990 & 7--25  & 32 \\
MFG      & 7B  & 13 & 0.7792 & 0.1082 & 0.6926 & 0.8658 & 7--26  & 32 \\
IFGOrb   & 7B  & 14 & 0.7415 & 0.1004 & 0.6612 & 0.8219 & 7--29  & 32 \\
PCC      & 7B  & 15 & 0.8835 & 0.1203 & 0.7872 & 0.9797 & 7--29  & 32 \\
dmPFC    & 7B  & 14 & 0.8353 & 0.1043 & 0.7519 & 0.9188 & 7--18  & 32 \\
EarlyAud & 7B  & 13 & 0.5940 & 0.0756 & 0.5336 & 0.6545 & 7--21  & 32 \\
\hline
AG       & 14B & 17 & 0.9861 & 0.1001 & 0.9061 & 1.0662 & 5--38  & 39 \\
ATL      & 14B & 16 & 0.7910 & 0.1175 & 0.6970 & 0.8851 & 6--38  & 39 \\
PTL      & 14B & 16 & 0.8204 & 0.1121 & 0.7308 & 0.9101 & 6--35  & 39 \\
IFG      & 14B & 16 & 0.8709 & 0.1055 & 0.7865 & 0.9553 & 8--35  & 39 \\
MFG      & 14B & 13 & 0.7340 & 0.1315 & 0.6288 & 0.8393 & 5--20  & 39 \\
IFGOrb   & 14B & 17 & 0.7772 & 0.0460 & 0.7404 & 0.8141 & 12--30 & 39 \\
PCC      & 14B & 16 & 0.9076 & 0.1142 & 0.8162 & 0.9990 & 8--35  & 39 \\
dmPFC    & 14B & 16 & 0.8389 & 0.1310 & 0.7341 & 0.9436 & 14--23 & 39 \\
EarlyAud & 14B & 17 & 0.5544 & 0.0720 & 0.4969 & 0.6120 & 5--20  & 39 \\
\hline
\end{tabular}
\end{table*}

\begin{table}[!ht]
\centering
\scriptsize
\caption{LLaMA3: Subject Variability at Optimal Layers}
\label{tab:llama3_subject_variability}
\begin{tabular}{|l| c| c| c| c| c| c|}
\hline
\textbf{Model} & \textbf{Layer} & \textbf{Mean $\pm$ SD} & \textbf{SEM} & \textbf{95\% CI} & \textbf{CV (\%)} & \textbf{Variability} \\
\hline
1B    & L8   & 0.8003 $\pm$ 0.1065 & 0.0435 & [0.688, 0.912] & 13.31\% & MODERATE \\
3B    & L13  & 0.9083 $\pm$ 0.0824 & 0.0337 & [0.822, 0.995] & 9.08\% & LOW \\
8B    & L10  & 0.8588 $\pm$ 0.1115 & 0.0455 & [0.742, 0.976] & 12.98\% & MODERATE \\
14B   & L16  & 0.8679 $\pm$ 0.0936 & 0.0382 & [0.770, 0.966] & 10.78\% & MODERATE \\
\hline
\end{tabular}
\end{table}

\begin{figure*}[!ht]
    \centering
    \includegraphics[width=0.52\linewidth]{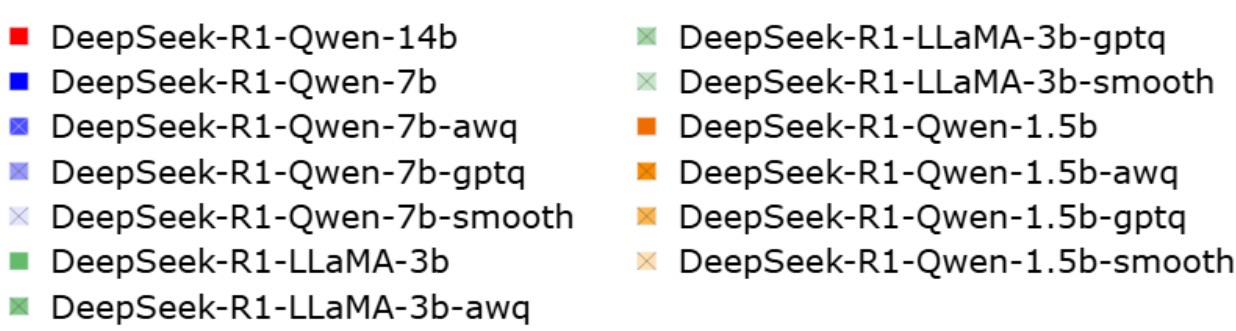} \\
    \includegraphics[width=0.33\linewidth]{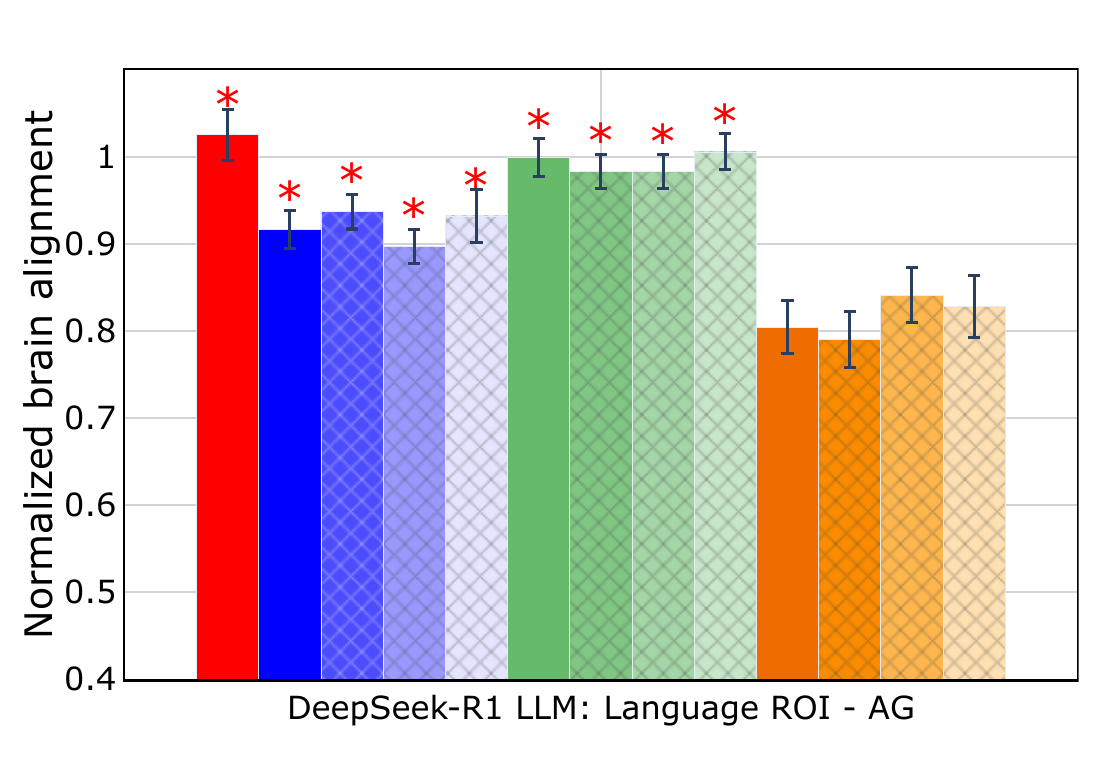}
    \includegraphics[width=0.33\linewidth]{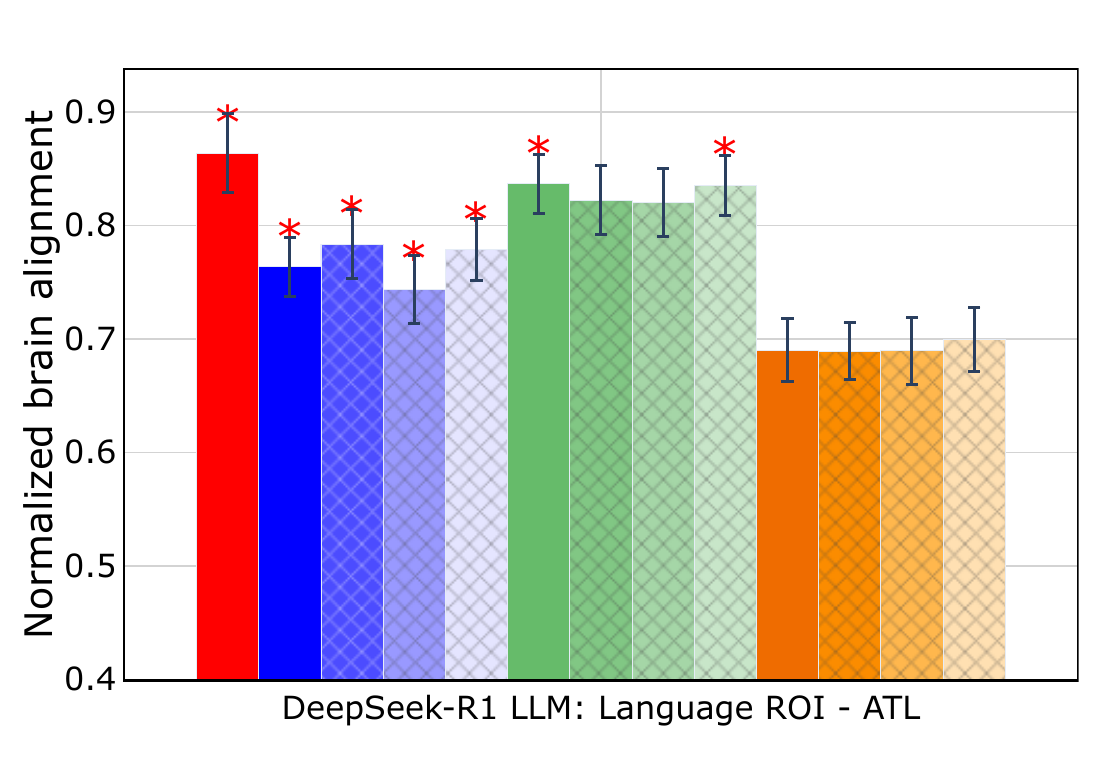}
    \includegraphics[width=0.33\linewidth]{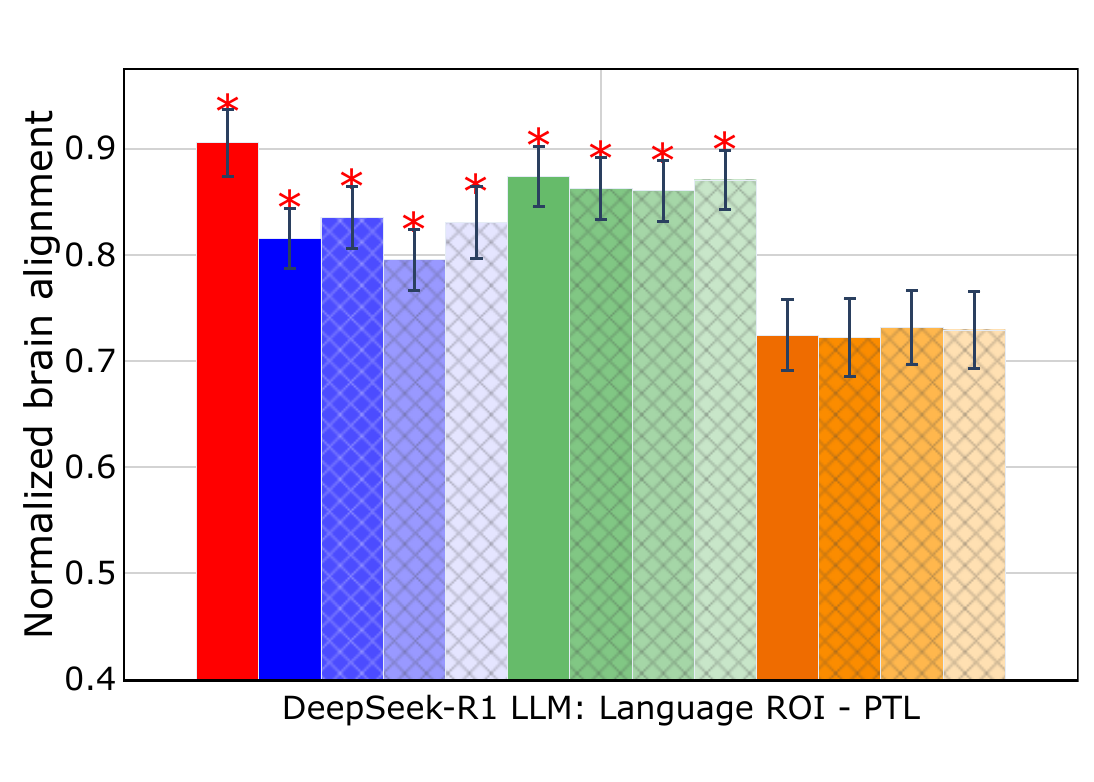}
    \includegraphics[width=0.33\linewidth]{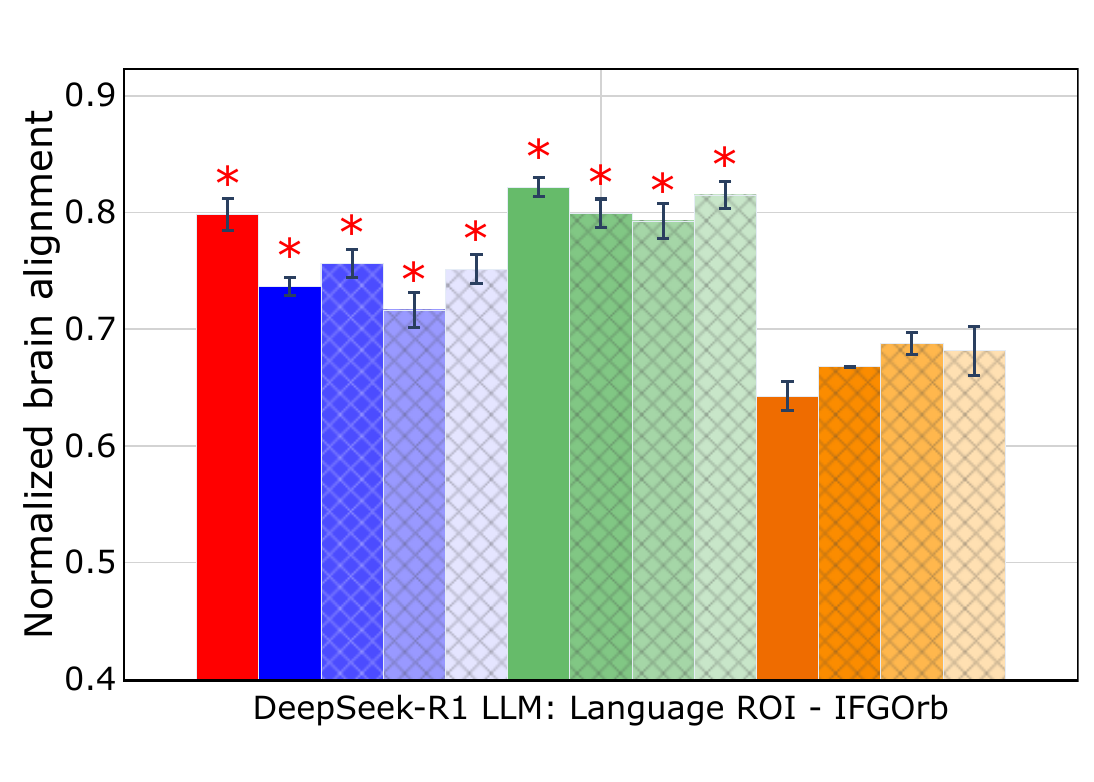}
    \includegraphics[width=0.33\linewidth]{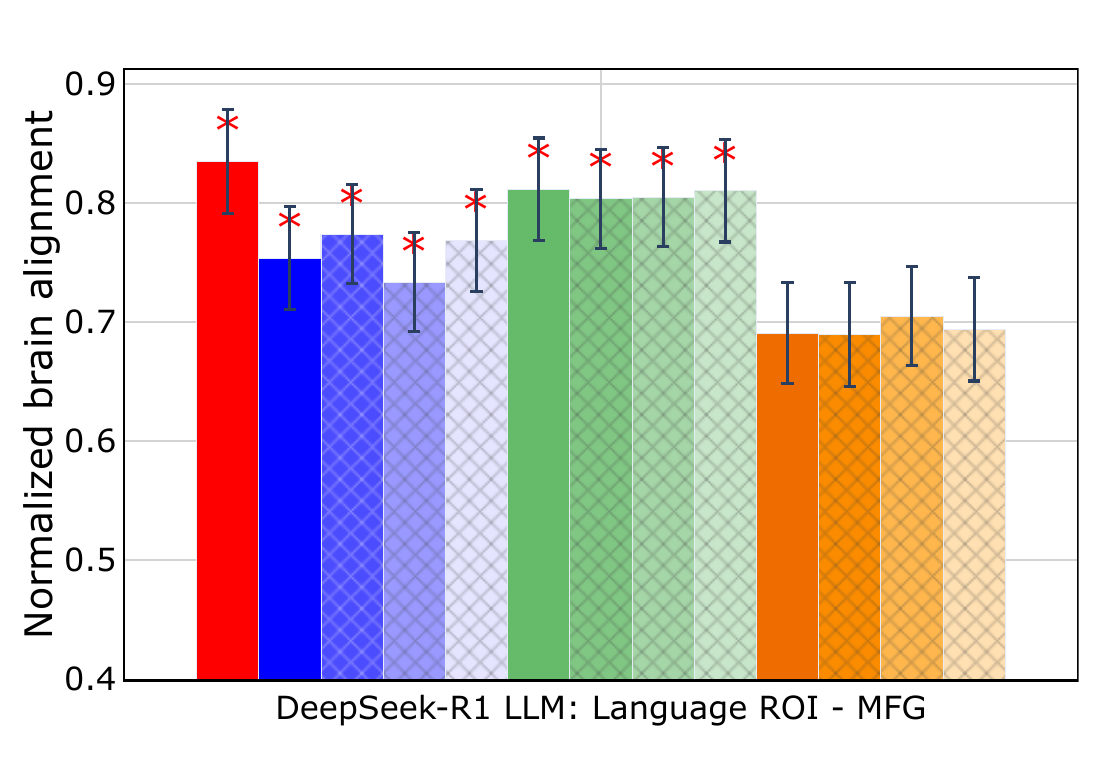}
    \includegraphics[width=0.33\linewidth]{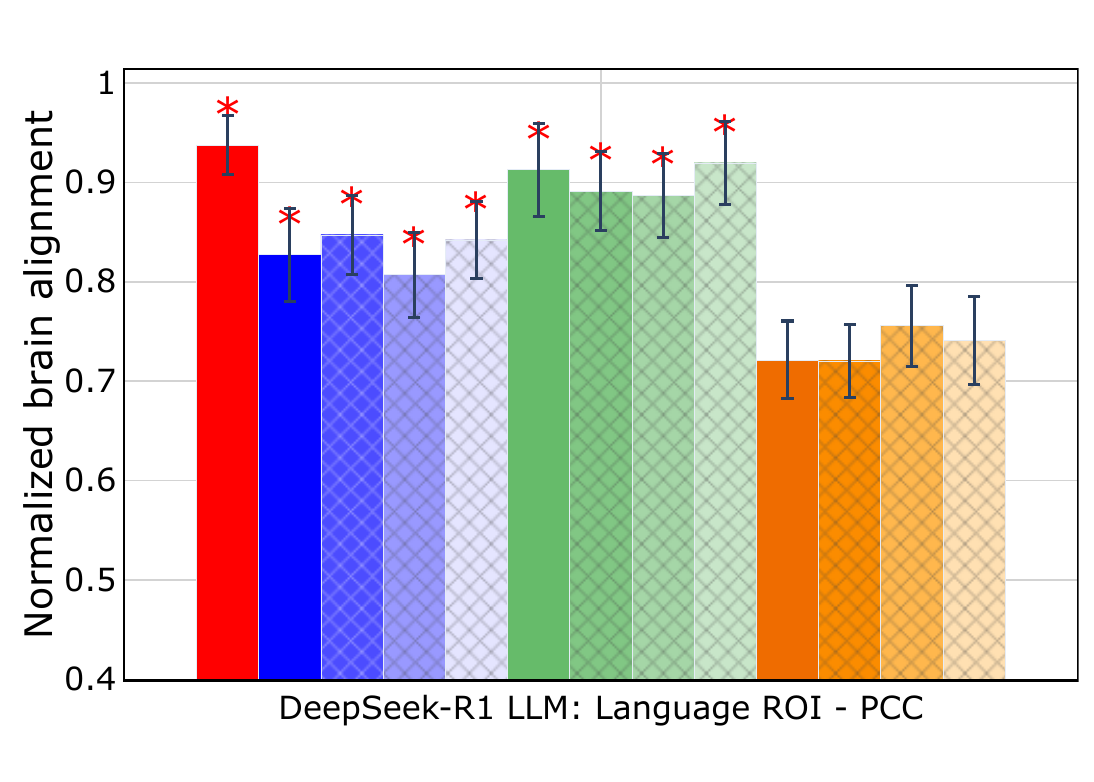}
    \includegraphics[width=0.33\linewidth]{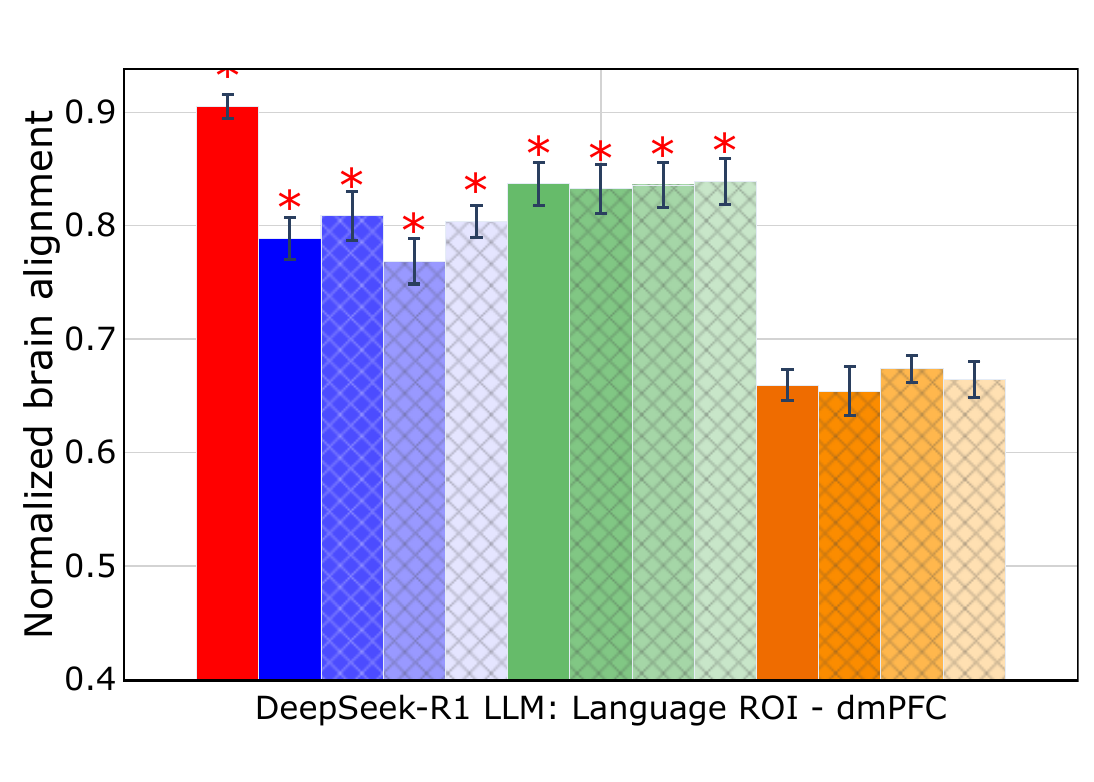}
    \includegraphics[width=0.33\linewidth]{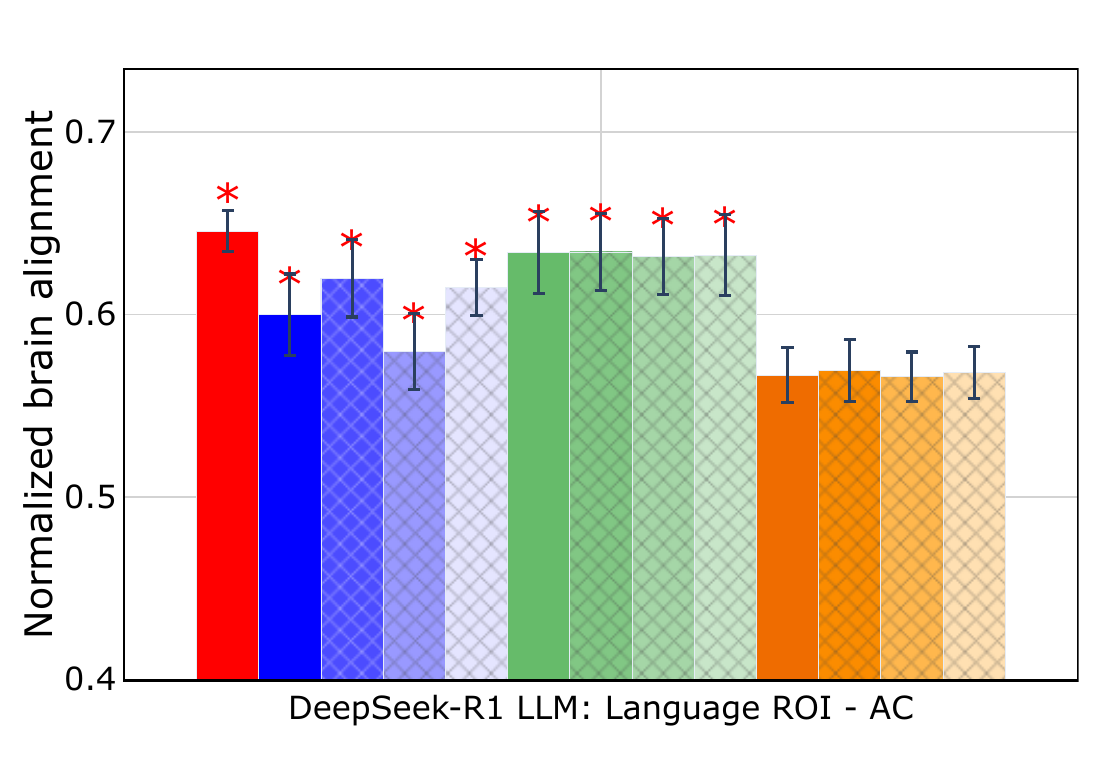}
    \caption{Normalized Predictivity of SLMs, LLMs, and Quantized Language Models for DeepSeek-R1 models.}
    \label{fig:deepseek_vem_language}
\end{figure*}

\begin{figure*}[!ht]
    \centering
    \includegraphics[width=0.52\linewidth]{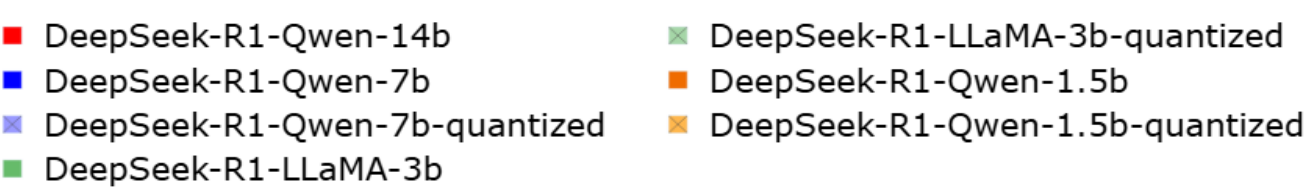} \\
    \includegraphics[width=0.33\linewidth]{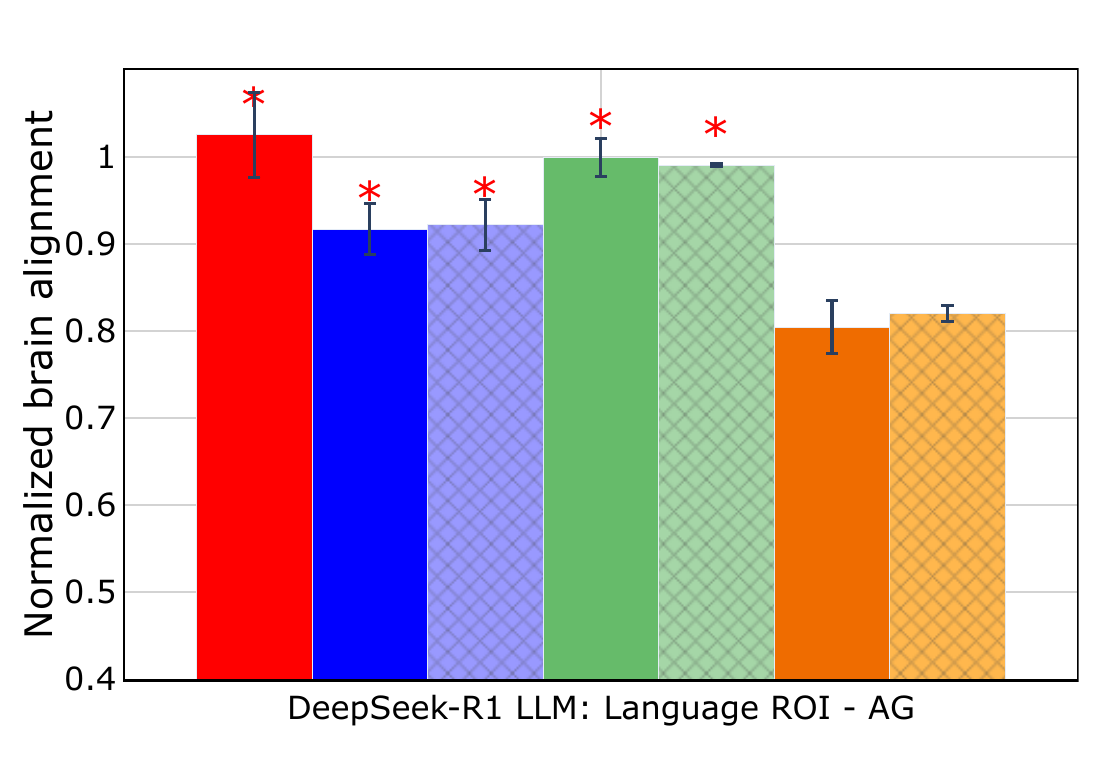}
    \includegraphics[width=0.33\linewidth]{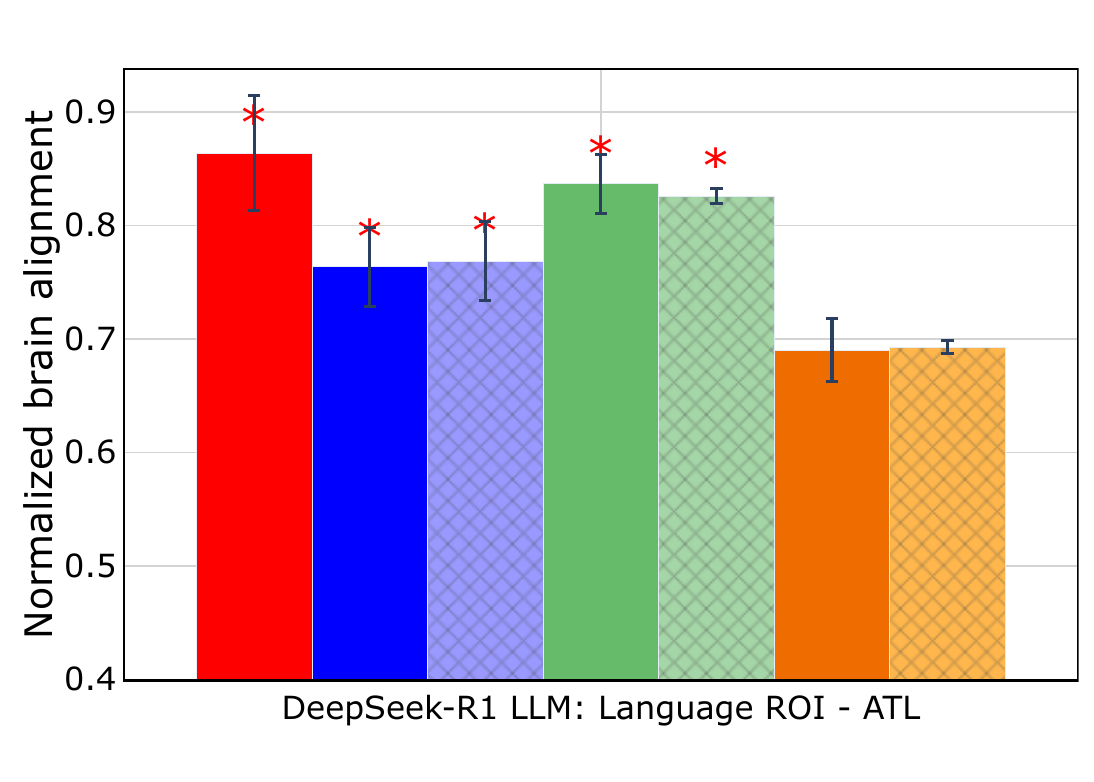}
    \includegraphics[width=0.33\linewidth]{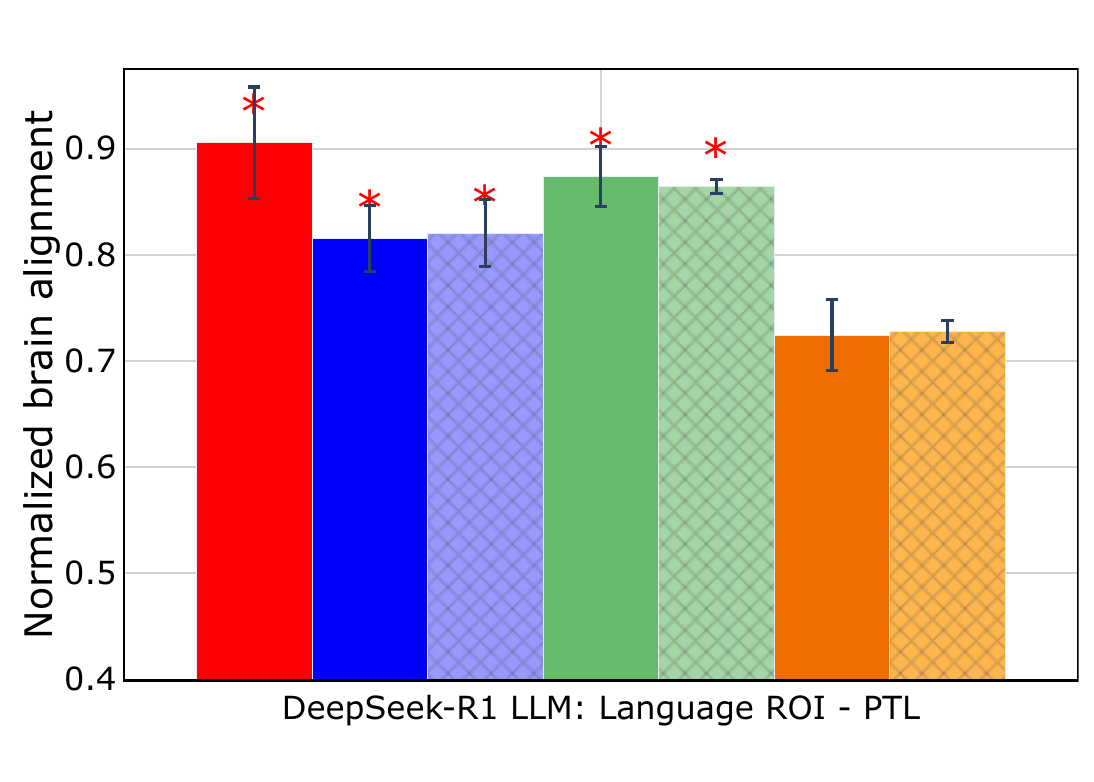}
    \includegraphics[width=0.33\linewidth]{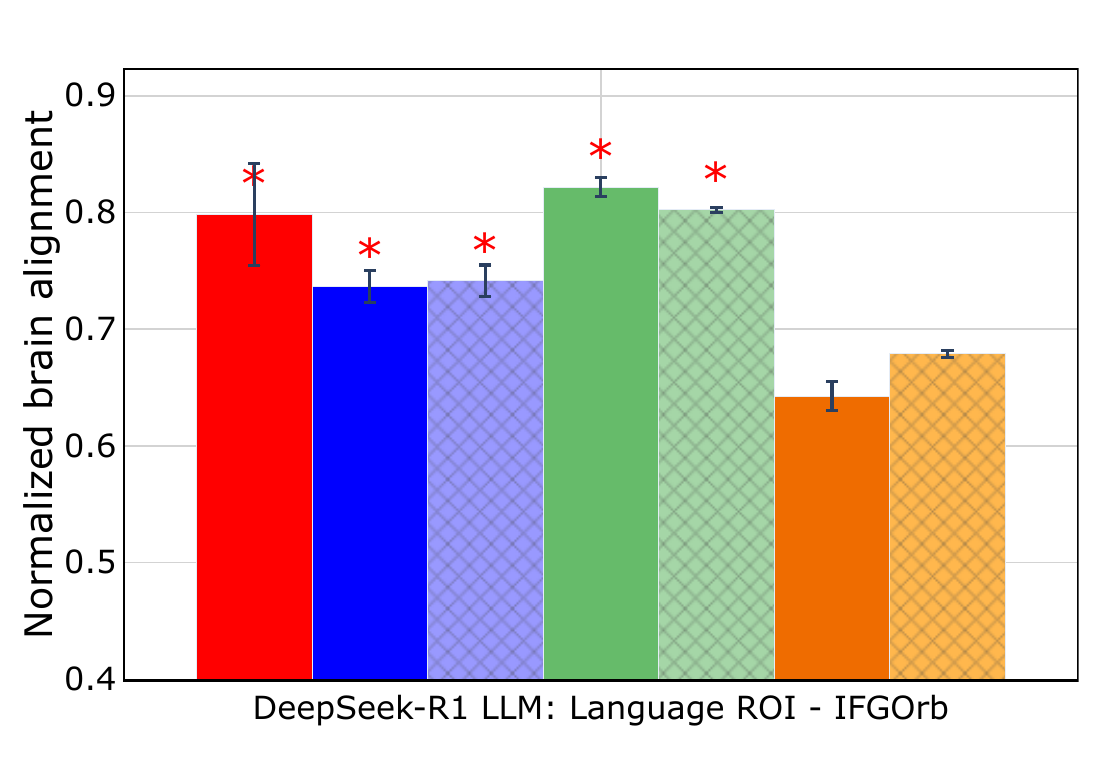}
    \includegraphics[width=0.33\linewidth]{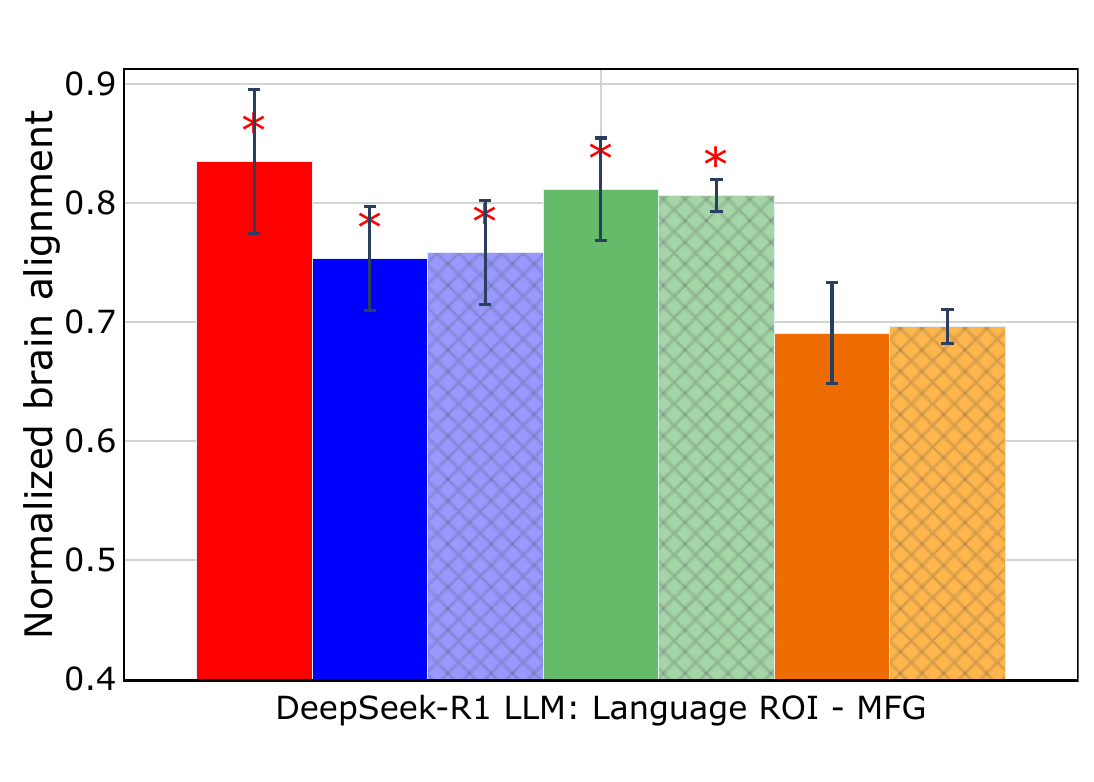}
    \includegraphics[width=0.33\linewidth]{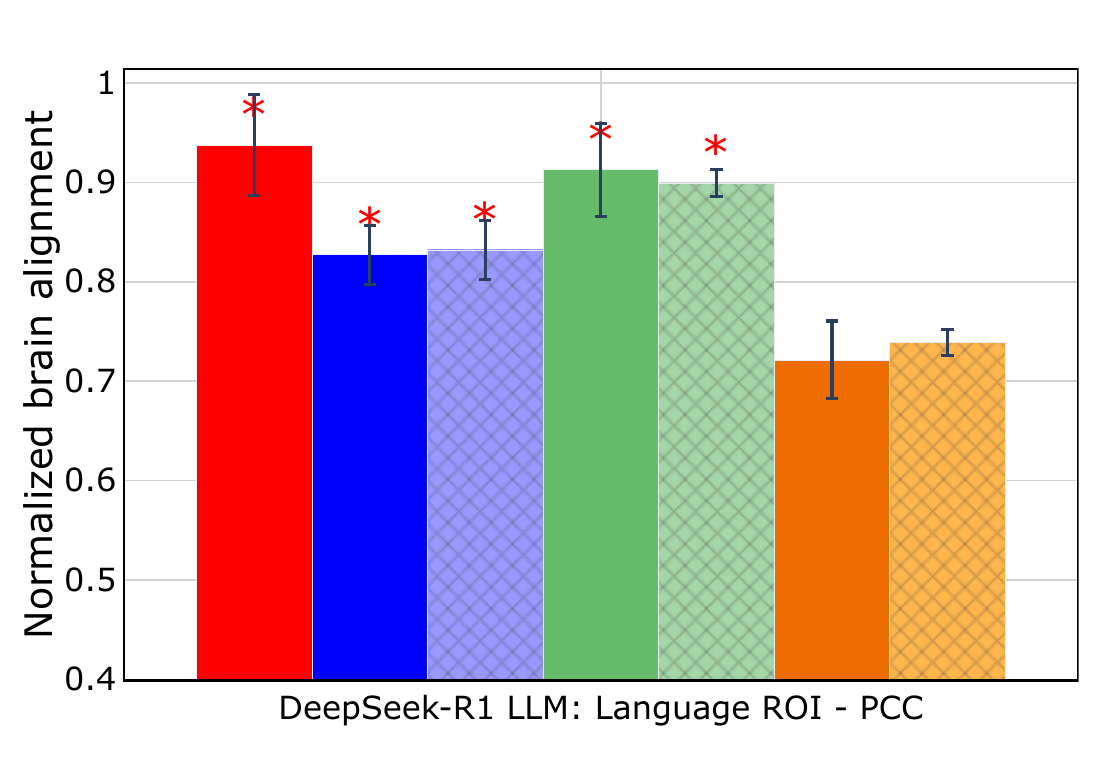}
    \includegraphics[width=0.33\linewidth]{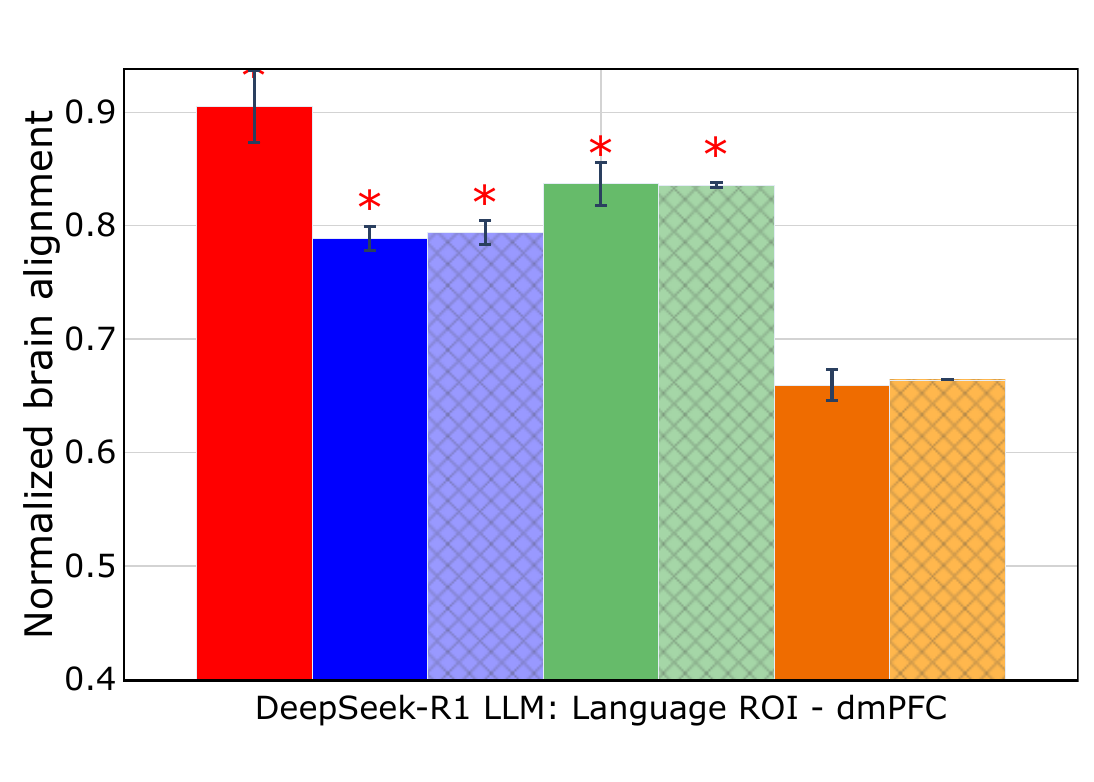}
    \includegraphics[width=0.33\linewidth]{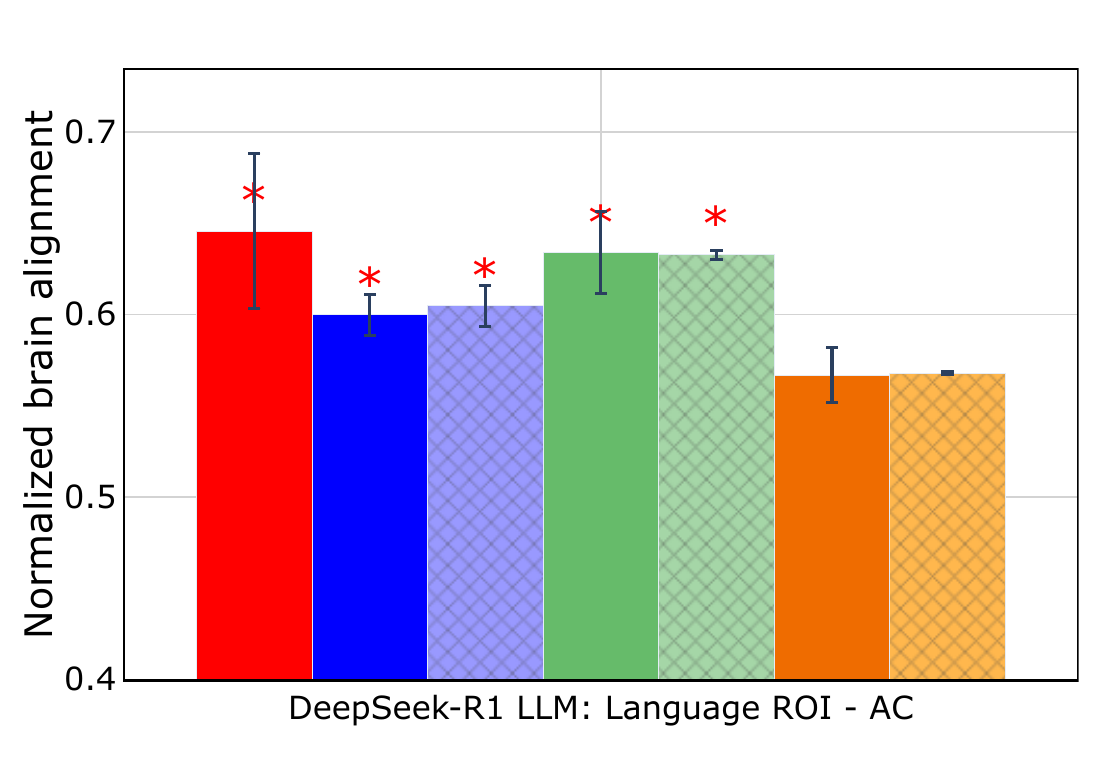}
    \caption{Normalized predictivity of DeepSeek-R1 SLMs and LLMs, including grouped comparisons
of the base and quantized variants.}
    \label{fig:deepseek_merged_quantized}
\end{figure*}

\section{Encoding Performance on Naturalistic Reading fMRI Dataset}
\label{app:reading_fmri}

\begin{figure}[!ht]
    \centering
    \includegraphics[width=\linewidth]{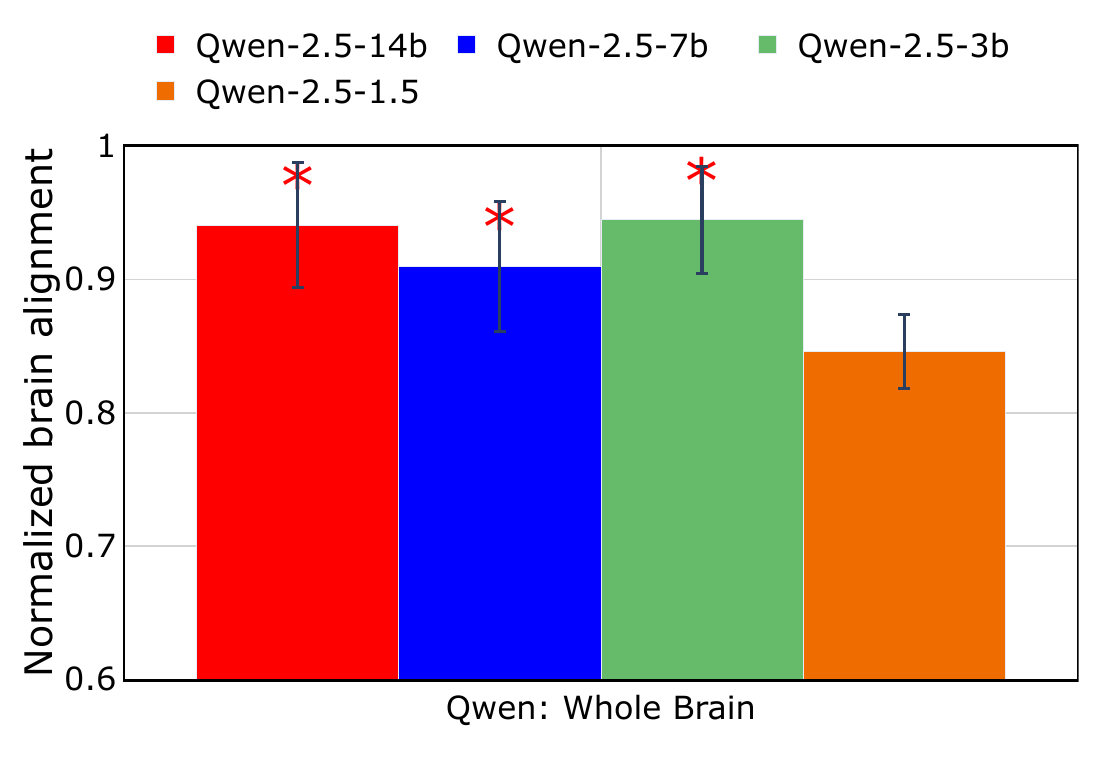}
    \caption{Reading brain dataset (Qwen-2.5): Normalized brain alignment was computed by averaging across participants, layers, and voxels. \textcolor{red}{Red: }14b, \textcolor{blue}{Blue: }7b, \textcolor{green}{Green: } 3b, \textcolor{orange}{Orange:} 1.5b, Solid: full-precision SLMs/LLMs. \textcolor{red}{*} at a particular bar indicates that the model's prediction performance is significantly better than 1.5b SLMs. The plot shows whole-brain normalized alignment.}
    \label{fig:reading_fmri}
\end{figure}

We have now extended our experiments to an additional dataset i.e. we performed voxelwise encoding on the Moth Radio Hour Reading fMRI dataset~\citep{deniz2019representation}, which contains the same nine subjects and large number of samples under a different linguistic paradigm (i.e. reading). This additional evaluation allows us to assess the generalizability of our findings across datasets and tasks. We use Qwen models (Qwen2.5-1.5b, Qwen2.5-3b, Qwen2.5-7b and Qwen2.5-14b) to validate the brain alignment to examine whether 3b SLMs maintain similar brain alignment to larger versions of the models. From Fig.~\ref{fig:reading_fmri}, we observe that 3B SLMs yield brain alignment comparable to the 7B and 14B Qwen2.5 models, whereas 1.5B SLMs exhibit a clear drop in brain alignment on the Reading Brain dataset.

\section{Decoder gap: Brain Decoding (Stimulus Reconstruction)}
\label{app:decoding_reconstruction}

We now perform brain decoding to reconstruct text stimuli from fMRI brain activity using LLaMA-3-8B, and two SLMS (LLaMA-3-3B and LLaMA-3-1B).

Inspired by BrainLLM (Ye et al. 2025), we perform end-to-end text stimulus reconstruction from fMRI brain activity. We follow the same BrainLLM methodology, where we use the same Moth Radio Hour dataset (11 stories) with the same train/test split, where ten stories are used for training and one held-out story is used for generation. Concretely, we train a brain-to-text decoder and report standard text-generation metrics-BLEU-1, WER, METEOR, and BERT-F1-for three models: LLaMA-3-8B, LLaMA-3.2-3B, and LLaMA-3.2-1B (Table~\ref{brain_decoding_metrics}). Across reconstructed segments per model on test dataset, LLaMA-3.2-3B achieves the best performance on all four metrics (BLEU-1 = 0.120, WER = 4.22, METEOR = 0.110, BERT-F1 = 0.825), slightly outperforming LLaMA-3.2-8B and clearly improving over the LLaMA-3.2-1B baseline (BLEU-1 = 0.070, METEOR = 0.055, BERT-F1 = 0.811). These BERT-F1 scores in the 0.81–0.83 range indicate that the decoded text reliably preserves the semantic content of the original stimulus, while BLEU-1 in the 0.07–0.12 range is in line with prior work where exact word-level recovery from fMRI is known to be challenging.

To make the reconstruction quality more interpretable, we also include qualitative examples comparing ground-truth text and decoded outputs (Table.~\ref{brain_decoding_examples}). These examples illustrate that the decoder often recovers the overall meaning, emotional tone, and discourse context, even when individual words differ-e.g., reconstructions that correctly express embarrassment, uncertainty, or interactions with children, despite not matching every token verbatim.

Overall, these new brain decoding experiments show that our SLMs are not only good encoders of brain activity, but also support meaningful decoding: they can reconstruct linguistically coherent text from fMRI with high semantic fidelity and reasonable word-level accuracy. We emphasize that decoding is not the primary focus of the present work, but the added section demonstrates that the brain-aligned representations we study are indeed rich enough to support stimulus reconstruction.


\begin{table*}[!ht]
\centering
\scriptsize
\caption{Example fMRI-based stimulus reconstructions for LLaMA-3-8B, LLaMA-3.2-3B, and LLaMA-3.2-1B.}
\label{brain_decoding_examples}
\begin{tabularx}{\linewidth}{|l| l| X| l|}
\toprule
Example & Variant & Text & Metrics \\
\hline

\multirow{4}{*}{1} 
  & Ground Truth 
  & \textit{little sad leaving here about} 
  & -- \\[0.25em]
  & LLaMA-3-8B
  & huge fan of the show so i'm really excited to be able to share this with you guys and i hope you'll enjoy it as much as i enjoyed watching 
  & BLEU-1 = 0.0000, WER = 6.0000, METEOR = 0.0000 \\[0.25em]
  & LLaMA-3.2-3B 
  & in this beautiful city. I started to walk in this beautiful in this beautiful city. I started to walk in this beautiful in this beautiful city. I started to 
  & BLEU-1 = 0.0000, WER = 5.8000, METEOR = 0.0000 \\[0.25em]
  & LLaMA-3.2-1B 
  & , you know what i'm talking about. You know what i'm talking about. You know what i'm talking about. You know what i'm talking about 
  & BLEU-1 = 0.0400, WER = 4.8000, METEOR = 0.0714 \\[0.5em]

\hline

\multirow{4}{*}{2} 
  & Ground Truth 
  & \textit{and that i had to hold onto him} 
  & -- \\[0.25em]
  & LLaMA-3-8B
  & i was going to have to take him to the hospital but i had no idea what to do with him so i went to a friend of mine and he told 
  & BLEU-1 = 0.3750, WER = 3.5000, METEOR = 0.2885 \\[0.25em]
  & LLaMA-3.2-3B 
  & , she's crying and then she sees me and she com, she's crying and then she sees me and she com, she's crying and then she sees 
  & BLEU-1 = 0.1852, WER = 3.2500, METEOR = 0.0505 \\[0.25em]
  & LLaMA-3.2-1B 
  & could see her face and she was smiling at me and then she said ``hi'' and i said ``hi'' back to her and then she turned around and 
  & BLEU-1 = 0.2500, WER = 3.1250, METEOR = 0.1500 \\[0.5em]

\hline

\multirow{4}{*}{3} 
  & Ground Truth 
  & \textit{to be a teacher with all these kids} 
  & -- \\[0.25em]
  & LLaMA-3-8B
  & for them to see their children i had no idea how hard it was for them to see their children i had no idea how hard it was for them to see 
  & BLEU-1 = 0.0938, WER = 3.8750, METEOR = 0.0962 \\[0.25em]
  & LLaMA-3.2-3B 
  & is a great place to live, work, and play. It's also home to some of the best restaurants, bars, and coffee shops in the city. 
  & BLEU-1 = 0.1200, WER = 3.0000, METEOR = 0.2635 \\[0.25em]
  & LLaMA-3.2-1B 
  & and i said you know like and i said you know like and i said you know like and i said you know like and i said you know like and i 
  & BLEU-1 = 0.0000, WER = 4.0000, METEOR = 0.0000 \\

\hline
\end{tabularx}
\end{table*}

\section{Effect of Pruning}
\label{app:unstructured_pruning}

We have now included unstructured pruning which is equivalent to quantization and now report preliminary results for Qwen-2.5 models. In particular, we perform unstructured magnitude pruning on the linear layers of Qwen2.5-3B and Qwen2.5-1.5B, at sparsity levels 0.1, 0.25, and 0.5. For Qwen2.5-3B, Table~\ref{pruning_effect_qwen2.5_3b} summarizes brain alignment (mean $\pm$ s.e.m. across subjects) for the base, quantized variants, and pruned models, showing that AWQ and SmoothQuant slightly improve over the FP16 baseline (0.933 $\pm$ 0.035 and 0.930 $\pm$ 0.035 vs. 0.924 $\pm$ 0.033), GPTQ is modestly lower (0.910 ± 0.037), and unstructured pruning up to 50\% keeps alignment in a narrow range (0.910–0.907 with s.e.m. 0.032–0.043).

These results suggest that, for Qwen2.5-3B, moderate unstructured pruning (10–25\%) preserves brain alignment at a level comparable to quantized or full-precision models, with little change in SER, while aggressive pruning (50\%) begins to degrade linguistic competence (SER increases) despite only a small drop in alignment. We view these pruning experiments as complementary to our main quantization results: they show that both post-training quantization and unstructured pruning can preserve brain alignment to a surprising degree, but they also highlight potential trade-offs with linguistic competence at high sparsity.

We acknowledge that we still do not systematically explore structured compression across model components, where different methods could be compared under matched compression ratios to assess their differential impact on both linguistic competence and brain alignment. We now explicitly note that this is an important direction for future work, but falls outside the scope of the current paper given the substantial additional experiments it would require.

Although our model suite includes the DeepSeek-R1-Distill family (which is itself a product of knowledge distillation), in this work we do not systematically study distillation as a compression method. We treat DeepSeek as an additional, pretrained model family for evaluation, and focus our controlled compression experiments on post-training quantization (and preliminary unstructured pruning). A careful comparison of different distillation strategies under matched compression ratios is therefore an important direction for future work.


\begin{table}[!ht]
\centering
\scriptsize
\caption{Comparison of quantization and pruning for Qwen2.5-1.5B (mean $\pm$ SEM across subjects).}
\label{pruning_effect_qwen2.5_1.5b}
\begin{tabular}{|l|c|c|c|}
\hline
Variant              & Mean $\pm$ SEM      & 95\% CI             & Notes \\
\hline
FP16 (baseline)      & 0.830 $\pm$ 0.025   & [0.781, 0.879]      & full precision \\
AWQ                  & 0.854 $\pm$ 0.031   & [0.793, 0.915]      & INT4 quantization \\
GPTQ                 & 0.828 $\pm$ 0.026   & [0.777, 0.879]      & INT4 quantization \\
SmoothQuant          & 0.836 $\pm$ 0.025   & [0.787, 0.885]      & INT4 quantization \\
Prune 10\%           & 0.847 $\pm$ 0.026   & [0.796, 0.898]      & unstructured pruning \\
Prune 25\%           & 0.824 $\pm$ 0.025   & [0.775, 0.873]      & unstructured pruning \\
Prune 50\%           & 0.608 $\pm$ 0.053   & [0.504, 0.712]      & unstructured pruning \\
\hline
\end{tabular}
\end{table}

\section{Model Size vs. Brain Alignment}
\label{app:model_size_brain_alignment}

\begin{figure*}[!ht]
    \centering
    \includegraphics[width=0.9\linewidth]{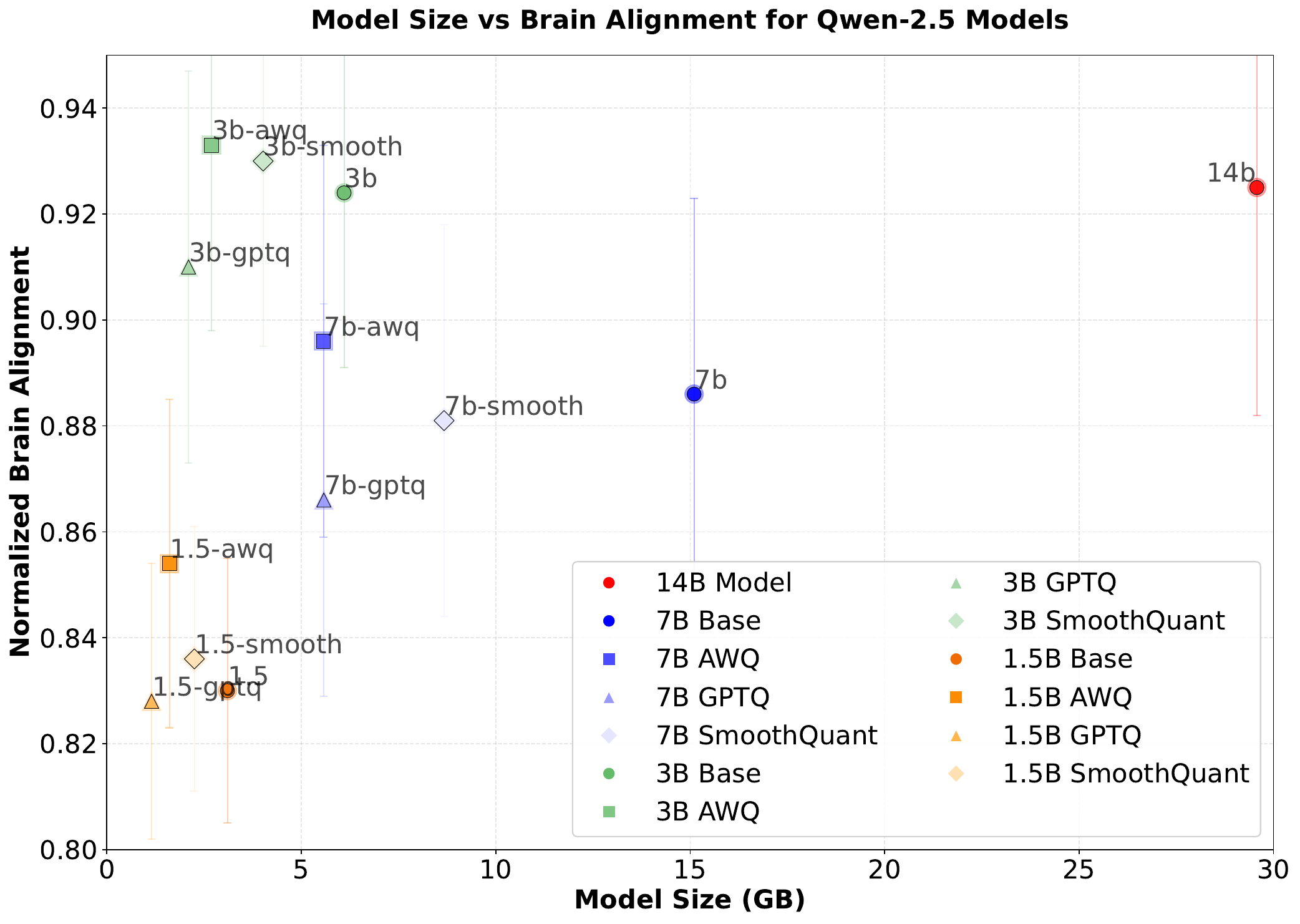}
    \caption{Qwen2.5: Plot of Model Size (x-axis) vs. Normalized Brain Alignment (y-axis).}
    \label{fig:qwen_model_alignment}
\end{figure*}

\begin{figure*}[!ht]
    \centering
    \includegraphics[width=0.9\linewidth]{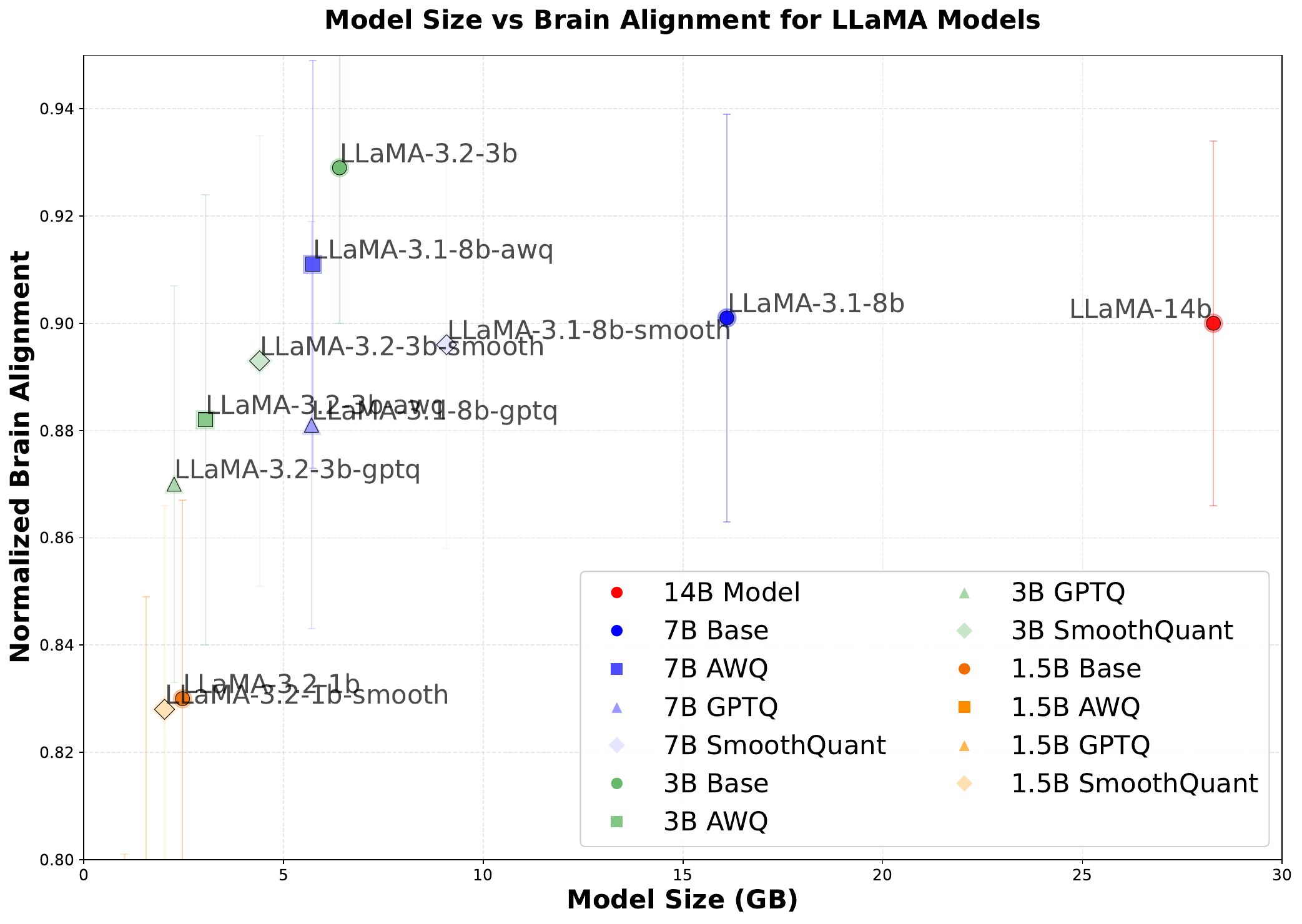}
    \caption{LLaMA-3: Plot of Model Size (x-axis) vs. Normalized Brain Alignment (y-axis).}
    \label{fig:llama_model_alignment}
\end{figure*}

\begin{figure*}[!ht]
    \centering
    \includegraphics[width=\linewidth]{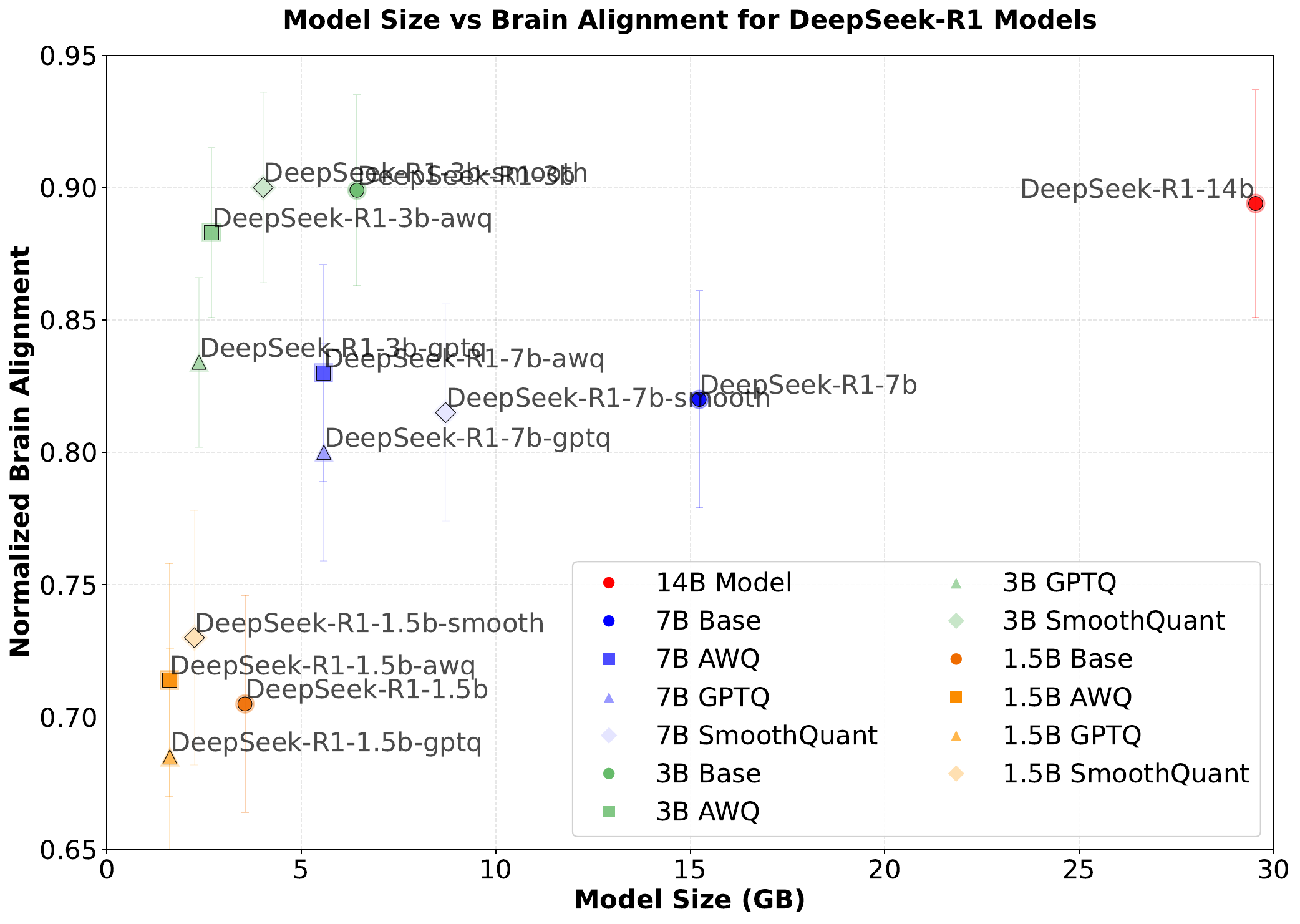}
    \caption{DeepSeek-R1: Plot of Model Size (x-axis) vs. Normalized Brain Alignment (y-axis).}
    \label{fig:deepseek_model_alignment}
\end{figure*}

Figs.~\ref{fig:qwen_model_alignment},~\ref{fig:llama_model_alignment}, and~\ref{fig:deepseek_model_alignment} plot Model size (GB) vs. Average Normalized brain alignment for Qwen2.5, LLaMA-3, and DeepSeek-R1 models (1.5B/3B/7B/14B) and their AWQ, GPTQ, and SmoothQuant variants. Across all three families, the 3B SLMs and their AWQ/SmoothQuant variants generally lie slightly above or very close to their FP16 counterparts at substantially reduced size, whereas GPTQ variants tend to fall slightly below the FP16 models despite achieving stronger compression.

\end{document}